\documentclass[longauth]{aa}  
\usepackage{graphicx}
\usepackage{amsmath}
\usepackage{wasysym}
\usepackage{mathtools} 
\usepackage{booktabs} 
 \usepackage{rotating,multirow} 
\usepackage[UKenglish]{isodate}
\cleanlookdateon 

\usepackage{graphicx}

\usepackage{natbib,twoopt}
\bibpunct{(}{)}{;}{a}{}{,} 
\makeatletter

\usepackage{color}
\usepackage{soul}

\def\kms{km\,s$^{-1}$}

\def\teff{$T_{\rm eff}$}
\def\logg{$\log g$}
\def\feh{[Fe/H]}

\def\vsini{$v\sin i$}
\def\halpha{${\rm H}\alpha$}
\def\hbeta{${\rm H}\beta$}
\def\wha{$W({\rm H}\alpha$)}
\def\wli{$W({\rm Li})$}
\def\ali{${\log \epsilon({\rm Li})}$}
\def\abund{${\log \epsilon({\rm X})}$}
\def\ha10{${\rm H}\alpha\,10\%$}
\def\mdot{$\dot{M}_{\rm acc}$}

\def\lif{Li\,(6707.84\,\AA)}
\def\li{Li}
\def\hachr{$\Delta W({\rm H}\alpha$)$_{\rm chr}$}
\def\hbchr{$\Delta W({\rm H}\beta$)$_{\rm chr}$}
\def\fachr{$F({\rm H}\alpha$)$_{\rm chr}$}
\def\fbchr{$F({\rm H}\beta$)$_{\rm chr}$}
\newcommand\abs[1]{\left|#1\right|}

\begin{document} 
\title{Gaia-ESO Survey: The analysis of pre-main sequence stellar spectra.}
\titlerunning{Gaia-ESO Survey: The analysis of pre-main sequence stellar spectra.}

\author{
A.~C. Lanzafame\inst{1,2} \and 
\\
A. Frasca\inst{2}         \and
F. Damiani\inst{3}                \and
E. Franciosini\inst{4}            \and
M. Cottaar\inst{5}                \and
S.~G. Sousa\inst{6,7}                  \and
H.~M. Tabernero\inst{8}  \and 
\\
A. Klutsch\inst{2}                \and
L. Spina\inst{4}                  \and
K. Biazzo\inst{2}         \and
L. Prisinzano\inst{3}             \and
G.~G. Sacco\inst{4}           \and
S. Randich\inst{4}           \and     
E. Brugaletta\inst{1}     \and
E. Delgado Mena\inst{6}                \and
V. Adibekyan\inst{6}              \and
D. Montes\inst{8}                 \and
R. Bonito\inst{9,3}                 \and
J.~F. Gameiro\inst{6}             \and
\\
J.~M. Alcal{\'a}\inst{10}                \and
J.~I. Gonz{\'a}lez Hern{\'a}ndez\inst{11,25}  \and 
R. Jeffries\inst{12}               \and
S. Messina\inst{2}        \and
M. Meyer\inst{5}                  \and 
\\
G. Gilmore\inst{13}    \and            
M. Asplund\inst{14}           \and     
J. Binney\inst{15}             \and    
P. Bonifacio\inst{16}          \and    
J.~E. Drew\inst{17}            \and    
S. Feltzing\inst{18}           \and    
A.~M.~N. Ferguson\inst{19}     \and    
G. Micela\inst{3}             \and    
I. Negueruela\inst{20}         \and    
T. Prusti\inst{21}             \and    
H-W. Rix\inst{22}              \and    
A. Vallenari\inst{23}             \and 
E.~J. Alfaro\inst{24}            \and  
C. Allende Prieto\inst{11,25}       \and  
C. Babusiaux\inst{16}        \and      
T. Bensby\inst{18}             \and    
R. Blomme\inst{26}             \and    
A. Bragaglia\inst{27}          \and    
E. Flaccomio\inst{3}         \and     
P. Francois\inst{16}          \and     
N. Hambly\inst{19}           \and      
M. Irwin\inst{13}            \and      
S.~E. Koposov\inst{13,28}      \and       
A.~J. Korn\inst{29}         \and       
R. Smiljanic\inst{31}         \and     
S. Van Eck\inst{32}          \and      
N. Walton\inst{13}          \and             
A. Bayo\inst{24,34}            \and       
M. Bergemann\inst{13}       \and       
G. Carraro\inst{35}         \and  
M.~T. Costado\inst{24}               \and     
B. Edvardsson\inst{29}       \and      
U. Heiter\inst{29}           \and      
V. Hill\inst{30}            \and       
A. Hourihane\inst{13}       \and       
R.~J. Jackson\inst{12}      \and      
P. Jofr\'e\inst{13}         \and       
C. Lardo\inst{27}            \and      
J. Lewis\inst{13}            \and         
K. Lind\inst{13}             \and            
L. Magrini\inst{4}          \and      
G. Marconi\inst{35}         \and       
C. Martayan\inst{35}        \and       
T. Masseron\inst{13}        \and       
L. Monaco\inst{35}           \and      
L. Morbidelli\inst{4}       \and      
L. Sbordone\inst{33}         \and              
C.~C. Worley\inst{13}              \and
S. Zaggia\inst{23}             
}

\authorrunning{A.~C. Lanzafame et al.}

\offprints{A.~C. Lanzafame \\ \email{a.lanzafame@unict.it}}

\institute{
Universit\`a di Catania, Dipartimento di Fisica e Astronomia, Sezione Astrofisica, Via S. Sofia 78, I-95123 Catania, Italy \\
 \email{Alessandro.Lanzafame@oact.inaf.it}
\and
INAF-Osservatorio Astrofisico di Catania, Via S. Sofia 78, I-95123 Catania, Italy
\and
INAF - Osservatorio Astronomico di Palermo, Piazza del Parlamento 1, 90134, Palermo, Italy
\and
INAF - Osservatorio Astrofisico di Arcetri, Largo E. Fermi 5, 50125, Florence, Italy
\and
Institute for Astronomy, ETH Zurich, Wolfgang-Pauli-Strasse 27, 8093, Zurich, Switzerland
\and
Centro de Astrof\'isica, Universidade do Porto, Rua das Estrelas, 4150-762 Porto, Portugal Departamento de F\'isica e Astronomia, Faculdade de Ci\^encias, Universidade do Porto, Rua do Campo Alegre, 4169-007 Porto, Portugal
\and
Departamento de F\'isica e Astronomia, Faculdade de Ci\^encias, Universidade do Porto, Rua do Campo Alegre, 4169-007 Porto, Portuga
\and
Universidad Complutense de Madrid, Departamento de Astrof\'isica, E-28040 Madrid, Spain
\and
Universit\`a di Palermo, Dipartimento di Fisica e Chimica, Viale delle Scienze, Ed. 17, I-90128 Palermo, Italy
\and
INAF - Osservatorio Astronomico di Capodimonte, via Moiariello 16, 80131, Napoli, Italy
\and
Instituto de Astrof\'{\i}sica de Canarias, E-38205 La Laguna, Tenerife, Spain
\and
Astrophysics Group, Research Institute for the Environment, Physical Sciences and Applied Mathematics, Keele University, Keele, Staffordshire ST5 5BG, United Kingdom   
\and
Institute of Astronomy, University of Cambridge, Madingley Road, Cambridge CB3 0HA, United Kingdom
\and
Research School of Astronomy \& Astrophysics, Australian National University, Cotter Road, Weston Creek, ACT 2611, Australia
\and
Rudolf Peierls Centre for Theoretical Physics, Keble Road, Oxford, OX1 3NP, United Kingdom
\and
GEPI, Observatoire de Paris, CNRS, Universit\'e Paris Diderot, 5 Place Jules Janssen, 92190 Meudon, France
\and
Centre for Astrophysics Research, STRI, University of Hertfordshire, College Lane Campus, Hatfield AL10 9AB, United Kingdom
\and
Lund Observatory, Department of Astronomy and Theoretical Physics, Box 43, SE-221 00 Lund, Sweden
\and
Institute of Astronomy, University of Edinburgh, Blackford Hill, Edinburgh EH9 3HJ, United Kingdom
\and
Departamento de F\'{i}sica, Ingenier\'{i}a de Sistemas y Teor\'{i}a de la Se$\tilde{\rm n}$al, Universidad de Alicante, Apdo. 99, 03080, Alicante, Spain
\and
ESA, ESTEC, Keplerlaan 1, Po Box 299 2200 AG Noordwijk, The Netherlands
\and
Max-Planck Institut f\"{u}r Astronomie, K\"{o}nigstuhl 17, 69117 Heidelberg, Germany
\and
INAF - Padova Observatory, Vicolo dell'Osservatorio 5, 35122 Padova, Italy
\and
Instituto de Astrof\'{i}sica de Andaluc\'{i}a-CSIC, Apdo. 3004, 18080, Granada, Spain
\and
Universidad de La Laguna, Dept. Astrof\'{\i}sica, E-38206 La Laguna, Tenerife, Spain
\and
Royal Observatory of Belgium, Ringlaan 3, 1180, Brussels, Belgium
\and
INAF - Osservatorio Astronomico di Bologna, via Ranzani 1, 40127, Bologna, Italy
\and
Moscow MV Lomonosov State University, Sternberg Astronomical Institute, Moscow 119992, Russia
\and
Department of Physics and Astronomy, Uppsala University, Box 516, SE-75120 Uppsala, Sweden
\and
Laboratoire Lagrange (UMR7293), Universit\'e de Nice Sophia Antipolis, CNRS,Observatoire de la C\^ote d'Azur, CS 34229,F-06304 Nice cedex 4, France
\and
Department for Astrophysics, Nicolaus Copernicus Astronomical Center, ul. Rabia\'{n}ska 8, 87-100 Toru\'{n}, Poland
\and
Institut d'Astronomie et d'Astrophysique, Universit\'{e} libre de Brussels, Boulevard du Triomphe, 1050 Brussels, Belgium
\and
European Southern Observatory, Karl-Schwarzschild-Str. 2, 85748 Garching bei M\"unchen, Germany
\and
Instituto de F\'isica y Astronomi\'ia, Universidad de Valparai\'iso, Chile
\and
European Southern Observatory, Alonso de Cordova 3107 Vitacura, Santiago de Chile, Chile
}

\date{Received 6 August 2014 / Accepted 14 January 2015}

\abstract{
The Gaia-ESO Public Spectroscopic Survey is obtaining high quality spectroscopy of some 100,000 Milky Way stars using the FLAMES spectrograph at the VLT.
}{
This paper describes the analysis of UVES and GIRAFFE spectra acquired in the fields of young clusters whose population includes pre-main sequence (PMS) stars.
}{
Both methods that have been extensively used in the past and new ones developed in the contest of the Gaia-ESO survey enterprise are available and used.
The internal precision of these quantities is estimated by inter-comparing the results obtained by such different methods, while the accuracy is estimated by comparison with independent external data, like effective temperature and surface gravity derived from angular diameter measurements, on a sample of benchmarks stars.
A validation procedure based on such comparisons is applied to discard spurious or doubtful results and produce {\it recommended} parameters.
Specific strategies are implemented to deal with fast rotation, accretion signatures, chromospheric activity, and veiling.
}{
The analysis carried out on spectra acquired in young clusters' fields during the first 18 months of observations, up to June 2013, is presented in preparation of the first release of advanced data products.
These include targets in the fields of the
\object{$\rho$\,Oph}, 
\object{Cha\,I}, 
\object{NGC2264}, 
\object{$\gamma$ Vel}, 
and \object{NGC2547} clusters. 
Stellar parameters obtained with the higher resolution and larger wavelength coverage from UVES 
are reproduced with comparable accuracy and precision using the smaller wavelength range and lower resolution of the GIRAFFE setup adopted for young stars, which allows us to provide with confidence stellar parameters for the much larger GIRAFFE sample.
Precisions are estimated to be $\approx$ 120\,K r.m.s. in \teff, $\approx$0.3 dex r.m.s. in \logg, and $\approx$0.15 dex r.m.s. in \feh, for both the UVES and GIRAFFE setups.
}
{}

\keywords{open clusters and associations: general -- surveys -- methods: data analysis -- stars: pre-main sequence -- stars: fundamental parameters -- open clusters and associations: individual: 
\object{$\rho$\,Oph}, 
\object{Cha\,I}, 
\object{NGC2264}, 
\object{$\gamma$ Vel}, 
\object{NGC2547} 
\vspace{-3mm}}

\maketitle


\section{Introduction}
\label{sec:Introduction}

Spectrum analyses of pre-main sequence stars (PMS) require special techniques, notably for dealing with peculiarities of cool, low-mass members of young clusters.
Optical spectra of such stars may include the presence of veiling, large broadening due to fast rotation, emission lines due to accretion and/or chromospheric activity, and molecular bands.
The subtraction of inhomogeneous and variable nebular emission may also be problematic and some residual features can remain in spectra of some young clusters' members after the sky-background removal.

One of the main objectives of the Gaia-ESO Survey is to provide radial velocities (RV) with a precision $\approx$ 0.2 - 0.25 \kms\ for stars in young open clusters, to complement Gaia proper motions with comparable accuracy for a statistically significant sample \citep{2012Msngr.147...25G,2013Msngr.154...47R} reaching also fainter targets.
This survey also complements Gaia by deriving metallicity and detailed abundances for several elements, including lithium, which is particularly relevant in the studies of the evolution of low-mass stars and in the determination of clusters' age. 
This requires a derivation of all fundamental parameters (effective temperatures \teff, metallicity \feh, surface gravity \logg, and projected rotational velocity \vsini) independently of the Gaia results.

The \halpha\ profile of such young low-mass stars bears information on their chromospheric activity, accretion rate, and mass loss.
Because of their common origin, strong accretion is expected to be correlated with veiling; this can be used for both checking our results, as no correlation would be indicative of large uncertainties, and exploring the extent and details of such a correlation.
Chromospheric activity is known to depend on stellar rotation and both evolve in time; the Gaia-ESO is also going to provide the possibility of exploring the activity-rotation relation and their evolution on a large sample of young stars.

The Gaia-ESO target selection aims at producing unbiased catalogues of stars in open clusters.
Selection criteria based mainly on photometry, supported, when possible, by kinematic memberships, have been adopted for this purpose, although this implies that a large number of non-members are also observed, which are identified {\it a posteriori} from the radial velocity measurements (Bragaglia et al., in prep.). 
In our case, the GIRAFFE targets are late-type (F to early-M) stars in the magnitude range 12$\le$V$\le$19 mag, in the PMS or main sequence (MS) phase. 
Based on available information, the selection of UVES targets tries to include only slowly rotating (\vsini$<$ 15 \kms) single G--K stars in the magnitude range 9$<$V$<$15 without or with weak accretion (\mdot$<$ 10$^{-10}$ M$_{\odot}$ yr$^{-1}$).
To optimise the throughput of the survey, observations of cool stars in the fields of young open clusters are only carried out in the  GIRAFFE/HR15N setup (R=17\,000, $\lambda$ from 6470 to 6790\,\AA) and the Red 580 UVES setup (R=47\,000 centred at $\lambda=5800$\,\AA\ with a spectral band of 2000\,\AA).
The Medusa mode of the fibre fed system is used throughout the survey, 
allowing the simultaneous allocation of 132 and 8 fibres feeding GIRAFFE and UVES, respectively,
with about 20 (GIRAFFE) and 1 (UVES) fibres used to observe the sky background spectrum.

The GIRAFFE/HR15N setup covers both H$\alpha$ and \lif\ lines, and it is therefore particularly useful for the study of young stars.
However, \teff, \logg, and \feh\ diagnostics in this wavelength range are poorer than in other settings and still not satisfactorily reproduced by theoretical models.
For example, the paucity of \element{Fe} lines in the HR15N spectral range makes it difficult to derive both \logg\ and \feh\ in G-type stars from the analysis of the equivalent widths of \ion{Fe}{I} and \ion{Fe}{II} lines.

This paper presents the analysis of the Gaia-ESO spectra in the fields of young open clusters (age $<$ 100 Myr) and is one of a series presenting a description of the Gaia-ESO survey in preparation of its first release of advanced data products.
The Gaia-ESO scientific goals, observations strategies, team organisation, target selection strategy, data release schedule, data reduction, analysis of OBA-type and FGK-type stars not in the fields of young open clusters, non-standard objects and outliers, external calibration, and the survey-wide homogenisation process are discussed in other papers of this series.

The paper is organised as follows.
In Sect.\,\ref{sec:data} the data analysed in the first two Gaia-ESO internal data releases are presented.
In Sect.\,\ref{sec:GeneralStrategy} the principles and general strategies of the Gaia-ESO PMS analysis are outlined. 
Methods and validation for the initial raw measurements, fundamental parameters ($T_{\rm eff}, \log g$, [Fe/H], micro-turbulence velocity, veiling, and $v\sin i$), and derived parameters (chromospheric activity, accretion rate, and elemental abundances) are presented in Sects.\,\ref{sec:RawMeasurements}, \ref{sec:FundamentalParameters}, and \ref{sec:DerivedParameters}.
The conclusions are in Sect.\,\ref{sec:Conclusions}.

\section{Data}
\label{sec:data}

\begin{table*}[ht]
\centering
\caption{
Young open clusters (age $<$ 100 Myr) observed by the Gaia-ESO survey in the first 18 months of observations, whose analysis is discussed in this paper. 
The cluster NGC6705 (M11) has also been included for validation and comparison across the survey.
}
\begin{tabular}{rccrrrrr}
\hline

Cluster        &   Approximate Age   &  Distance  &  \multicolumn{2}{c}{GIRAFFE} & \multicolumn{2}{c}{UVES} \\
               &  (Myr)       &  (pc)  &  All  & WTTS/CTTS  &  All & WTTS/CTTS  & iDR \\
\hline
$\rho$ Oph     &   1          &   120   &   200 &  30  &  23  &   5 & 2 \\
Cha\,I         &   2          &   160   &   674 &  93  &  49  &  14 & 1, 2\\
NGC2264        &   3          &   760   &  1706 & 446  & 118  &  23 & 2\\
$\gamma$ Vel   & $\sim$ 5--10 &   350   &  1242 & 200  &  80  &   2 & 1, 2\\
NGC2547        &   35         &   361   &   450 &  44  &  26  &   1 & 2\\
NGC6705        &   250        &  1877   &  1028 &   0  &  49  &   0 & 2\\
\hline
\end{tabular}
\label{tab:targets}
\end{table*}

The survey's analysis is performed in cycles, following the data reduction of newly observed spectra.
Each new analysis cycle improves upon the last one with updated input data (e.g. atomic and molecular data), improved analysis methods, and improved criteria to define the final recommended parameters.
At the end of each cycle an internal data release (iDR) is produced and made available within the Gaia-ESO consortium for scientific validation.
The description of the methods and recommended parameters criteria given in this paper applies to the analysis of the first 2 years of observations, which will form the basis of the first release of advanced data products to ESO.
The validation procedures presented in this paper consider the first 18 months of observations (iDR1 and iDR2).

The young open clusters observed in the first 18 months of observations are listed in Table\,\ref{tab:targets}, along with the total number of observed stars for each cluster and the number of stars identified as T\,Tauri from the properties of \halpha\ emission, spectral type, and \li\ absorption (see Sect.\,\ref{sec:halpha}). 
A total of 813 and 45 T\,Tauri stars have been identified in the GIRAFFE and UVES spectra, respectively.
The memberships of these young clusters, including stars not clearly showing the T\,Tauri distinctive features, will be discussed in other Gaia-ESO science verification papers \citep[e.g.,][]{2014A&A...563A..94J}.
The 200 Myr cluster NGC6705 (M11) has been observed in different setups \citep[see e.g.][]{2014A&A...569A..17C} to allow inter-comparison and validation of the analysis methods across the survey and, for this purpose, is included in our analysis.

\section{General analysis strategy}
\label{sec:GeneralStrategy}

The Gaia-ESO consortium is structured in several working groups (WGs).
The analysis of PMS stars is carried out by the WG12, to which six nodes contributed: 
INAF--Osservatorio Astrofisico di Arcetri, Centro de Astrofisica de Universidade do Porto (CAUP), Universit\`a di Catania and INAF--Osservatorio Astrofisico di Catania (OACT), INAF--Osservatorio Astronomico di Palermo (OAPA), Universidad Complutense de Madrid (UCM), and Eidgen\"ossische Technische Hochschule Z\"urich (ETH).

The main input to the Gaia-ESO PMS spectrum analysis consists of UVES and GIRAFFE spectra of cool stars in the field of young open clusters. 
The preliminary selection criteria are briefly outlined in Sect.\,\ref{sec:Introduction} and will be detailed in one of the papers of this series (Bragaglia et al., in prep.). 

The data reduction is performed as described in \cite{Sacco_etal:2014} for the UVES spectra, and Lewis et al. (in prep.) for the GIRAFFE spectra.
These are put on a wavelength scale and shifted to a barycentric reference frame.
Sky-background subtraction, as well as a normalisation to the continuum, is also performed in the data reduction process.
Multi-epoch spectra of the same source are combined in the {\it co-added} spectrum.
Quality information is provided including variance spectra, S/N, non-usable pixels, etc.
Additional inputs are the radial and rotational velocities, as described in Gilmore et al. (in prep.)
and \cite{Sacco_etal:2014}, and photometric data.
Clusters' distance and reddening are also considered as input to the spectrum analysis validation.

Double-lined spectroscopic binaries (SB2) and multiple systems are identified by looking at the shape of the cross-correlation function.
These stars are excluded from the current analysis and a multiplicity flag is reported in the final database. 

To ensure the highest homogeneity as possible in the quantities derived, all the different Gaia-ESO spectrum analysis methods adopt the same atomic and molecular data (Heiter et al., in prep.), as well as the same set of model atmospheres \citep[MARCS, ][]{2008A&A...486..951G}.

The output parameters of the Gaia-ESO PMS spectrum analysis are listed in Table\,\ref{tab:PMS-parameters}.
To apply a detailed quality control on the output parameters and optimise the analysis according to the star's characteristics, these are divided into three groups: {\em raw}, {\em fundamental}, and {\em derived}.
Raw parameters are the \halpha\ emission and \li\ equivalent widths (\wha\ and \wli), and the H$\alpha$ width at 10\% of the line peak \citep[\ha10, see, e.g.,][]{2004A&A...424..603N}. 
These are directly measured on the input spectra and do not require any prior information.
Their measurement is carried out before any other procedure to identify PMS stars  and their values are used for optimising the evaluation of the {\it fundamental} parameters in one of the methods used (see Sect.\,\ref{sec:FundamentalParameters}).
Besides \teff, \logg, and \feh\footnote{The solar Fe abundance of \cite{2007SSRv..130..105G}, $\log\epsilon({\rm Fe})_{\odot} = 7.45$, is adopted.}, the fundamental parameters derived include also micro-turbulence velocity ($\xi$), projected rotational velocity (\vsini), veiling \citep[$r$, see, e.g., ][]{1988BAAS...20R1092H}, and a gravity-sensitive spectral index \citep[$\gamma$, see][]{Damiani_etal:2014}.
Finally, the {\it derived} parameters are those whose derivation requires prior knowledge of the {\it fundamental} parameters, i.e. elemental abundances ($\log\epsilon({\rm X})$\footnote{$\log\epsilon({\rm X})=\log[N({\rm X})/N({\rm H})]+12$, i.e. a logarithmic abundance by number on a scale where the number of hydrogen atoms is 10$^{12}$.}), mass accretion rate (\mdot), chromospheric activity indices (\hachr\ and \hbchr), and chromospheric line fluxes (\fachr\ and \fbchr).

\begin{table*}[ht]
\centering
\caption{Gaia-ESO PMS analysis output parameters. Columns 2--13 list the number of stars in each cluster for which the parameter was derived from GIRAFFE (G) and UVES (U) spectra separately in iDR2. Lithium parameters and \vsini\ counts include upper limit estimates. Accretion and chromospheric activity parameter counts include only non negligible values. For the elemental abundances, the maximum number of derived values for each star/element is reported. See text for an explanation of the notation used.}
\label{tab:PMS-parameters}
\begin{tabular}{lrrrrrrrrrrrr}
\hline
Parameter      & \multicolumn{2}{c}{$\rho$ Oph} & \multicolumn{2}{c}{Cha\,I} & \multicolumn{2}{c}{NGC2264} & \multicolumn{2}{c}{$\gamma$ Vel} & \multicolumn{2}{c}{NGC2547} & \multicolumn{2}{c}{NGC6705} \\
\hline
 & G & U & G & U & G & U & G & U & G & U & G & U\\
\hline
\hline
\multicolumn{13}{c}{raw} \\
\hline
\wha                &  25&5  &  87&14 &  387&24  &  203&2  & 106&1  &   0&0  \\
\wli                & 189&23 & 633&47 & 1610&114 & 1186&75 & 404&25 & 708&48 \\
\ha10               &  33&5  & 103&14 &  807&23  &  264&2  & 239&1  &   0&0  \\
\hline
\multicolumn{13}{c}{fundamental} \\
\hline
\teff               & 170&21 & 572&39 & 1324&70  & 1104&51 & 361&24 & 394&32 \\
\logg               & 170&21 & 156&39 &  226&70  &  350&51 & 106&24 & 150&32 \\
$\gamma$            & 156& \dots  & 508& \dots  & 1199& \dots   & 1043& \dots  & 337& \dots  &   0&\dots  \\
\feh                & 170&21 & 515&39 & 1203&70  & 1018&51 & 311&24 & 360&32 \\
$\xi$               &  \dots & 14 &   \dots&23 &    \dots&42  &    \dots&46 &   \dots&15 &   \dots&30 \\
\vsini              & 154&23 & 521&42 & 1192&83  & 1004&75 & 332&25 & 107&33 \\
$r$                 &   4&3  &  20&7  &   77&6   &    5&0  &   5&0  &   0&0  \\
\hline
\multicolumn{13}{c}{derived} \\
\hline
\ali                & 154&23 & 514&40 & 1203&80  & 1017&57 & 311&25 & 356&31 \\
$\log\epsilon(X)$   &  \dots &15 &  \dots &28 &    \dots &39  &   \dots &46 &   \dots &14 &   \dots &31 \\
\mdot               &  14&4  &  56&7  &  212&11  &   40&1  &  21&0  &   0&0 \\
\hachr              &  21&12 &  69&29 &  267&50  &  205&18 & 115&16 &  61&0 \\
\hbchr              &   \dots &10 &  \dots &18 &   \dots &42  &   \dots &14 &   \dots &12 &   \dots &0 \\
\fachr              &  21&12 &  65&28 &  265&47  &  199&17 & 105&16 &  47&0 \\
\fbchr              &   \dots &10 &  \dots &17 &    \dots &41  &    \dots &14 &   \dots &12 &   \dots &0 \\
\hline
\end{tabular}
\end{table*}

Most parameters listed in Table\,\ref{tab:PMS-parameters} are derived from both UVES and GIRAFFE spectra, with the exception of 
$\xi$, \hbchr, \fbchr, and \abund, that are derived from UVES spectra only, and the gravity-sensitive spectral index $\gamma$, which is derived from the GIRAFFE spectra only (see Sect.\,\ref{sec:FundamentalParameters}).

In general, whenever possible, the same parameter is derived by different methods; this allows a thorough check of the derived parameters by inter-comparing the results and flagging discrepant results, which are then used to outline possible weaknesses of the methods and discard unreliable results.
In the absence of significant biases, the results from different methods are combined taking  a $\sigma$-clipped average to obtain the {\em recommended} parameters.
If significant biases are present, all results obtained with a method that can give rise to inaccurate or unreliable results in some ranges of parameters are rejected before combining the results as above.
These general criteria, whose application is discussed in details in 
Sects.\,\ref{sec:RawMeasurements}--\ref{sec:DerivedParameters},
are firstly applied on the {\em raw} parameters, then on the {\em fundamental} parameters, and finally on the {\it derived} parameters.
The {\em fundamental} parameters are also validated by comparing the results of the analysis methods applied to our spectra against fundamental parameters from angular diameter and parallax measurements (Sect.\,\ref{sec:benchmarks}).
A comparison with \teff\ derived from photometry for objects that are not affected by photometric excesses is reported in Appendix\,\ref{sec:photometry}.
The {\em recommended raw} and {\em fundamental} parameters are then used to produce the {\em recommended derived} parameters.
When satisfactory comparisons cannot be achieved, recommended parameters are not provided and only results from individual nodes are made available.
Recommended parameter uncertainties are estimated as both node-to-node dispersion and as average of individual node's uncertainties.
The final results minimise -- as much as possible -- biases that can affect individual methods and the associated uncertainties take differences that may arise from the use of different methods and algorithms into account.

Final results are further validated by a general analysis of the output \logg-\teff\ diagram, consistency of the parameters, and overview of the results based on the comparison of different clusters.

\section{Raw measurements}
\label{sec:RawMeasurements}

Measuring the raw parameters before carrying out any other analysis allows us to: 
(a) identify stars with strong accretion whose spectra may be affected by veiling; 
(b) perform a quality control on the raw parameters before they are used in the subsequent analysis; and
(c) apply the appropriate masks to the spectra for the determination of fundamental parameters. 

To derive raw parameters from a large dataset of spectra it is convenient to use procedures that are as automatic as possible.
However, in the case of PMS sources extending to M spectral type, such procedures must also be capable of dealing with large rotational broadening and the presence of molecular bands.
Here different methods are used, with different levels of automatism, which allows to examine the presence of biases, eliminate systematic discrepancies, and combine the results with a $\sigma$-clipping to disregard casual mistakes and outliers.

In the following we briefly describe the methods used to derive the raw parameters.

\subsection{\halpha\ equivalent width and \halpha\ width at 10\% of the line peak}
\label{sec:halpha}

Spectra with \halpha\ in emission are examined to identify stars with strong accretion, and therefore likely to be affected by veiling, using their \wha\ and \ha10\ measurements.

The Arcetri node measures \wha\ and \ha10\ on the continuum-normalised co-added
spectra of all stars that clearly show \halpha\ emission, using a semi-automatic procedure.
After manually defining the wavelength range and level of continuum, \wha\ is calculated by a direct integration of the flux above the continuum, while \ha10\ is derived by considering the level corresponding to 10\% of the maximum flux above the continuum in the selected wavelength range.
All measurements are visually checked and repeated in case of miscalculation (e.g. due to the presence of multiple peaks).
Uncertainties are estimated using multi-epoch observations of stars belonging to the first two young clusters that have been observed (i.e. $\gamma$ Vel and Cha\,I). 
Specifically, \wha\ and \ha10\ are first measured on each spectrum before co-adding, then the relative uncertainty for each star is estimated as $\Delta W = 2\abs{W_{1} - W_{2}} / (W_{1} + W_{2})$, where $W_{1}$ and $W_{2}$ are two measurements for the same star from spectra observed at different epochs.
A similar formula is used for $\Delta$\ha10.
Finally, the median of $\Delta$\wha\ and $\Delta$\ha10\ are assumed as the relative uncertainties for all stars\footnote{This may also be linked to \halpha\ variability.}

The CAUP node makes use of an automatic IDL\footnote{IDL\textregistered (Interactive Data Language) is a registered trademark of Exelis Visual Information Solutions.} procedure to first select stars with \halpha\ in emission and then measure \ha10\ and \wha\ on the normalised spectra.
Measurement uncertainty is evaluated from the spectrum S/N.  

The OACT node pre-selects spectra with \halpha\ in emission by visual inspection. 
Then, \wha\ and \ha10\ are measured using an IDL procedure.
\wha\ is measured by direct integration of the \halpha\ emission profile and its uncertainty evaluated by multiplying the integration range by the mean error in two spectra regions close to the \halpha\ line. 
\ha10\ uncertainty is evaluated by assuming an error of 10\% in the position of the continuum level.

The OAPA node employed two methods, one based on DAOSPEC \citep{2008PASP..120.1332S} and an IDL procedure, the other on a combination of IRAF and IDL tools.
In the first method, DAOSPEC is used to perform a continuum fit of the spectral region around \halpha. 
The \halpha\ profile in the unnormalised input spectrum is masked by giving as input to DAOSPEC a variable FWHM that takes the rotational and instrumental profiles into account. 
The fitted continuum is then used to normalise the input spectrum.
Such a continuum normalised spectrum is used to measure \ha10\ by an automatic IDL procedure. 
Since the uncertainties are assumed to be dominated by the fitting of the continuum, this is repeated four times using different orders (10, 15, 20 and 25) of the polynomial fitting in DAOSPEC.
The resulting \ha10\ values are then averaged to produce the final result.  
In the second method the normalisation is performed through IRAF with three different orders of the polynomial fitting (2, 5, and 10), then \wha\ and \ha10\ are measured with an automatic IDL routine and uncertainties derived as above.
A final visual inspection is performed to check the results and identify broad emission and P~Cygni-like profiles.
For the first data release both methods where used, while in the second data release only the second method was used.

The final spectra of NGC2264 are affected by some residual nebular emission, and a good subtraction of this contribution to the \halpha\ emission line cannot be achieved as the nebular emission is concentrated in the region near the \halpha\ line peak and is spatially variable \citep[see a detailed description of this topic for the analogous case of the cluster NGC6611 in][]{2013A&A...556A.108B}.
In this case, additional visual inspection of the spectra was necessary to ensure that the narrow nebular emission does not affect significantly the measurements.

\begin{figure}[ht]
\centering
\includegraphics[width=90mm]{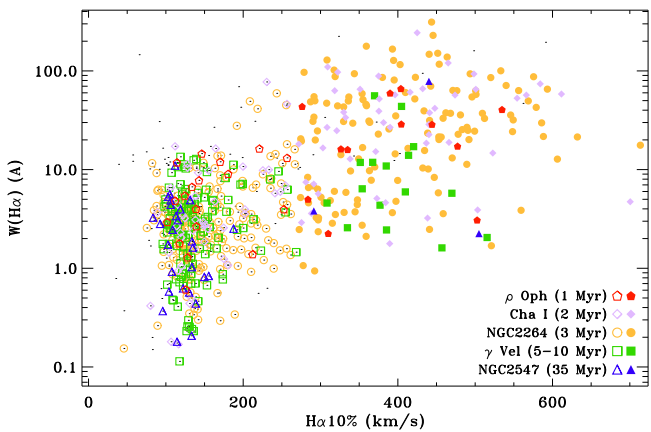}
\caption{\wha\ vs. \ha10\ for all young open clusters observed in the first 18 months of observations. Filled symbols are used for stars classified as CTTS, open symbols for stars classified as WTTS.}
\label{fig:giraffe_Ha10_EWHaAcc_pms}
\end{figure}

In the node-to-node comparison of the \wha\ results, average differences and dispersions $\sim 5$ \AA\ were found in the analysis of both UVES and GIRAFFE spectra, with only a few outliers.
Average differences in \ha10\ in the node-to-node comparison was $\sim$10 \kms, with a dispersion $\sim$50 \kms.

Only a $1\sigma$-clipping was therefore applied before computing the average \wha\ and \ha10\ as recommended values.
The recommended uncertainty was given, conservatively, as the largest amongst the average of individual uncertainties and the standard deviation of the mean.

The recommended \ha10\ is used, together with the recommended \wli\ (Sect.\,\ref{sec:WLi}), in our WTTS/CTTS classification.
If the \halpha\ is in emission and \wli$>$100\,m\AA, the star is identified as a T\,Tauri.
Following \cite{2003ApJ...582.1109W}, the T\,Tauri star is then classified as CTTS if \ha10$\ge$270 \kms.

A comparison of \wha\ vs. \ha10\ for all young open clusters observed in the first 18 months of observations is shown in Fig.\,\ref{fig:giraffe_Ha10_EWHaAcc_pms}. 
Note that the correlation of the two parameters is as expected from other works \citep[e.g.,][]{2003ApJ...582.1109W} and the fraction of CTTS consistently decreases with the age of the cluster.

\subsection{\li\ equivalent width}
\label{sec:WLi}

\begin{figure*}[ht]
\centering
\includegraphics[width=8.0cm]{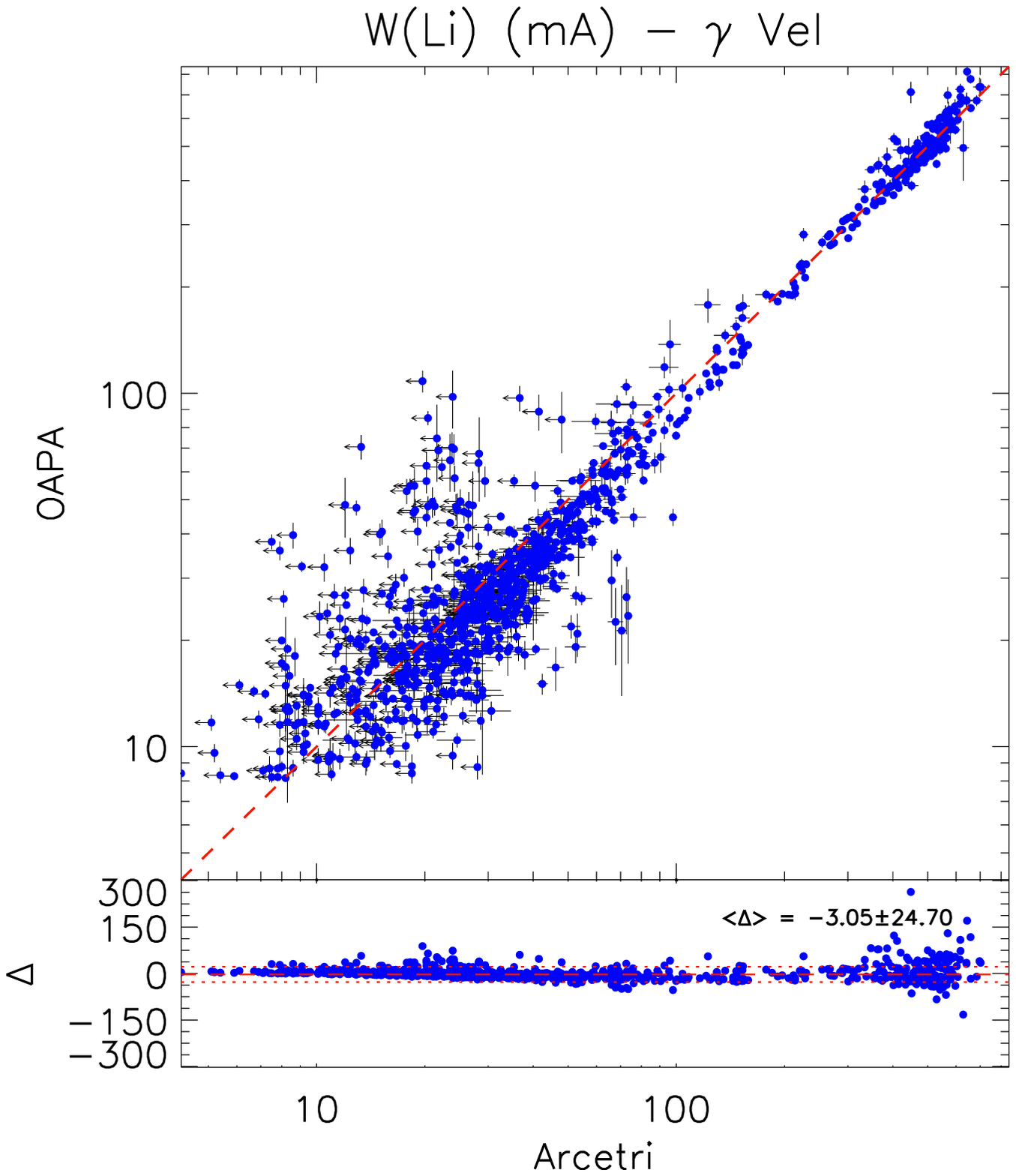}
\includegraphics[width=8.0cm]{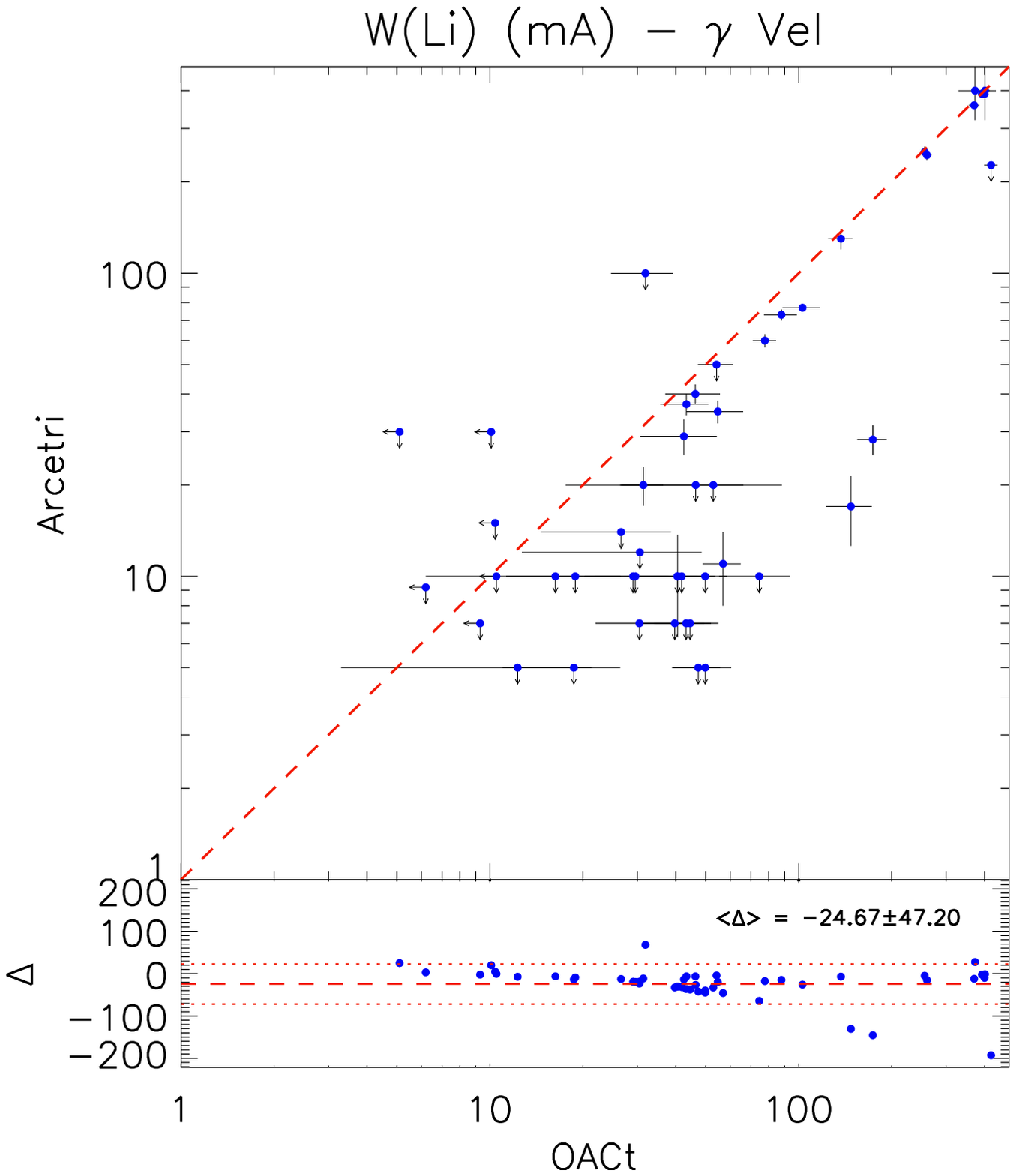}
\caption{Illustrative node-to-node \wli\ comparison for $\gamma$~Vel. Left panel: Comparison between OAPA (DAOSPEC, iDR1) and Arcetri (iDR2) for GIRAFFE spectra. Right panel: Comparison between Arcetri code and OACT (IRAF) for UVES spectra. Arrows indicate upper limits.}
\label{fig:ewlicomparison_gammaVel}
\end{figure*}

At young ages, the \li\ doublet line is often in the saturation regime and the rotational broadening frequently dominates.
As a consequence, in general, a direct profile integration of the \li\ line is to be preferred to a Gaussian or a Voigt profile fitting in deriving \wli. 
Furthermore, due to rotational broadening, the integration wavelength interval is very different from one spectrum to another. 
The \li\ doublet is also superimposed to molecular bands in spectra of M-type stars, which makes the placement of the continuum difficult, particularly when using automatic procedures. 
In such cases, despite being slow and prone to human error and subjective choices, interactive procedures, like those available in \texttt{IRAF}, remain one of the best options for measuring \wli, at least for comparison purposes. 
Weak \li\ lines in slow-rotating stars, on the other hand, can reliably be fitted with a Gaussian or a Voigt profile and integrated analytically, a method that can be easily implemented in automatic procedures and is more accurate than low-order numerical integration at low S/N. 

The Gaia-ESO PMS analysis makes use of three independent methods for deriving \wli\ from the GIRAFFE spectra: the direct profile integration available in the \texttt{IRAF-splot} procedure (OACT node); DAOSPEC \citep[][OAPA node]{2008PASP..120.1332S}; and a semi-automatic IDL procedure specifically developed for the Gaia-ESO by the Arcetri node.

The \texttt{IRAF-splot} task was applied by the OACT node to the unnormalised spectra to make use of the built-in Poisson statistics model of the data.
Such measurements are only performed in those cases where the \li\ line and the nearby continuum are clearly identifiable, which implies that, in general, small \wli\ ($\apprle 10$ m\AA), low S/N ($\apprle 20$), and spectra with very high \vsini\ ($\apprge 200$ \kms) are not considered.  

The DAOSPEC (OAPA) measurements were applied to all iDR1 spectra with S/N$>20$, and spectra with S/N$<20$ showing a strong lithium line. 
The spectra have been re-normalised prior to the equivalent width determination using high order Legendre polynomial fitting, which allows to follow the shape of molecular bands in M-type stars still maintaining a good agreement with the continuum of earlier type stars.
The typical width of absorption lines in each spectrum has been estimated by convolving the instrumental and rotational profile using \vsini\ from the data reduction pipeline.
Relative internal uncertainties are always better than 5\% for large equivalent widths ($>200$\,m\AA) and degrade up to $\sim$ 50\% for very low equivalent widths ($\sim 10$\,m\AA). 

The semi-automatic IDL procedure developed by the Arcetri node performs a spline fitting of the continuum over a region of $\pm 20$~\AA\ around the Li line using an iterative $\sigma$-clipping, and masking
both the Li line and the nearby \ion{Ca}{I} line at 6717.7\,\AA. 
When the automatic continuum fitting is not satisfactory (generally for poor S/N spectra or M-type stars), the fit is repeated by setting manually the continuum level. 
The \wli\ is then computed by direct integration of the line within a given interval, which depends on the stellar rotation and was determined by measuring the line widths on a series of rotationally broadened synthetic spectra. 
Errors are derived using the \cite{1988IAUS..132..345C} formula; when no Li line (including blends) is visible, the upper limit is set as three times the error.

The contribution of lines blended with \li\ in the GIRAFFE spectra was estimated, after the determination of the fundamental parameters (Sect.\,\ref{sec:FundamentalParameters}), by a spectral synthesis using Spectroscopy Made Easy \citep[SME,][]{1996A&AS..118..595V} with MARCS model atmospheres as input, taking  the star's \teff, \logg, and \feh\ into account.
For solar metallicity dwarfs above 4000\,K the estimated blends are in agreement with the \cite{Soderblom_etal:1993} relation.

Four nodes (Arcetri, CAUP, OACT, and UCM) calculated \wli\ in the UVES spectra. 
At the UVES resolution, when the star is slow rotating (\vsini$\apprle 25$ \kms) and the S/N is sufficiently high (S/N$\apprge 60$), it is possible to de-blend the \li\ line from the nearby features.

Both the CAUP and UCM nodes employed the \texttt{splot} task in IRAF on the unnormalised UVES spectra.
When the \li\ line and the nearby blends, mainly with \element{Fe} lines, are distinguishable, these are de-blended, in which case a Gaussian fitting to the line profile is adopted.
On the contrary, when the lines are indistinguishable, the blends contribution is estimated using the \texttt{ewfind} driver within \texttt{MOOG} code \citep{sneden}, and a direct integration of the line is adopted.

The Arcetri node adopted the same method used for GIRAFFE (see above), except in those cases where it was possible to de-blend the line using IRAF as done by the CAUP and UCM nodes.
When this was not possible, the blends were estimated using SME.

For iDR1, the OACT node employed IRAF as for the GIRAFFE spectra, using SME to estimate the blends.
For iDR2, \wli\  was derived by spectra subtraction with the template having the closest fundamental parameters but no (or negligible) Li absorption.
In this latter case the blends are removed by the spectra subtraction itself.

It is worth stressing that the PMS analysis output includes both blends-corrected and -uncorrected \wli. 
When a node does not provide blends-corrected \wli, this is estimated using SME and the node's fundamental parameters if available.
Note that, in the analysis of GIRAFFE spectra, blends are estimated using SME in all cases; the recommended blends-corrected \wli\ are calculated from the recommended blends-uncorrected \wli\ using the recommended fundamental parameters.
Conversely, in the analysis of UVES spectra the recommended blends-corrected \wli\ are derived by averaging the node values, as discussed below.

The blends-uncorrected \wli\ obtained by the three different methods from GIRAFFE spectra were first compared to check for systematic differences before combining them to produce the final results (see Fig.\,\ref{fig:ewlicomparison_gammaVel} for an illustrative comparison). 
After discarding results of one node not consistent with the other two, no significant bias remained and the relative standard deviation of the $W$ difference was at 20\% level.
Also, no trend of the node-to-node differences with respect to S/N nor \vsini\ was present in the selected results.
As a conservative uncertainty estimate on the {\it recommended} \wli\ we adopted the larger of the standard deviation and the mean of the individual method uncertainties.
In iDR1, 90\% of the \wli\ measurements have relative uncertainties better than 14\% and 28\% in $\gamma$ Vel and Cha\,I, respectively, the differences being due to the higher fraction of stars of low \teff\ and spectra with lower S/N in Cha\,I with respect to $\gamma$ Vel.
About 90\% of all the iDR2-GIRAFFE \wli\ measurements have uncertainty better than 16\,m\AA. 

In the \wli\ UVES measurements no
systematic deviation nor trends of the node-to-node differences with S/N nor \vsini\ were found from the node-to-node comparison (see Fig.\,\ref{fig:ewlicomparison_gammaVel} for an illustrative comparison) and the recommended values were derived by taking the mean with a 1$\sigma$-clipping.
In iDR1
the median uncertainties are 3\,m\AA\ (4\%) and 10\,m\AA\ (3\%) for $\gamma$ Vel and Cha\,I, respectively\footnote{Note that, given the small number of measurements in UVES compared to GIRAFFE, the median uncertainty is chosen to characterise the internal precision achieved rather than the distribution.}.
About 90\% of all the iDR2-UVES \wli\ measurements have uncertainty better than 22\,m\AA. 

When all \wli\ measurements for a given star are flagged as upper limits, the recommended \wli\ is also flagged as an upper limit and the lowest measurement is adopted. 
Conversely, when at least one \wli\ measurement for a given star is not flagged as upper limit, all upper limit estimates for that star are disregarded and the recommended value is derived as above.

\begin{figure}[ht]
\centering
\includegraphics[width=8.0cm]{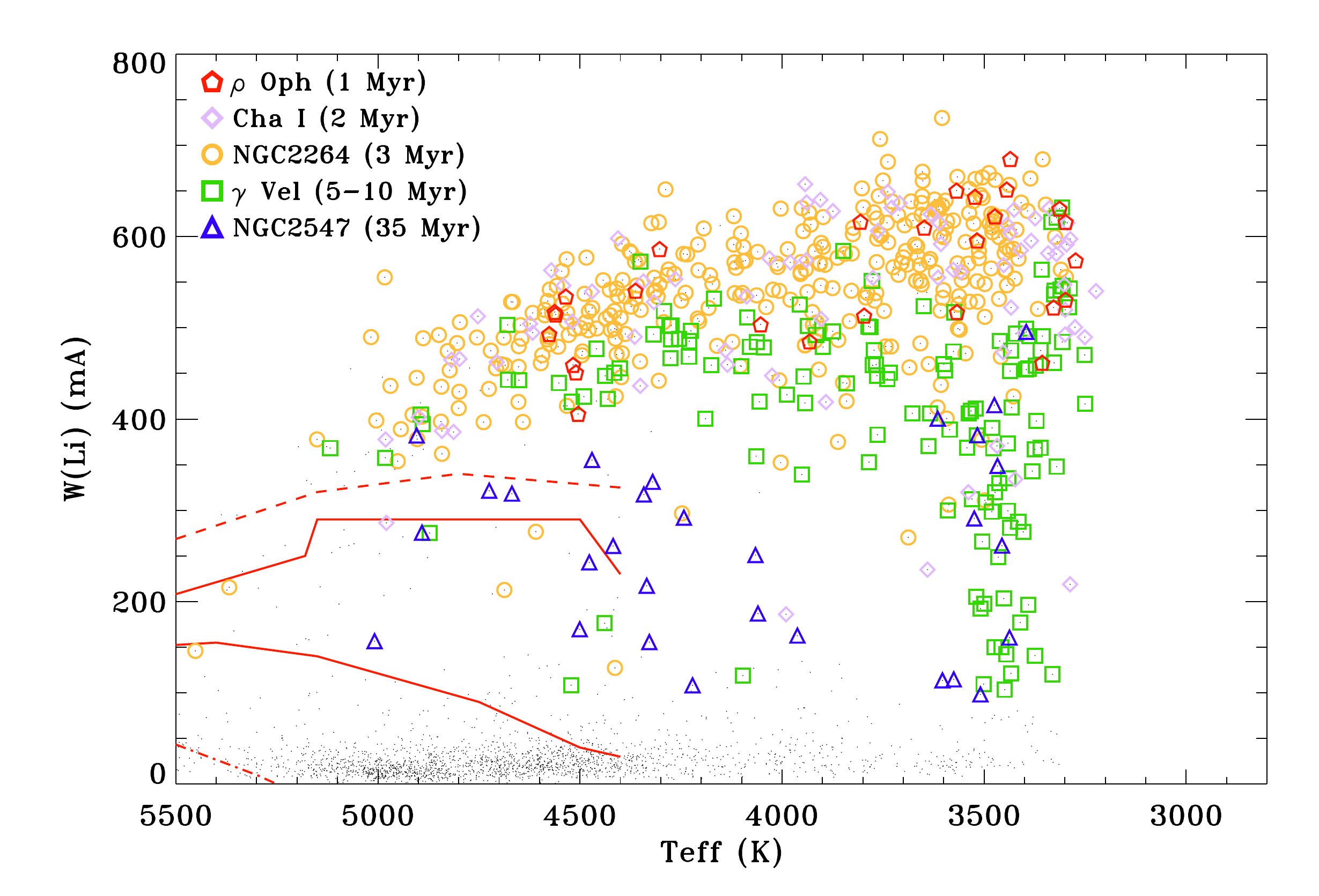}
\caption{Blends-corrected \wli\ vs. \teff\ for all young open clusters observed in the first 18 months of observations. For comparison, the lower and upper envelope in the Pleiades are shown as solid lines, while dashes and dotted lines are the upper envelopes of IC2602 and the Hyades, respectively. Dots are used for non-members. Upper limits are not included for clarity.}
\label{fig:giraffe_Teff_EWcLi_pms}
\end{figure}

{\em Recommended} blends-corrected \wli\ vs. \teff\ are shown in Fig.\,\ref{fig:giraffe_Teff_EWcLi_pms}, compared with the Pleiades upper and lower envelope and the upper envelopes of IC2602 (30 Myr) and the Hyades \citep[][and references therein]{2005A&A...442..615S}.
Note the increasing Li depletion with age at \teff$\gtrsim$ 3500\,K and the lack of depletion at lower \teff\ as expected.
A comparison with theoretical models can be found in \cite{2014A&A...563A..94J} for $\gamma$ Vel.
The NGC2547 Li depletion pattern is found in remarkable agreement with \cite{2005MNRAS.358...13J}.

\section{Fundamental parameters}
\label{sec:FundamentalParameters}

Two nodes (OACT and OAPA) provide fundamental parameters from the analysis of GIRAFFE spectra, and four nodes (Arcetri, CAUP, OACT, and UCM) from the analysis of UVES spectra. 
With the exception of OACT, the nodes analysing UVES spectra use similar, classical procedures, i.e. measurement of equivalent widths and MOOG \citep{sneden}, enforcing the usual equilibrium relations.
However, different strategies are adopted for the selection of the lines to be used and in the automatisation of the procedures, as described in Sects.\,\ref{sec:FunparsArcetri}--\ref{sec:FunparsUCM}.

As anticipated in Sect.\,\ref{sec:GeneralStrategy}, the validation of fundamental parameters is carried out internally by a node-to-node comparison (Sect.\,\ref{sec:internal}), and externally by comparisons with \teff\ and \logg\ derived from angular diameter measurements on a sample of stars taken as benchmark (Sect.\,\ref{sec:benchmarks})

\subsection{OACT}

A code that has been extensively used for determining fundamental parameters in PMS stars is \texttt{ROTFIT} \citep[e.g.,][]{2006A&A...454..301F}, which compares the target spectrum with a set of template spectra from ELODIE observations of slowly-rotating non-active stars \citep{2001A&A...369.1048P} artificially rotation-broadened and {\it veiled} at varying \vsini\ and~$r$.
In the following we report a brief summary of the method implemented by \texttt{ROTFIT}, together with some adaptations to the case at hand.

In \texttt{ROTFIT}, the template spectra that most closely reproduce the target spectrum when broadened and {\it veiled} are selected, and their weighted average \teff, \logg, and \feh\ assigned to the target star.
As a figure of merit the $\chi^2$ calculated on the target spectrum and the rotational-broadened and veiled template spectrum is adopted.
The weight used in the average is proportional to $\chi^{-2}$.
A discussion on the \texttt{ROTFIT} templates, together with the homogenisation with the Gaia-ESO spectrum analysis, is reported in Appendix\,\ref{sec:ROTFIT_templates}.
The \texttt{ROTFIT} analysis requires different wavelength masks for different type of objects.
The masking criteria for the application to the Gaia-ESO survey are reported in Appendix\,\ref{sec:ROTFIT_masks}.

For the GIRAFFE/HR15N spectra the whole spectral range from 6445 to 6680\,\AA\ and the ten best templates (i.e. with the lowest $\chi^2$) are considered. 
The analysis of the UVES 580 spectra is independently performed on segments of 100\,\AA\ each (excluding the parts containing strong telluric lines and the core of Balmer lines), by considering only the best five templates for each segment. 
The final parameters \teff, \logg, \feh, and \vsini\ are obtained by taking the weighted averages of the mean values for each segment, with the weight being proportional to $\chi^{-2}$ and to the amount of information contained in the segment, quantified by the total line absorption $f_i=\int(F_{\lambda}/F_{\rm C}-1)d\lambda$
(where $F_{\lambda}$/$F_{\rm C}$ is the continuum normalised flux).

The \teff, \logg, \feh, and \vsini\ uncertainties are given as the standard errors of the weighted means, to which the average uncertainties of the templates' stellar parameters are added quadratically.
These are estimated to be $\pm$\,50\,K, $\pm$\,0.1\,dex, $\pm$\,0.1\,dex, and $\pm$\,0.5\,\kms\ for \teff, \logg, \feh, and \vsini, respectively.
The target's spectral type corresponds to that of the best template.

The code also provides an estimate of the veiling by searching for the $r$ value that gives the lowest $\chi^2$.
The determination of the fundamental parameters for a star with veiling is more uncertain than in the non-veiled case because: 
(a) lines' and molecular bands' depth are smaller in veiled spectra; and
(b) the determination of the veiling parameter $r$ implies the introduction of an additional degree of freedom in the parameters' fitting, degrading the overall accuracy with respect to the non-veiled stars.
However, a preliminary identification of stars with mass accretion, whose spectra are expected to be affected by veiling, can be done based on the values of \ha10\ or \wha\ \citep{2003ApJ...582.1109W,2004A&A...424..603N}.
It is therefore possible to restrict the veiling calculation to likely accreting stars only, thus preserving the accuracy for stars for which no veiling is expected.
Following \cite{2003ApJ...582.1109W}, we assume that stars with \ha10$>$270 \kms\ can be optically veiled, with an extra margin to take uncertainties into account (see Sect.\,\ref{sec:halpha}).

Within the Gaia-ESO analysis, \vsini\ is also provided by WG8 for GIRAFFE spectra (Koposov et al., in prep.),
together with radial velocities and a first estimate of fundamental parameters, using a conceptually similar approach but with a different fitting strategy and different templates.
The comparison between the results of these two methods turned out to be useful in identifying WG8 unsuccessful fitting for some stars with strong veiling and emission lines. 
An illustrative comparison of the results of these two methods for $\gamma$ Vel can be found in \cite{Frasca_etal:2014}, who found a mean difference of $\approx$ 2.88 \kms\ and $\sigma \approx 6.27$ for stars in the field of $\gamma$ Vel.
An investigation on the lower limit imposed by the resolution of the instruments by means of Monte-Carlo simulations is also reported in \cite{Frasca_etal:2014}, according to which the lower \vsini\ limit has been set to 7 \kms\ in GIRAFFE spectra and 3 \kms\ in UVES spectra.

\subsection{OAPA}

An alternative approach for GIRAFFE/HR15N spectra, proposed by \cite{Damiani_etal:2014}, is based on  spectral indices in different wavelength ranges of the spectrum.
The derived spectral indices are calibrated against stars with known parameters, 
yielding quantitative estimates of \teff, \logg, and \feh.
This type of approach is usually applied to spectra with lower resolution than GIRAFFE. 
These narrow-band indices are affected by fast rotation: \teff\ becomes unreliable for \vsini$>$90 \kms, \feh\ above 70 \kms, and \logg\ above 30 \kms.
Therefore, depending on the \vsini\ of the star, not all parameters can be provided.
Using an appropriate combination of flux ratios, this method is also capable of producing an independent estimate of the veiling parameter $r$ \citep[see,][for details]{Damiani_etal:2014}.

\subsection{Arcetri}
\label{sec:FunparsArcetri}

The Arcetri node adopted \texttt{DOOp} \citep[DAOSPEC Option Optimiser pipeline,][]{2014A&A...562A..10C} for measuring line equivalent widths and \texttt{FAMA} \citep{2013A&A...558A..38M} for determining the fundamental parameters.
Line equivalent widths are measured using Gaussian fitting after a re-normalisation of the continuum; 
$W$ in the range between 20--120 m\AA, for the \ion{Fe}{i} and \ion{Fe}{ii} lines, and in the range between 5--120 m\AA, for the other elements, were used. The \texttt{FAMA} code makes use of the \ion{Fe}{i} and \ion{Fe}{ii} equivalent widths to derive the fundamental stellar parameters. 
The stellar parameters are obtained by searching iteratively for the three equilibria (excitation, ionisation, and trend between $\log n$(\ion{Fe}{i}) and $\log (W/\lambda)$), i.e. with a series of recursive steps starting from a set of initial atmospheric parameters and arriving at a final set of atmospheric parameters which fulfils the three equilibrium conditions. 
The convergence criterion is set using the information on the quality of the $W$ measurements, i.e. the minimum reachable slopes are linked to the quality of the spectra, as expressed by the dispersion around the average $<\log n$(\ion{Fe}{i})$>$. 
This is correct in the approximation that the main contribution to the dispersion is due to the error in the $W$ measurement rather than to inaccuracy in atomic parameters, as e.g. the oscillator strengths ($\log gf$).

\subsection{CAUP}

The fundamental parameters are automatically determined with a method used in previous works \citep[e.g.][]{2008A&A...487..373S,2011A&A...526A..99S} now adapted to the Gaia-ESO survey. 
The method is based on the excitation and ionisation balance of iron lines using \feh\ as a proxy for the metallicity. 
The iron lines constraining the parameters were selected from the Gaia-ESO line list using a new procedure detailed in \cite{2014A&A...561A..21S}.
The equivalent widths are automatically measured using the \texttt{ARES}\footnote{\texttt{ARES} is available for download at http://www.astro.up.pt/~sousasag/ares/} code \citep{2007A&A...469..783S} following the approach of \cite{2008A&A...487..373S,2011A&A...526A..99S} that takes the S/N of each spectrum into account. 
The stellar parameters are computed assuming LTE using the 2002 version of \texttt{MOOG} \citep{sneden} and the MARCS grid of atmospheric models. 
For this purpose, the interpolation code provided with the MARCS grid was modified to produce an output model readable by \texttt{MOOG}. 
Moreover, a wrapper program was implemented to the interpolation code to automatise the method.
The atmospheric parameters are inferred from the previously selected \ion{Fe}{i}-\ion{Fe}{ii} line list. 
The downhill simplex \citep{1992nrca.book.....P} minimisation algorithm is used to find the best parameters. In order to identify outliers caused by incorrect $W$ values, a 3$\sigma$-clipping of the \ion{Fe}{i} and \ion{Fe}{ii} lines is performed after a first determination of the stellar parameters. 
After this clipping, the procedure is repeated without the rejected lines. 
The uncertainties in the stellar parameters are determined as in previous works \citep{2008A&A...487..373S,2011A&A...526A..99S}.

\subsection{UCM}
\label{sec:FunparsUCM}

The UCM node employed the \texttt{StePar} code \citep{2012A&A...547A..13T,2013hsa7.conf..673T}, modified to operate with the spherical and non-spherical MARCS models.
For iDR1 the $W$ measurements were carried out with the \texttt{ARES} code \citep{2007A&A...469..783S}\footnote{The approach of \cite{2008A&A...487..373S} to adjust the \texttt{ARES} parameters according to the S/N of each spectrum was followed.}. 
For iDR2 the UCM node adopted the \texttt{TAME} code \citep{2012MNRAS.425.3162K}\footnote{\texttt{TAME} is a tool that can be run in automated or manual mode.} 
The manual mode has an interface that allows the user control over the $W$ measurements to verify problematic spectra when needed. 
The \cite{2012MNRAS.425.3162K} approach to adjust the \texttt{TAME} continuum $\sigma$ rejection parameter according to the S/N of each spectrum was followed.
The \texttt{StePar} code computes the stellar atmospheric parameters using \texttt{MOOG} \citep[][]{sneden}. 
The 2002 and 2013 versions of \texttt{MOOG} were used in iDR1 and iDR2, respectively.
Five line lists were built-up for different regimes: metal rich dwarfs, metal poor dwarfs, metal rich giants, metal poor giants and extremely metal poor stars.
The code iterates until it reaches the excitation and ionisation equilibrium and minimises trends of abundance vs. $\log (W/\lambda)$. 
The downhill simplex method  \citep{1992nrca.book.....P} was employed to minimise a quadratic form composed of the excitation and ionisation equilibrium conditions. 
The code performs a new simplex optimisation until the metallicity of the model and the iron abundance are the same.
Uncertainties for the stellar parameters are derived as described in \cite{2012A&A...547A..13T,2013hsa7.conf..673T}. 
In addition, a 3$\sigma$ rejection of the \ion{Fe}{i} and \ion{Fe}{ii} lines is performed after a first determination of the stellar parameters; 
\texttt{StePar} is then re-run without the rejected lines.

\subsection{Comparison with benchmark stars}
\label{sec:benchmarks}

\begin{figure*}[htp]
\centering
\includegraphics[width=80mm]{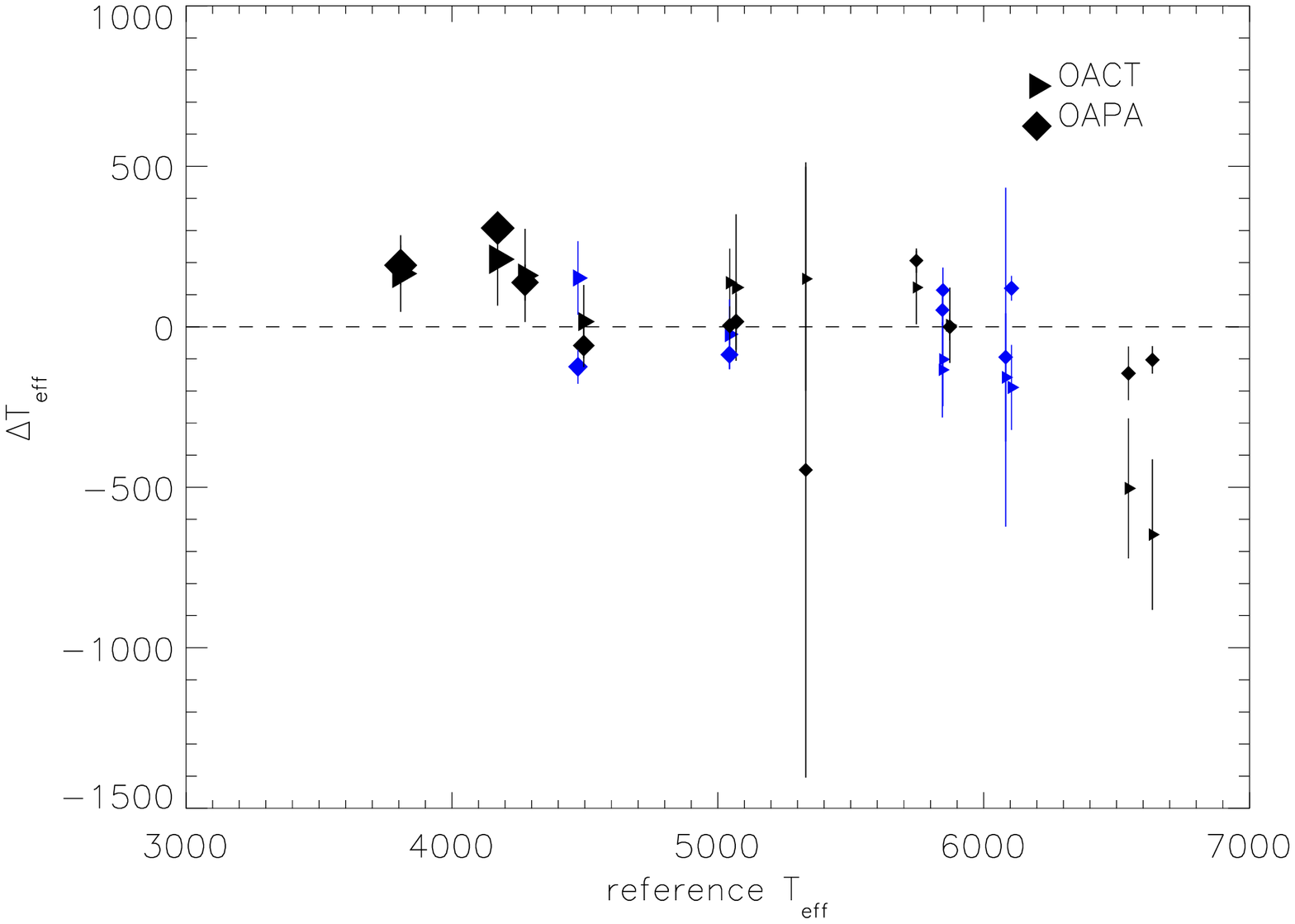}
\hspace{-0.5cm}
\includegraphics[width=80mm]{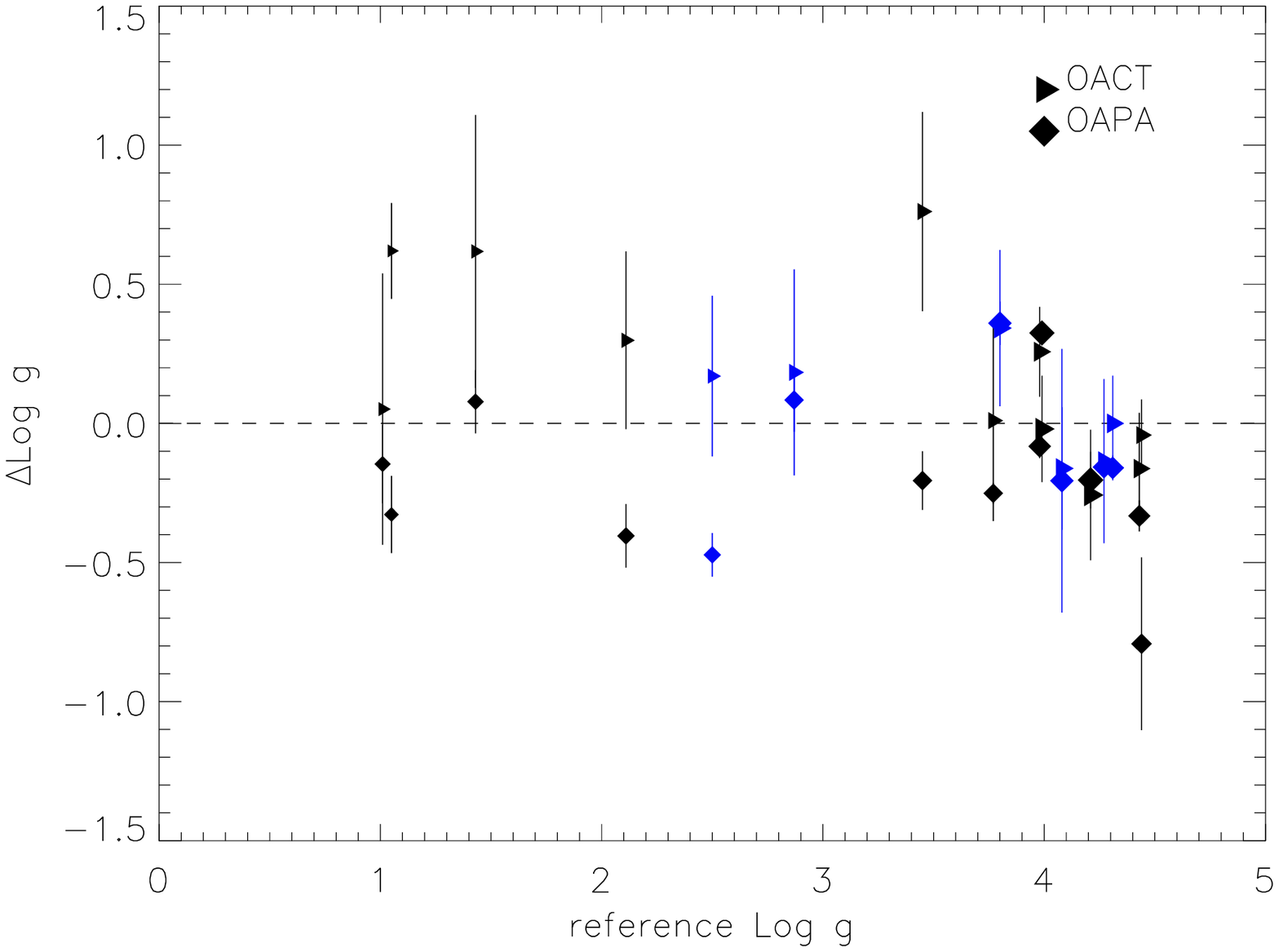}
\hspace{-0.5cm}
\includegraphics[width=80mm]{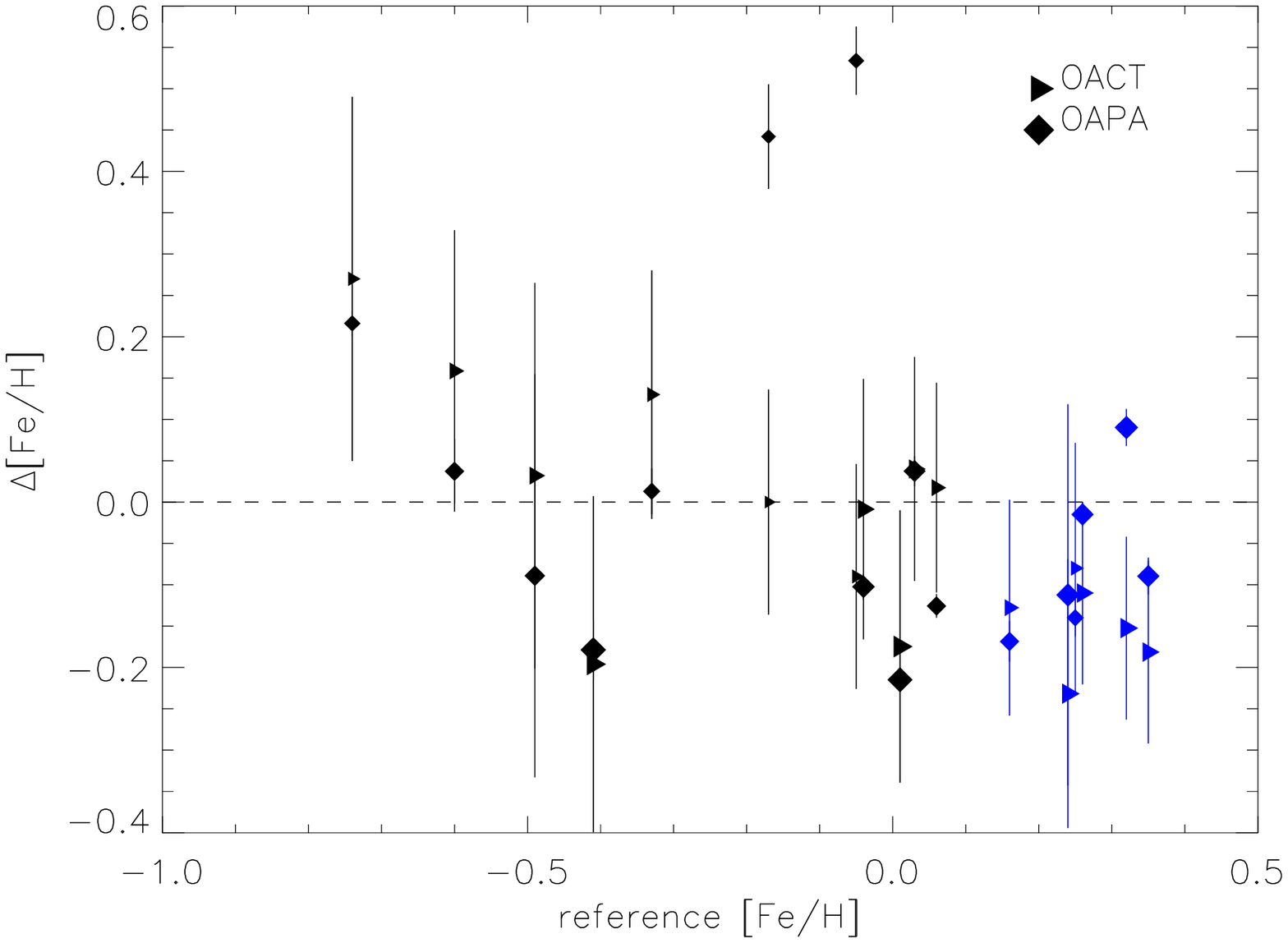}
\caption{GIRAFFE benchmarks comparison. Black for \feh $<$ 0.1, blue for \feh $>0.1$. Size is inversely proportional to \logg\ in the \teff\ plot, proportional to \teff\ in the \logg\ and \feh\ plot.}
\label{fig:GIRAFFE_benchmarks_comparison}
\end{figure*}

\begin{table*}[ht]
\centering
\caption{Average differences from benchmark reference values (see Fig.\,\ref{fig:GIRAFFE_benchmarks_comparison} and \ref{fig:UVES_benchmarks_comparison}). The Arcetri's \feh\ results are offset by 0.09 dex before computing $\Delta$\feh\ and  $\sigma$\feh\  (see text for details).}
\begin{tabular}{rrrrrrrrr}
\hline 
           &  $\langle\Delta T_{\rm eff}\rangle$  & $\langle\sigma (T_{\rm eff})\rangle$ & $\langle\Delta \log g\rangle$ & $\langle\sigma  (\log g)\rangle$  & $\langle\Delta$ [Fe/H]$\rangle$ &  $\langle\sigma($[Fe/H]$)\rangle$ &  $\langle\Delta \xi\rangle$  &  $\langle\sigma ({\xi})\rangle$ \\ 
\hline
\multicolumn{9}{c}{GIRAFFE} \\
\hline
      OACT  &    50.  &     124. &    0.19 &      0.29 &  $-$0.04 &   0.14 & \dots & \dots    \\
      OAPA  &    18.  &     120. & $-$0.15 &      0.28 &  $-$0.03 &   0.16 & \dots & \dots    \\
\hline
\multicolumn{9}{c}{UVES} \\
\hline
      OACT  &    38.  &     124. &    0.15 &      0.26 &     0.05 &   0.18 &      \dots &     \dots   \\
   Arcetri  &    55.  &      95. &    0.14 &      0.19 &     0.10 &   0.12 &      0.00  &     0.31    \\
      CAUP  &    34.  &      96. & $-$0.02 &      0.28 &  $-$0.03 &   0.08 &      0.04  &     0.33    \\
       UCM  &    56.  &      90. &    0.09 &      0.25 &  $-$0.01 &   0.08 &      0.04  &     0.38    \\
\hline
\end{tabular}
\label{tab:benchmarks}
\end{table*}

The precision of the fundamental parameters can be assessed by comparison with results from accurate independent methods like interferometric angular diameter measurements \citep[e.g.][]{2012ApJ...746..101B,2012ApJ...757..112B} which, in combination with the \textsc{Hipparcos} parallax and measurements of the star's bolometric flux, allow the computation of absolute luminosities, linear radii, and effective temperatures.
As part of the Gaia-ESO activities, and also in support of the Gaia mission, a list of stars with accurate fundamental parameters derived from such independent methods is being compiled (\citealt{2014A&A...564A.133J}; Heiter et al., in prep.) and included in the Gaia-ESO target list.
For the range of parameters of interest to the PMS analysis, however, only very few benchmark stars spectra are available in iDR1 and iDR2.

A comparison of the iDR2-GIRAFFE fundamental parameters of benchmark stars with those compiled from the literature is shown in Fig.\,\ref{fig:GIRAFFE_benchmarks_comparison} and Table\,\ref{tab:benchmarks}, in the range of interest.
In this case the \teff\ deviations are mostly within $\approx200$\,K. 
There is a systematic large deviation of OACT values above 6000\,K.
At lower temperatures, deviations larger than $\approx200$\,K are found in OAPA results for \object{HD\,10700}. 
Therefore, although the sample analysed is limited, good results are found for both nodes, except for OACT above 6000\,K.
Excluding the OACT values above this limit, the standard deviation is $\approx$ 120\,K for both datasets.
The OACT \teff\ upper limit for the GIRAFFE analysis was further lowered to 5500\,K based on comparison with \teff\ from photometry (Appendix\,\ref{sec:photometry}; see also Sect.\,\ref{sec:internal}).

Deviations as large as almost 0.7 dex  in \logg\ are found in the comparison with the benchmarks, with standard deviation $\approx$ 0.3 dex for both datasets.
From the comparison with benchmarks alone it is not possible to identify a range in which one method performs better than the other.
Indeed the node-to-node comparisons for each cluster outlined a rather complex situation that leads to the parameters selection described in Sect.\,\ref{sec:internal}.

In the parameters range of interest
(i.e. excluding very metal-poor stars), 
\feh\ is approximately reproduced with a maximum deviation of 0.3 dex and a standard deviation of $\approx$ 0.15 dex.

\begin{figure*}[htp]
\centering
\includegraphics[width=80mm]{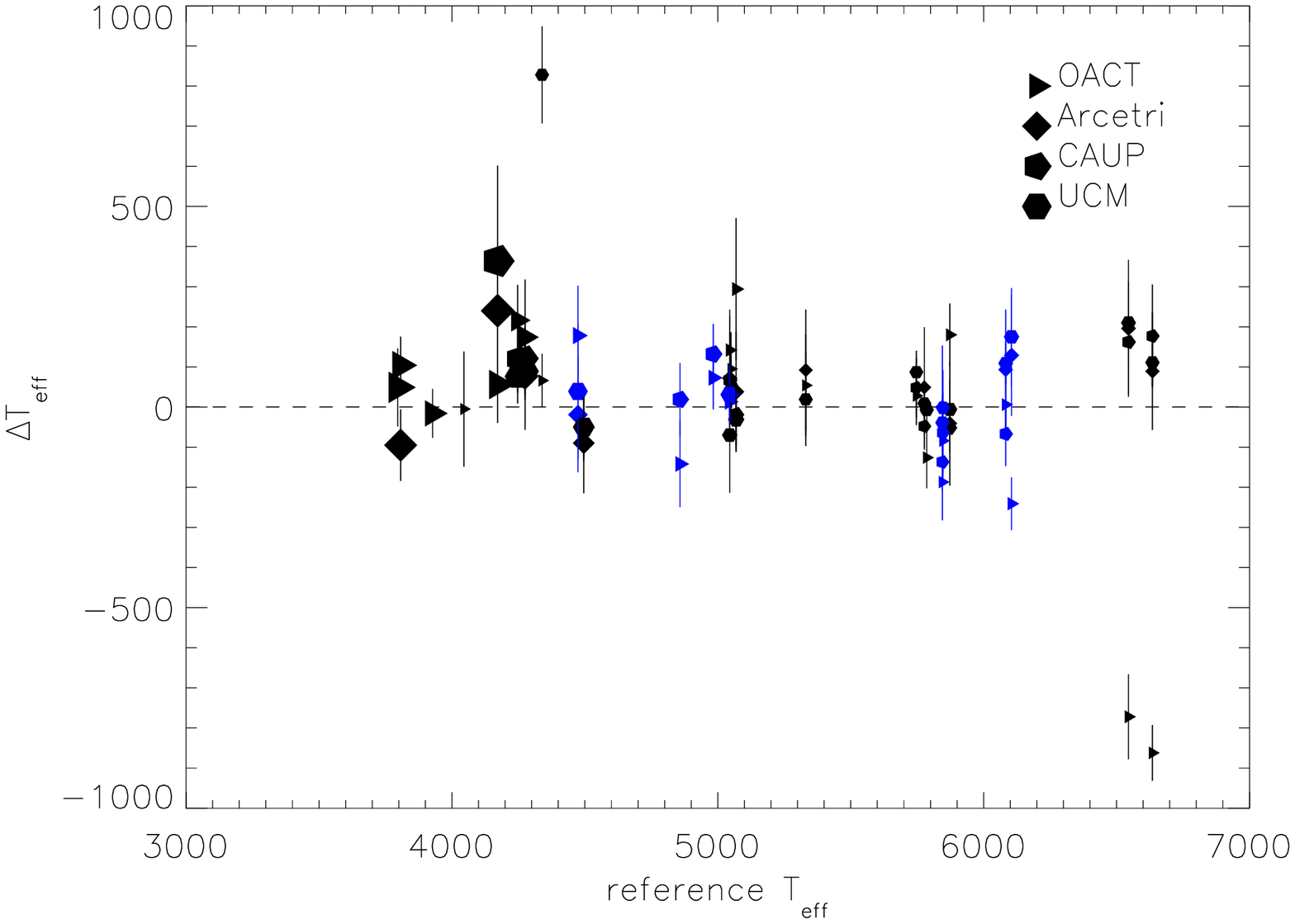}
\includegraphics[width=80mm]{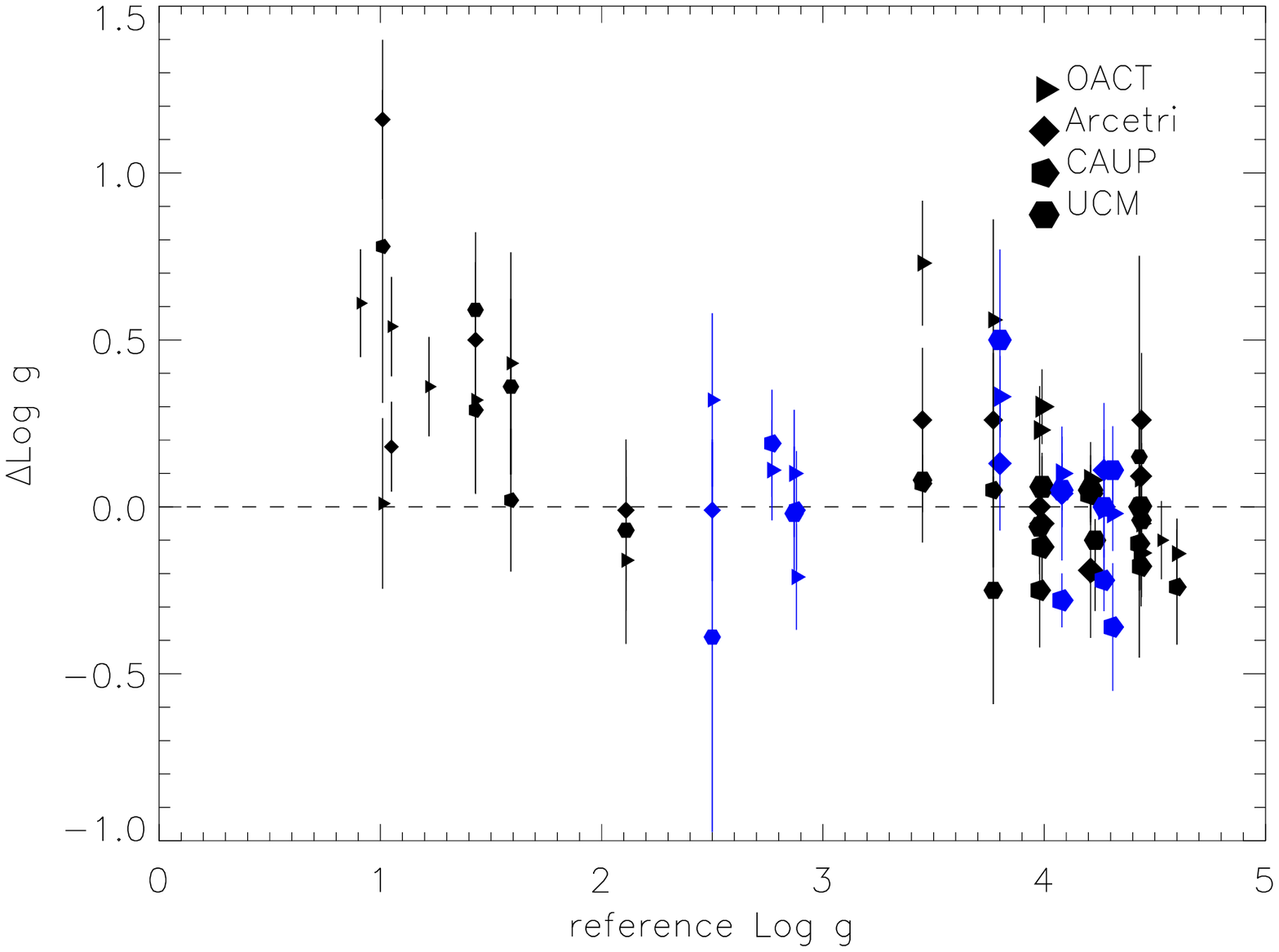}
\includegraphics[width=80mm]{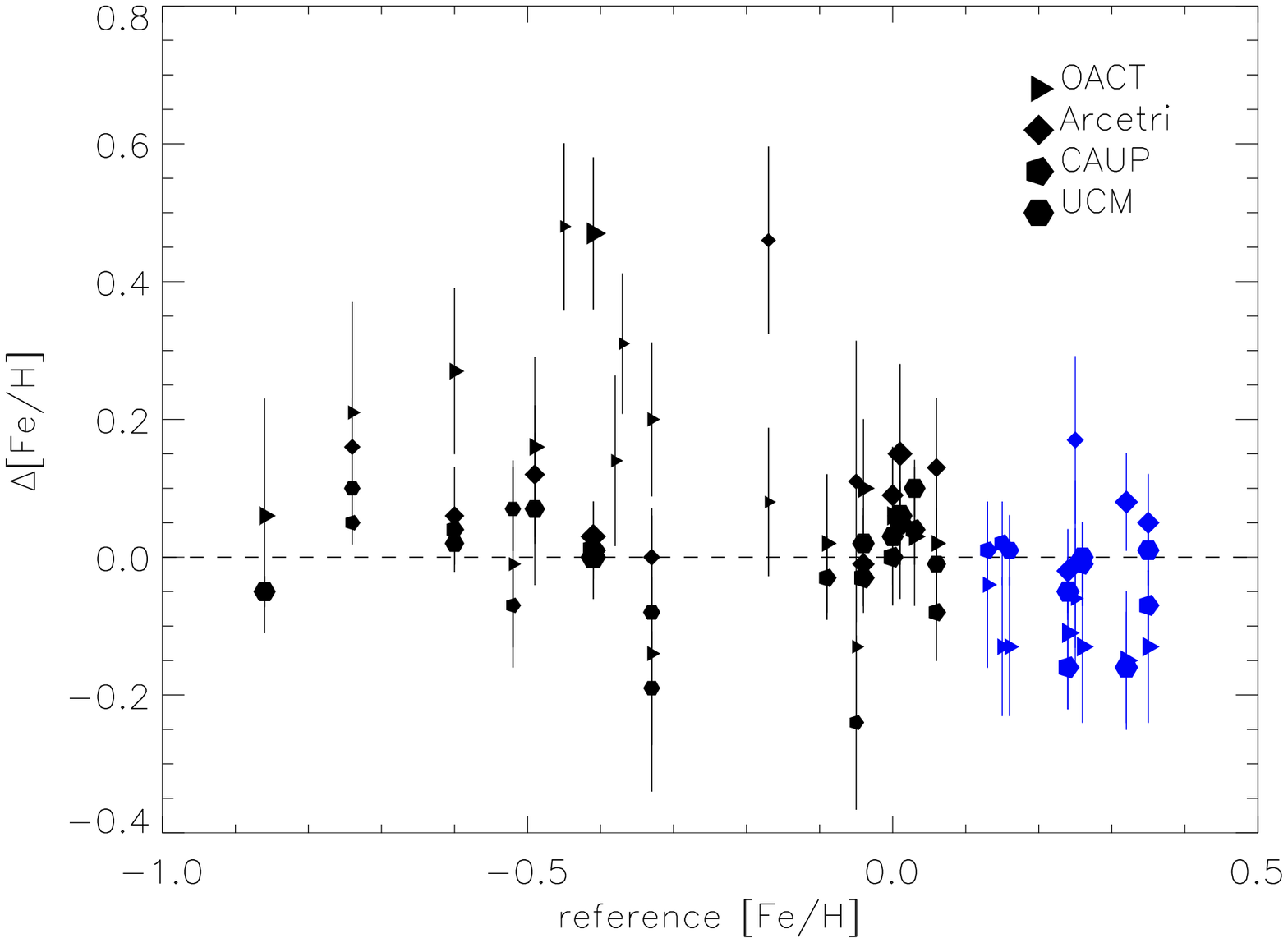}
\includegraphics[width=80mm]{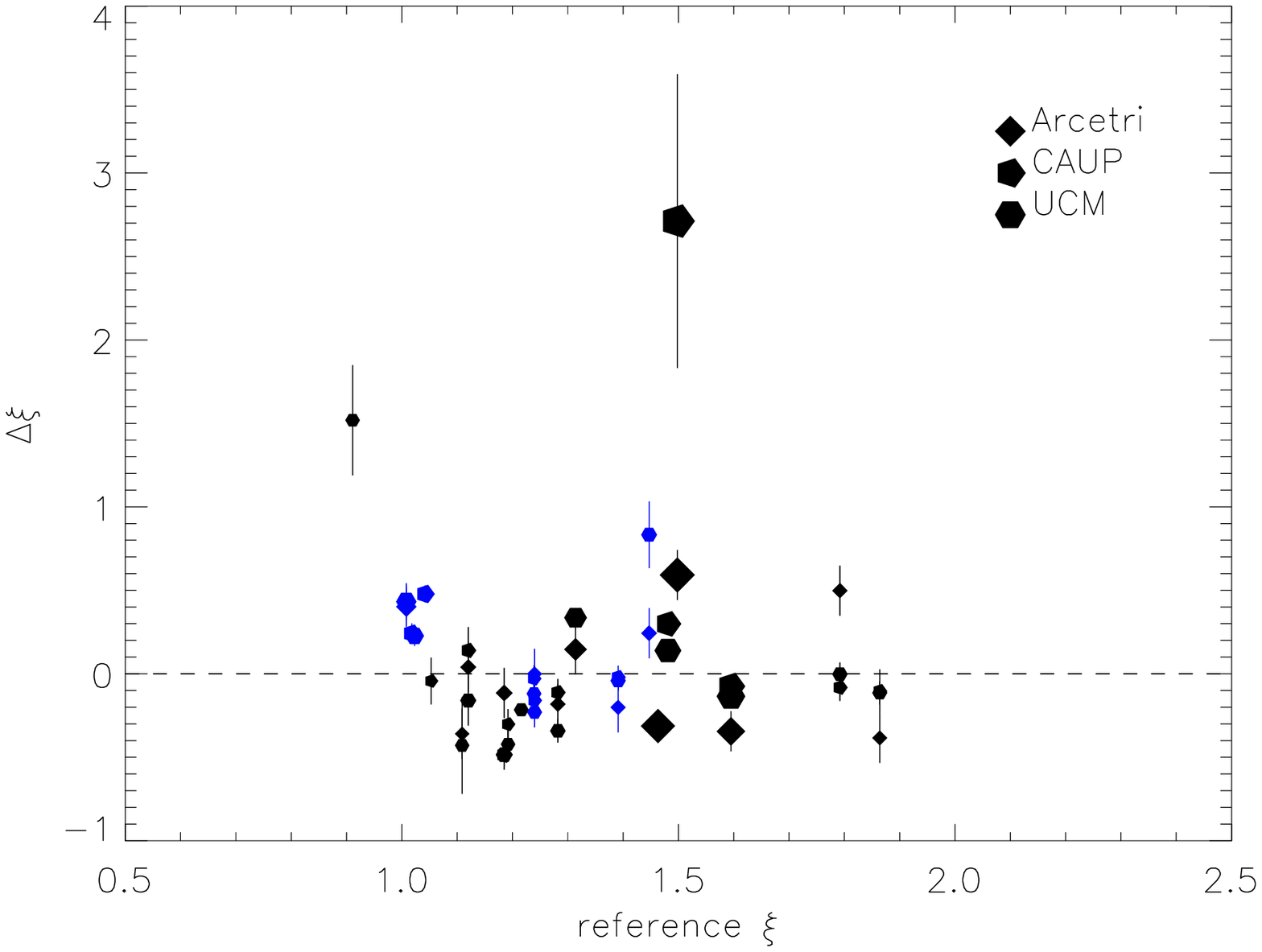}
\caption{UVES benchmarks comparison. Black for \feh $<$ 0.1, blue for \feh $>0.1$. Size inversely proportional to \logg\ in the \teff\ and $\xi$ plot, proportional to \teff\ in the \logg\ and \feh\ plot. Reference $\xi$ from Bergemann et al. (2014, in prep.) \citep[see also][]{Smiljanic_etal:2014}.}
\label{fig:UVES_benchmarks_comparison}
\end{figure*}

The comparison of the UVES fundamental parameters of benchmarks with those compiled from the literature is shown in Fig.\,\ref{fig:UVES_benchmarks_comparison} and Table\,\ref{tab:benchmarks}.
The results for the solar spectrum are outlined in Table\,\ref{tab:UVES_solar}.

In general \teff\ deviations from benchmarks are all within 300\,K (maximum) with a few outliers.
Amongst these, UCM \teff\ for \object{61\,Cyg\,A} differs by about 800\,K, 
but this large deviation does not point to particular problems in some parameters' range as verified through the node-to-node comparison.

The OACT systematic discrepancies in \teff\ above 6500\,K, on the other hand, indicate a \teff\ upper limit also for the validity of \texttt{ROTFIT} UVES analysis.
This discrepancy is seen also in the node-to-node comparison, on which we estimate an upper limit of 6200\,K for the validity of the OACT results.

Excluding such outliers, the standard deviation is $\approx$ 100\,K for the Arcetri, CAUP, and UCM nodes, and $\approx$ 120\,K for the OACT node, with average difference of 34\,K for CAUP, 38\,K for OACT, and 55\,K for Arcetri and UCM.

Within the UVES dataset of young open clusters, very few sources have \teff\ $<$4000\,K.
In this range recommended data are based on OACT results only, 
as the presence of molecular bands prevents to carry out analysis based on MOOG.

\begin{table*}[ht]
\centering
\caption{UVES results on the solar spectrum.}
\begin{tabular}{rrrrrrrrr}
\hline
           &  $\Delta T_{\rm eff}$  & $\sigma T_{\rm eff}$ & $\Delta \log g$ & $\sigma  \log g$  & $\Delta$ [Fe/H] &  $\sigma $ [Fe/H] &  $\Delta \xi$  &  $\sigma_{\xi}$ \\
\hline
      OACT  &  $-$1.  &      67. & $-$0.14 &      0.11 &   0.06 &   0.10 &     \dots  &     \dots   \\
   Arcetri  &    49.  &     150. &    0.09 &      0.20 &   0.09 &   0.07 &      1.00  &     0.15    \\
      CAUP  & $-$48.  &      59. & $-$0.18 &      0.12 &   0.00 &   0.07 &      0.87  &     0.08    \\
       UCM  &     9.  &      48. &    0.00 &      0.11 &   0.03 &   0.04 &      0.75  &     0.08    \\
\hline
\end{tabular}
\label{tab:UVES_solar}
\end{table*}

The agreement in \logg\  is approximately at the same level for all nodes. 
Benchmarks' \logg\ is generally reproduced within a maximum deviation of $\approx0.7$ dex and a standard deviation of $\approx0.3$, only one Arcetri value deviating more than that.

\feh\ is generally reproduced within a maximum deviation of $\approx0.3$ dex , except one and two measurements by the Arcetri and OACT nodes, respectively, with deviations of $\approx0.5$ dex.
The standard deviation is $\lesssim 0.1$ dex for the CAUP and UCM nodes, $\approx0.2$ dex for the Arcetri node, and $\approx0.3$ for the OACT node. 
The OACT node tends to overestimate (underestimate) the metallicity below (above) \feh=0.
However, this does not lead to significant systematic differences in individual clusters and the OACT results are therefore maintained. 
The node-to-node comparisons for individual clusters show that the Arcetri node systematically overestimates \feh, which is not evident in the comparison with the benchmarks possibly because of the large and coarse parameters' distribution of this latter.
To overcome this systematic behaviour, in iDR2 the value obtained by the Arcetri node for the solar spectrum (\feh=0.09, see Table\,\ref{tab:UVES_solar}) is subtracted in all measurements before computing the recommended \feh.
The recommended \feh\ agrees with the benchmarks within 0.15\,dex r.m.s.

Solutions with large uncertainties or large $\xi$ ($\gtrsim 2$ \kms) are disregarded by the nodes.
Differences in $\xi$ with respect to the values tabulated for the benchmarks are generally below 1 \kms.

The recommended fundamental parameters are therefore computed taking the average of the nodes' results with a 1$\sigma$-clipping when at least 3 values are provided.
As discussed above, below 4000\,K only the OACT values are given as recommended values.
In iDR2 we disregarded the OACT UVES values for \teff$>$6200\,K.

Note that, despite the large difference in resolution and spectral range, the comparison with benchmarks shows that the UVES \teff\ accuracy is only marginally better than GIRAFFE's, while \logg\ and \feh\ results from the two setups are of comparable accuracy.
Our recommended values include \teff\ and \feh\ for 11 stars and \logg\ for 3 stars (see Sect.\,\ref{sec:internal}) from both the UVES and GIRAFFE setups.
The comparison of our results for the same stars in the two setups shows that the \teff\ ratio (GIRAFFE/UVES) has a mean of 0.99 and a median of 1.00. 
The differences in \feh\ (GIRAFFE-UVES) have a mean of 0.13\,dex and a median of 0.16\,dex.
Among the 3 benchmark stars for which we give recommended \logg\ from both GIRAFFE and UVES setups according to the criteria described in Sect.\,\ref{sec:internal}, two are in the range of interest (\logg $\approx$ 4.0) and the maximum difference with the benchmark value is -0.09\,dex. 

\subsection{Internal comparison}
\label{sec:internal}

The node-to-node comparison for the UVES individual cluster results before data selection and calibration (see Sect.\,\ref{sec:benchmarks}) gives systematic differences in the ranges 80--160\,K in \teff, 0.1--0.3\,dex in \logg, and 0.06--0.17\,dex in \feh, while dispersions are in the ranges 160--260\,K in \teff, 0.1--0.3\,dex in \logg, 0.13--0.45\,dex in \feh.
The application of the data selection and calibration discussed in Sect.\,\ref{sec:benchmarks} reduces systematic differences below 100\,K in \teff, and below 0.15\,dex in \feh.
The final node-to-node mean dispersion in the recommended data is 110\,K in \teff, 0.21\,dex in \logg, and 0.10\,dex in \feh.
These values are very close to the median dispersion: 106\,K in \teff, 0.17\,dex in \logg, and 0.11\,dex in \feh.
Biases in the recommended data are therefore successfully reduced.

For the GIRAFFE results, systematic differences before data selection are in the ranges 110--200\,K in \teff, 0.4--0.8\,dex in \logg, and 0.01--0.03 in \feh, while dispersions are in the ranges 210--330\,K in \teff, 0.65--1.00\,dex in \logg, and 0.17--0.26\,dex in \feh.
The situation here is more complex than in the UVES case. The problems to address are:

\begin{enumerate}
\item[(1)]{The OACT (\texttt{ROTFIT}) \logg\ for PMS stars tend to be too high, clustering essentially on the MS\footnote{This is due to the basic criteria for defining the templates, identified as slow rotators, inactive stars and with no significant Li-absorption, which imply that no PMS star can be taken as template.};}
\item[(2)]{The OAPA \logg\ for PMS stars tends to be too low, often lower than suggested by models\footnote{An absolute calibration of the gravity-sensitive spectral index in the PMS is very difficult (or impossible with currently available data) because of the lack of suitable PMS calibrators.};}
\item[(3)]{The PMS domain is contaminated by non-members with spurious \logg\ in both \logg-\teff\ diagrams.}
\item[(4)]{The RGB in the OACT \logg-\teff\ diagram follows the calibrated relation taken from \cite{Cox:2000}, while in the OAPA diagram it doesn't.}
\item[(5)]{The OAPA \logg-\teff\ diagram outside the MS, PMS, and RGB domains is sparsely populated, with both some very low and very high values, which are indicative of possible presence of some large errors.}
\item[(6)]{The OAPA \teff, \logg, and \feh\ are valid for \vsini $<$ 90, 30, and 70 \kms, respectively.}
\item[(7)]{Because of a continuum normalisation problem on the \halpha\ wings, in iDR1 and iDR2 the OACT parameters need to be discarded for \teff $>$ 5500\,K.}
\end{enumerate}

In order to reduce biases as much as possible and provide reliable recommended results we adopt the following solution:

\begin{enumerate}

\item{The OAPA \teff\ are considered only for \vsini\ $<$ 90 \kms. The OACT \teff\ are considered only below 5500\,K. In cases where both the OACT and OAPA \teff\ are available these are averaged. In all other cases the remaining value, if any, is adopted as recommended \teff.}
\item{The OAPA \feh\ are considered only for \vsini\ 	$<$ 70 \kms. The OACT \feh\ are considered only below 5500\,K. In cases where both the OACT and OAPA \feh\ are available these are averaged. In all other cases the remaining value, if any, is adopted as recommended \feh.}
\item{The OAPA \logg\ are considered only for \vsini\ $<$ 30 \kms. The OACT \logg\ are considered only for \teff$<$5500\,K. In cases where both the OACT and OAPA \logg\ are available these are averaged if they differ by less than 0.3 dex. When they differ by more than 0.3 dex, if the OACT \logg\ $>$ 4.2  and the OAPA \logg\ $>$ 5, the OACT \logg\ is given as recommended value. In all other cases we do not give recommended \logg.}
\item{The OAPA  gravity--sensitive $\gamma$ index \citep{Damiani_etal:2014} is given as a recommended parameter for \vsini\ $<$ 30 \kms.}
\end{enumerate}

The application of such criteria leads to a final node-to-node mean dispersion in the recommended data of 98\,K in \teff, 0.23\,dex in \logg, and 0.14\,dex in \feh.
These values are very close to the median dispersion: 95\,K in \teff, 0.22\,dex in \logg, and 0.14\,dex in \feh.
Biases in the recommended data are therefore successfully reduced in the GIRAFFE case too.

When a recommended \logg\ is not given, it may be still possible to identify an approximate evolutionary status based on the OACT and OAPA results. Those stars for which a trustworthy \logg\ cannot be recommended are therefore flagged, when possible, as PMS, MS or post-MS stars according to the criteria listed in Table\,\ref{tab:evolutionary-status}.

\begin{table}[ht]
\centering
\caption{Criteria for the evolutionary status.}
\label{tab:evolutionary-status}
\begin{tabular}{cccl}
\hline\hline
\noalign{\smallskip}
\teff      & \logg$_{\rm OACT}$  & \logg$_{\rm OAPA}$ & Status\\
\noalign{\smallskip}
\hline
\noalign{\smallskip}
$< 5500$\,K		& $> 3$		& $3$--$4.2$	& PMS\\
$< 5500$\,K		& $3$--$4.2$	& \dots			& PMS\\
\dots			& \dots			& $> 4.2$		& MS\\
\dots			& $< 3$		& $< 3$		& post-MS\\
\noalign{\smallskip}
\hline
\end{tabular}
\end{table}

\begin{figure}[t]
\centering
\includegraphics[width=90mm]{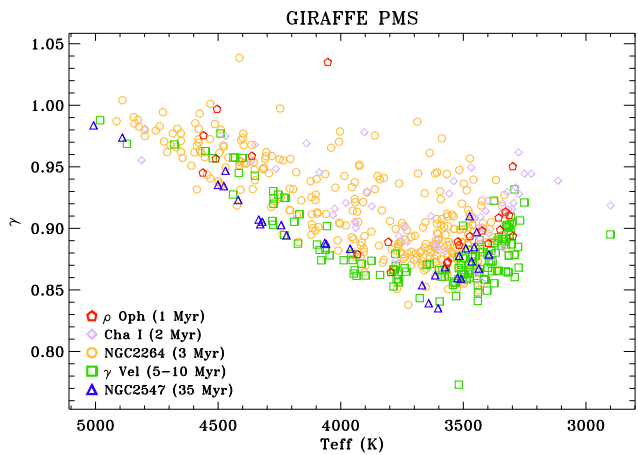}
\caption{The OAPA gravity-sensitive $\gamma$ index vs. \teff\ for all clusters analysed in iDR1 and iDR2. The group of younger clusters ($\rho$ Oph, Cha\,I, and NGC2264) are clearly distinguishable from the group of older clusters ($\gamma$ Vel and NGC2547).}
\label{fig:giraffe_gamma_teff_pms}
\end{figure}

The gravity-sensitive spectral index $\gamma$ obtained by the \cite{Damiani_etal:2014} approach can provide a rank order in age of the clusters.
This can be seen, for the clusters analysed to date, in Fig.\,\ref{fig:giraffe_gamma_teff_pms}, where values for the younger clusters group ($\rho$ Oph, Cha\,I, and NGC2264) are clearly separated from those of older clusters group ($\gamma$ Vel and NGC2547).
However, both the scatter in $\gamma$ and the small age differences between clusters in the younger or the older group still prevent a clear separation in age.

\subsection{Overview in the \logg--\teff\ plane}
\label{sec:LoggTeff}

\begin{figure}[htp]
\centering
\includegraphics[width=90mm]{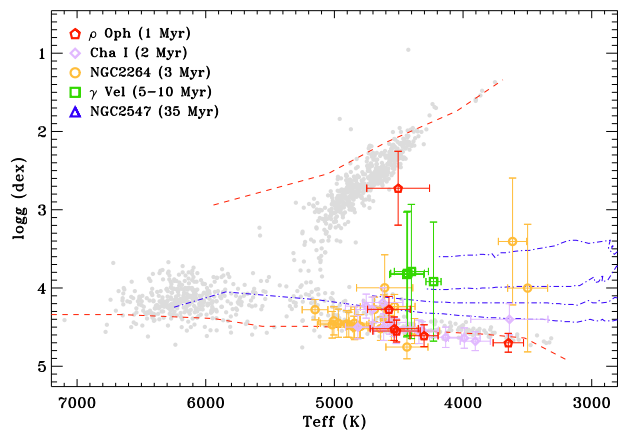}
\caption{\logg--\teff\ diagram for all targets. Grey filled circles are used for clusters' non-members and stars not classified as CTTS nor WTTS.  The red dashed lines are the dwarfs and giants sequences from \cite{Cox:2000}. The blue dot-dashed lines are the isochrones at 1, 5, 10, and 20 Myr from \cite{Allard_etal:2011}.}
\label{fig:giraffe_hr_diagram_pms}
\end{figure}

As a final check on our recommended fundamental parameters,  we examine the \logg--\teff\ diagram obtained with our data (Fig.\,\ref{fig:giraffe_hr_diagram_pms}) and compare it with the calibration of MK spectral classes reported in \cite{Cox:2000} and the theoretical PMS isochrones from \cite{Allard_etal:2011}.
We note the clustering of field stars on the MS and the RGB, as expected, while for the PMS clusters' members 
a residual bias towards the MS and the RGB remains.
In $\lesssim 1/2$ of the cases, \logg\ values for PMS stars are located approximately where predicted by the models, although with large uncertainties. 

\subsection{Comparison between fundamental parameters derived from GIRAFFE and UVES}

\begin{figure*}[htp]
\centering
\includegraphics[width=63mm]{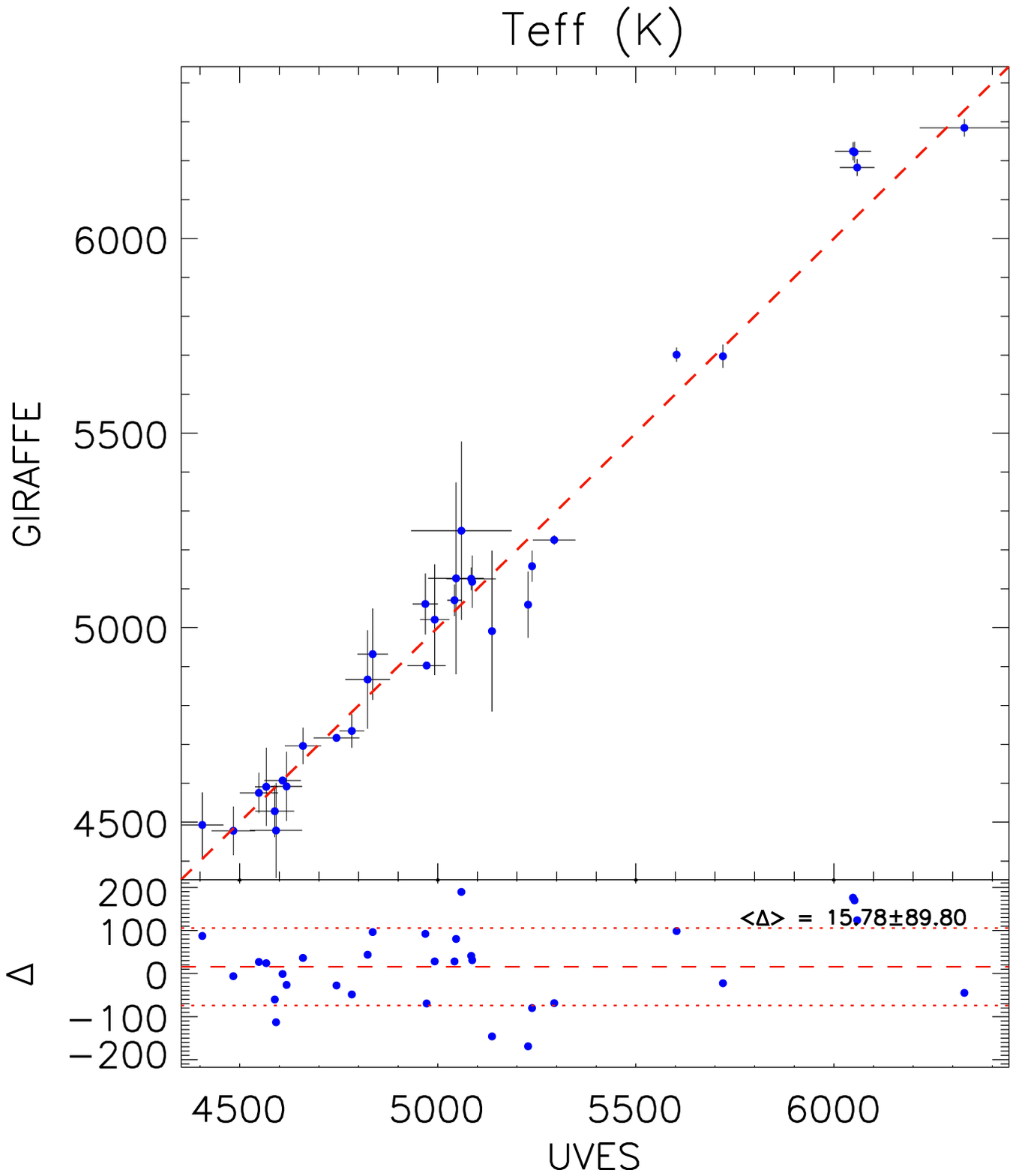}
\hspace{-0.5cm}
\includegraphics[width=63mm]{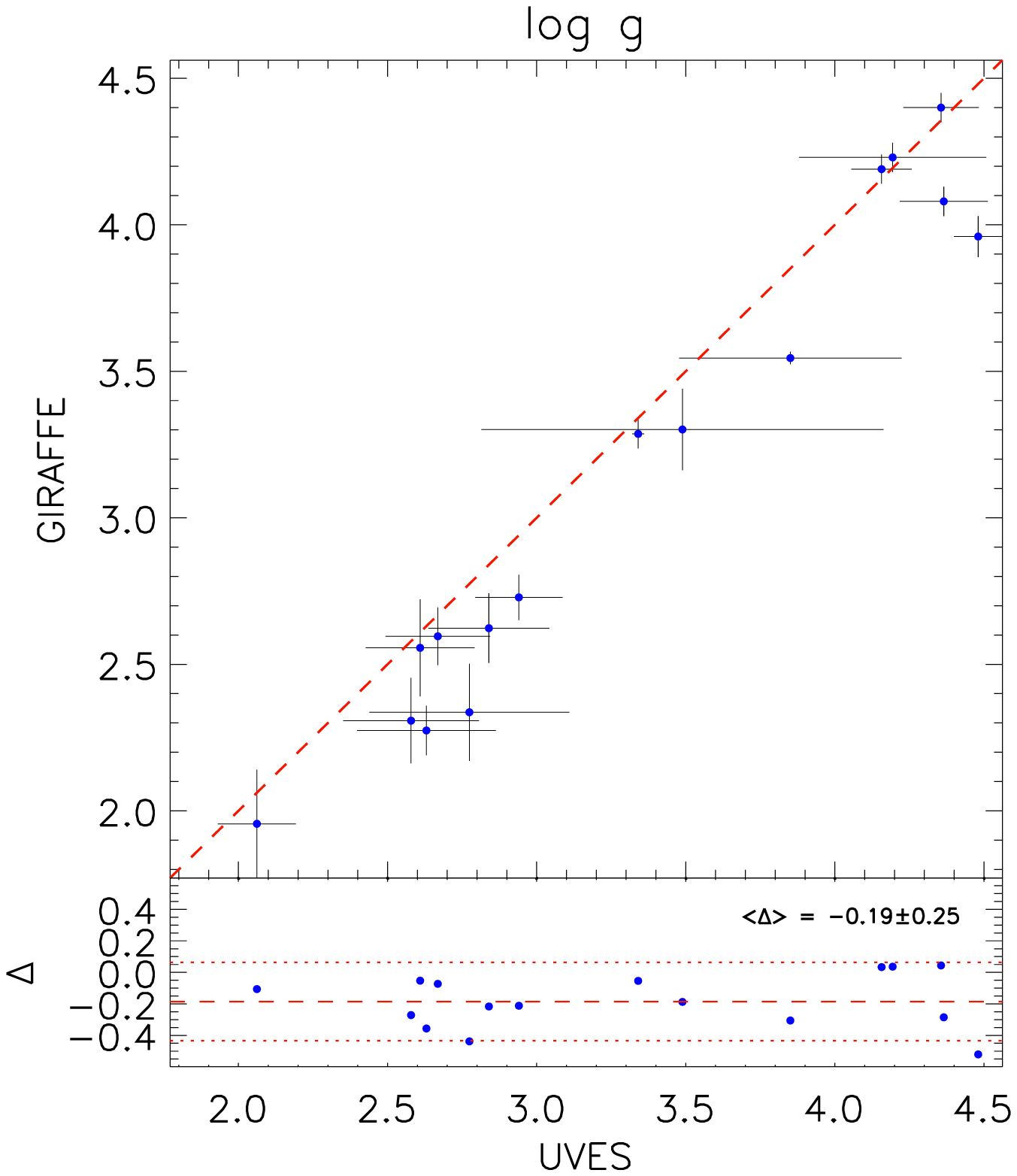}
\hspace{-0.5cm}
\includegraphics[width=63mm]{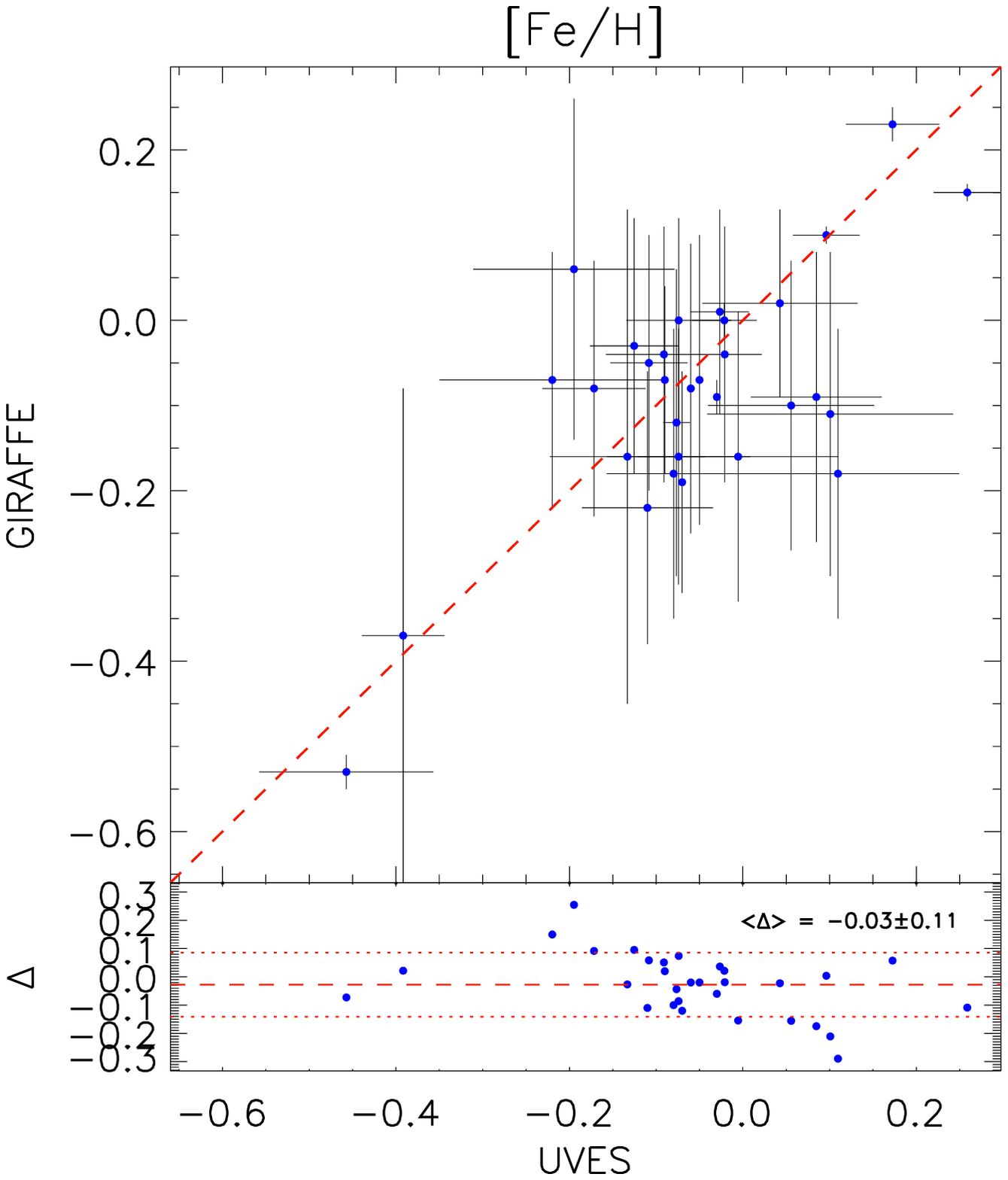}
\caption{Comparison between fundamental parameters derived from GIRAFFE and UVES spectra in the $\gamma$ Vel field (iDR2).}
\label{fig:best_giraffe_uves}
\end{figure*}

A number of stars in the $\gamma$ Vel field have been observed with both UVES and GIRAFFE. 
For iDR2, our analysis produced \teff\ and \feh\ values for 31 stars and \logg\ for 16 stars in this common sample.
Note that the lower number of \logg\ values is due to the application of the criteria described in Sect.\,\ref{sec:internal}, which were applied to iDR2 but not to iDR1.
The comparison of the recommended values for this sample is satisfactory (see Fig.\,\ref{fig:best_giraffe_uves} for iDR2) and support the validity of our approach both in the parameters determination and in the derivation of the recommended values.
A similar comparison is reported in \cite{2014A&A...567A..55S} for iDR1.
Indeed the reproducibility of the parameters obtained with the higher resolution and larger wavelength coverage from UVES using a much smaller wavelength range and a lower resolution as in GIRAFFE is a remarkable achievement and increases our confidence in our parameters determination from the much larger GIRAFFE sample.

\subsection{Veiling vs. \halpha\ emission}

For iDR1, the \texttt{ROTFIT} veiling parameter was adopted as the recommended one.
In iDR2, however, it was recognised that some residual nebular emission remained after sky-subtraction, particularly in NGC2264, which were not sufficiently masked in the \texttt{ROTFIT} calculations.
As a consequence, the \texttt{ROTFIT} veiling parameter for NGC2264 was clearly overestimated and the OAPA solution was adopted as recommended in iDR2.
Note that this does not invalidate the results of iDR1 as $\gamma$ Vel and Cha\,I spectra are not affected by residual sky emission in the reduced spectra.

\begin{figure}[t]
\centering
\includegraphics[width=80mm]{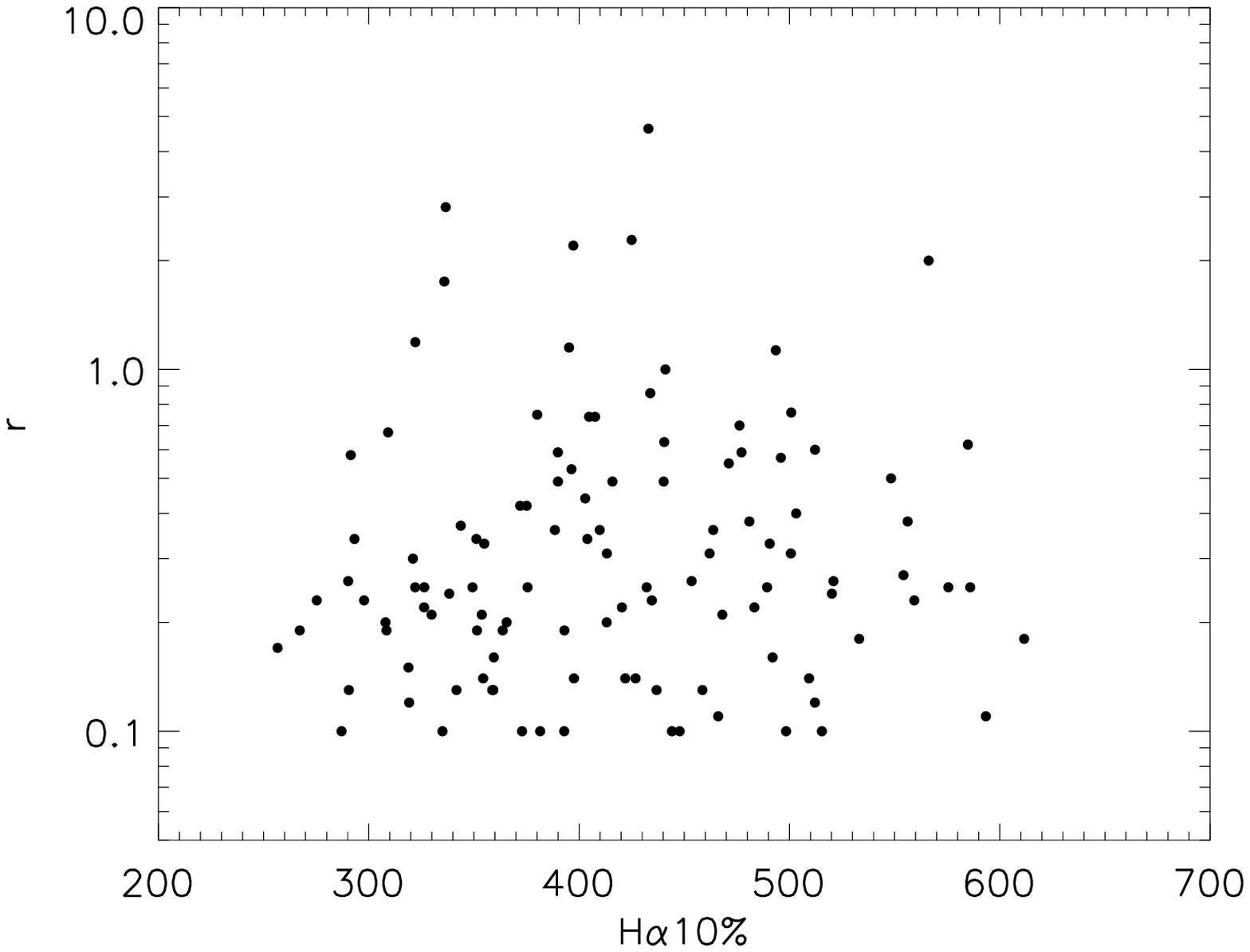}
\includegraphics[width=80mm]{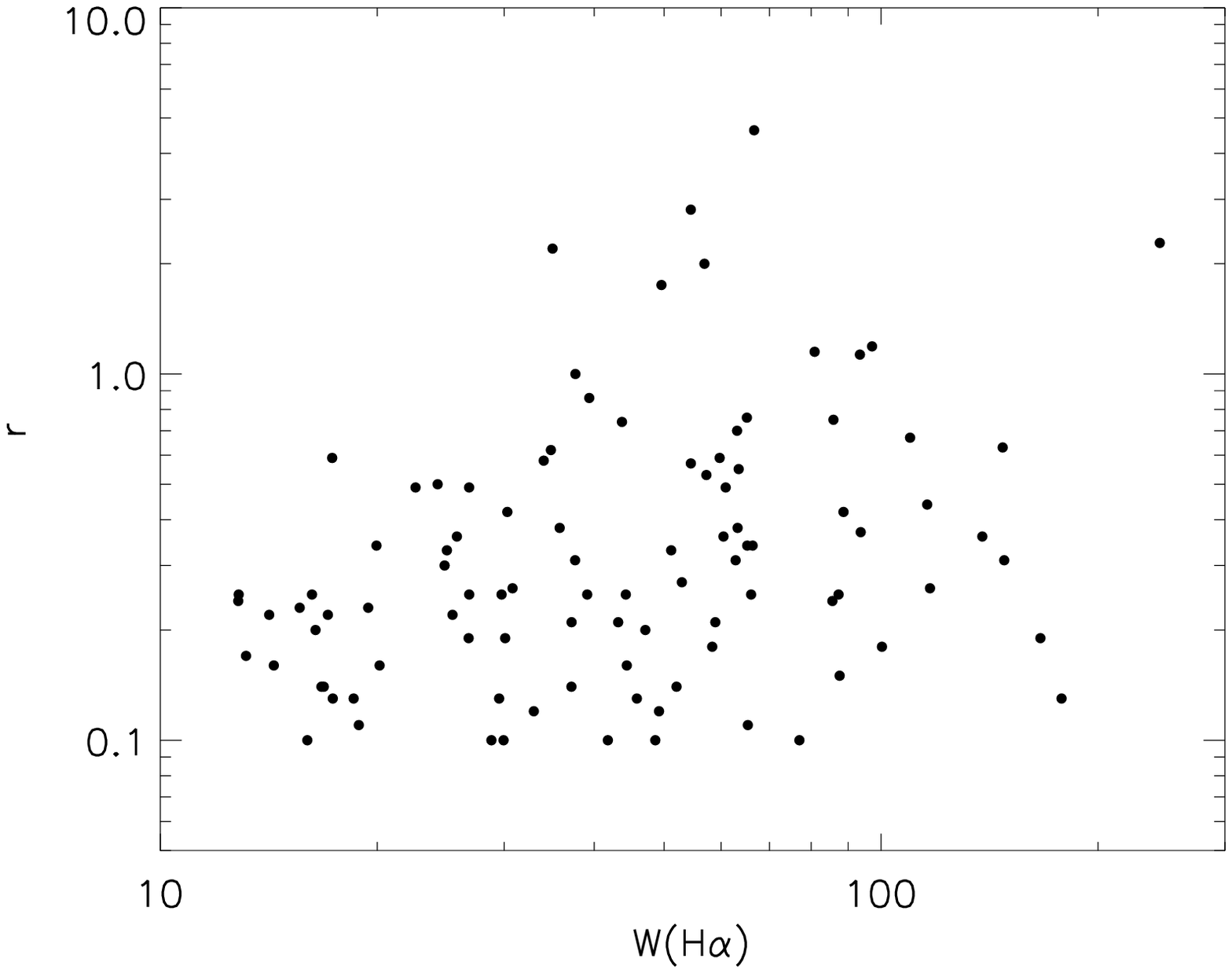}
\caption{Veiling parameter $r$ vs. \ha10\ (top panel) and vs. \wha\ (bottom panel) for iDR2. }
\label{fig:veiling_vs_ha}
\end{figure}

\cite{Frasca_etal:2014} found a positive correlation between \halpha\ flux and $r$ in the iDR1 data for Cha\,I objects with $r \ge 0.25$, for which the Spearman's rank analysis yielded a coefficient $\rho=0.58$ with a significance of $\sigma=0.003$.
The same analysis for all clusters in iDR2 gives a coefficient $\rho=0.39$ with a significance of $\sigma=0.004$.
However, a correlation between $r$ and \ha10\ or \wha\ is not evident in the iDR2 data (see Fig.\,\ref{fig:veiling_vs_ha}), where we do see an increase of the upper envelope with either \ha10\ or \wha, but the large scatter makes the correlation not significant. 
This is at variance with what expected from previous work \citep[e.g.,][]{2003ApJ...582.1109W} and therefore it outlines possible limitations in our veiling determination.
Further validation based on comparison with different methods is deferred to future work.

\section{Derived parameters}
\label{sec:DerivedParameters}

\subsection{Li abundance}
\label{sec:Li_abundance_methods}

In the whole GIRAFFE analysis, Li abundances, \ali, were computed from the fundamental parameters (Sect.\,\ref{sec:FundamentalParameters}) and the \wli\ measurements (Sect.\,\ref{sec:WLi}) using the curve of growth (COG) from \cite{Soderblom_etal:1993} and \cite{2007ApJ...659L..41P} above and below 4000\,K, respectively, with a linear interpolation between the tabulated values.
The recommended \ali\ is derived using the recommended fundamental parameters and recommended \wli\ as input.
Uncertainties were obtained by propagating the \teff\ and \wli\ uncertainties.

\begin{figure}[htp]
\centering
\includegraphics[width=80mm]{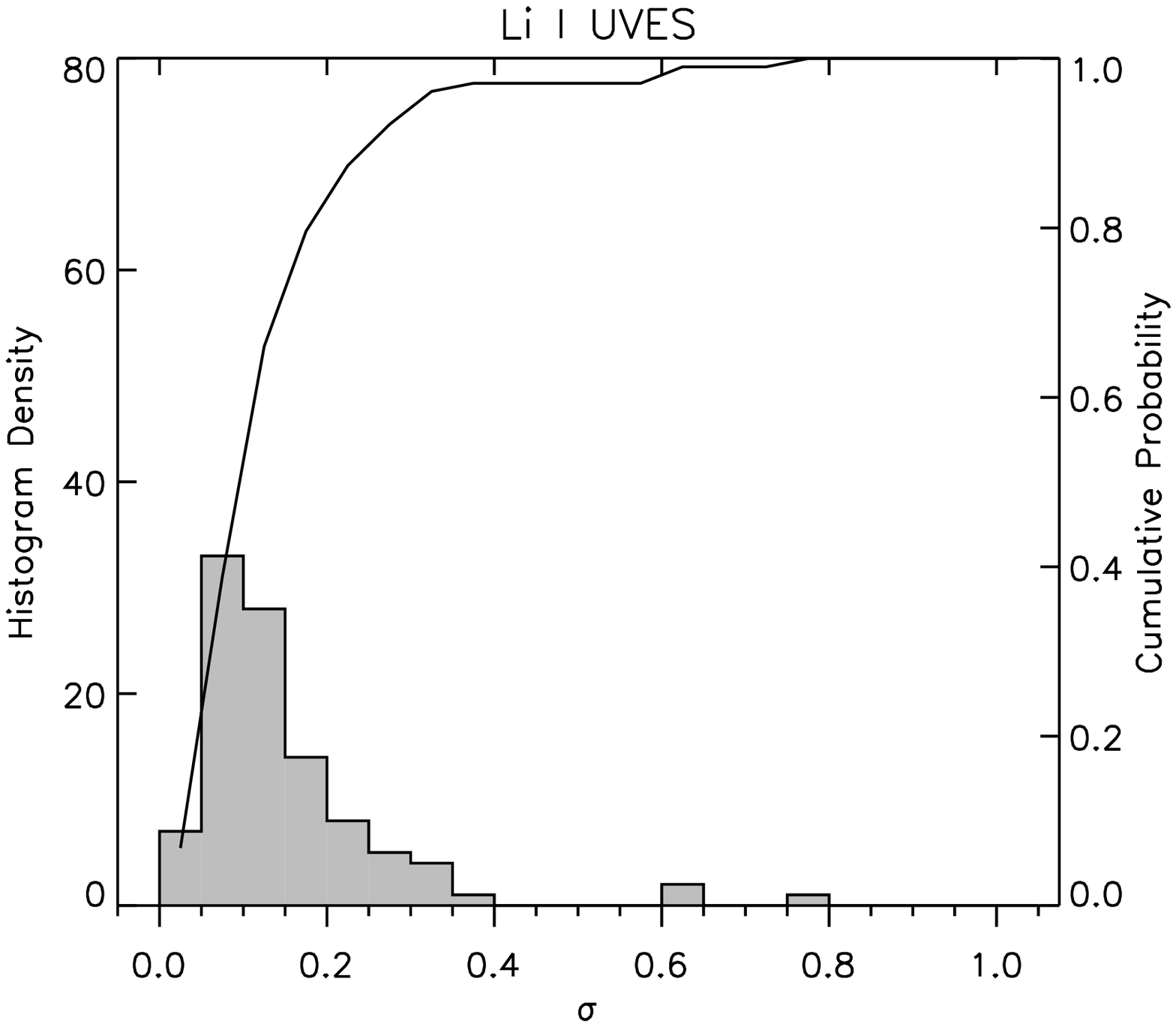}
\includegraphics[width=80mm]{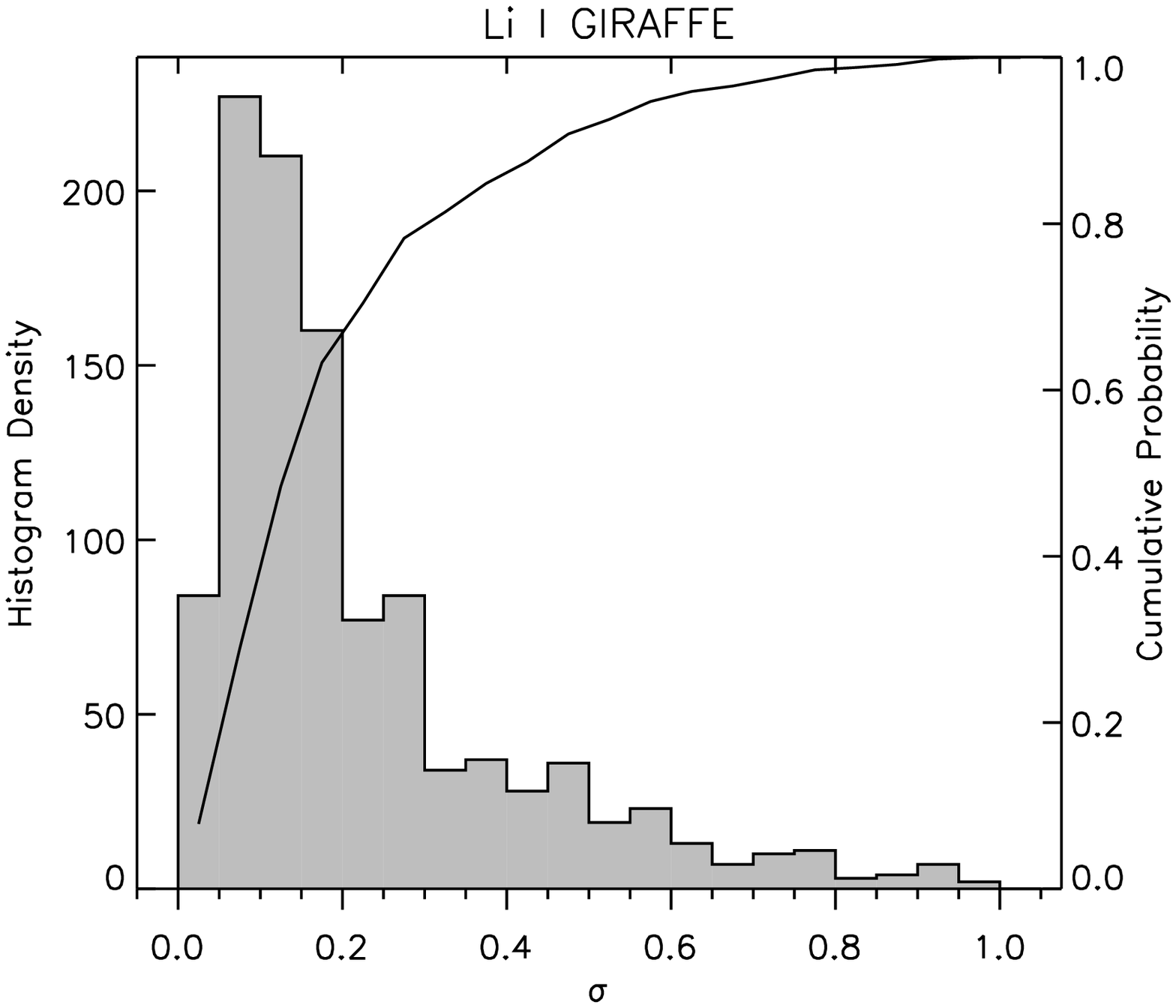}
\caption{Li abundance uncertainty histogram for all sources in iDR2. For the GIRAFFE spectra the \ali\ uncertainty is propagated from the uncertainty in \teff\ and \wli. For the UVES spectra the node-to-node dispersion is considered. A solid line is used for the cumulative probability (right ordinate axis). See text for details.}
\label{fig:sigma_Li1}
\end{figure}

The approach adopted in the GIRAFFE case has the advantage of allowing us to focus on the accuracy of the fundamental parameters and \wli, relying then on the best COG available to derive node-specific and recommended \ali.
Note that the two COGs adopted do not join smoothly at 4000\,K, but the interpolation scheme ensures a smooth transition between the two regimes.
A derivation of a self-consistent COG in the whole \teff\ range is planned as a future improvement.
The node-to-node dispersion in the GIRAFFE case (see Fig.\,\ref{fig:sigma_Li1} for the whole iDR2) then propagates only from the \wli\ measurements and shows a fairly random distribution wit a median of 0.17\,dex.

In the UVES analysis, the OACT and Arcetri \ali\ were derived as in the GIRAFFE case.
The CAUP and UCM nodes, on the other hand, derived \ali\ by a standard LTE analysis using the driver  \texttt{abfind} in the revised version of the spectral synthesis code \texttt{MOOG} \citep{sneden} (see also Sects.\, \ref{sec:FundamentalParameters} and \ref{sec:elemental_abundances_methods}). 
CAUP used the 2010 version of \texttt{MOOG}, while UCM used the 2002 and 2013 versions for iDR1 and iDR2, respectively.
Uncertainties were estimated by varying each atmospheric parameter within its uncertainty range to derive the propagated uncertainty in \ali. 
The propagated uncertainties where then combined quadratically.
In this case the recommended value is given as the average of all nodes' estimates available with a $\sigma$-clipping when at least three measurements are available.
Figure\,\ref{fig:sigma_Li1} shows that also in this case the node-to-node dispersion have a fairly random distribution, with a median uncertainty of 0.12 dex.

Possible $^{6}$Li contribution was neglected in all cases.

\subsection{Other elemental abundances}
\label{sec:elemental_abundances_methods}

Elemental abundances were computed by three nodes (Arcetri, CAUP, and UCM) when good quality UVES spectra were available in stars not affected by veiling and/or large \vsini.

The Arcetri node computed abundances using \texttt{FAMA}. We refer the reader to \citet{2013A&A...558A..38M} for a description of the method and the way in which lines are selected for the abundance analysis. 

The CAUP node derived individual abundances using the driver  \texttt{abfind} in the 2010 version
of \texttt{MOOG} \citep[see][for details]{2009A&A...497..563N,2012A&A...545A..32A} 
and equivalent widths measured with the \texttt{ARES} code.
The line list for elements other than Fe (with atomic number $A\leq$28) was selected through the cross-matching between the line list used by \cite{2012A&A...545A..32A} and the line list provided by Gaia-ESO.
For elements with $A>$28, lines that were suitable for $W$ measurements (as tested by the Gaia-ESO line-list working group) were first selected and from these the ones that \texttt{ARES} was able to measure were used.
The atomic data from the Gaia-ESO Survey was adopted.
CAUP considered hyperfine splitting in the analysis of Cu, Ba, Nd,
Sm and Eu abundances, i.e. for all the elements affected with
A $>$ 28 (using the driver \texttt{blends} in \texttt{MOOG}).
The errors of the abundances is given as the line-to-line scatter (when more than one line is measured).

The UCM node adopted an approach similar to CAUP.
For iDR1, two line-lists were prepared: one for dwarfs ($\log g \ge 4.0$) and one for giants ($\log g \le 4.0$). 
For iDR2 five line lists were used as done for the stellar parameters (see Sect.\,\ref{sec:FundamentalParameters}).
A total of 13 elements were analysed: Fe, the $\alpha$-elements (Mg, Si, Ca, and Ti), the Fe-peak elements
(Cr, Mn, Co, and Ni), and the odd-Z elements (Na, Al, Sc, and V).
To obtain individual abundances, the equivalent widths are fed into \texttt{MOOG} and then a 3$\sigma$-clipping for each chemical element was applied.

The elements for which at least two nodes derived abundances for at least one star and that were considered in the recommended results are:
\element{Na},
\element{Mg},
\element{Al},
\element{Si},
\element{Ca},
\element{Sc},
\element{Ti},
\element{V},
\element{Cr},
\element{Mn},
\element{Fe},
\element{Co},
\element{Ni},
\element{Zn},
\element{Zr},
\element{Mo},
\element{Ce}.
Only one node results were considered for:
\element{Cu},
\element{Y},
\element{Ba},
\element{La},
\element{Pr},
\element{Nd},
\element{Sm},
\element{Eu}.
Abundances are from the neutral species except for
\element{Ba},
\element{La},
\element{Ce},
\element{Pr},
\element{Nd},
\element{Sm},
\element{Eu},
for which they are from the ionised species.

\begin{figure*}[htp]
\centering
\includegraphics[width=40mm]{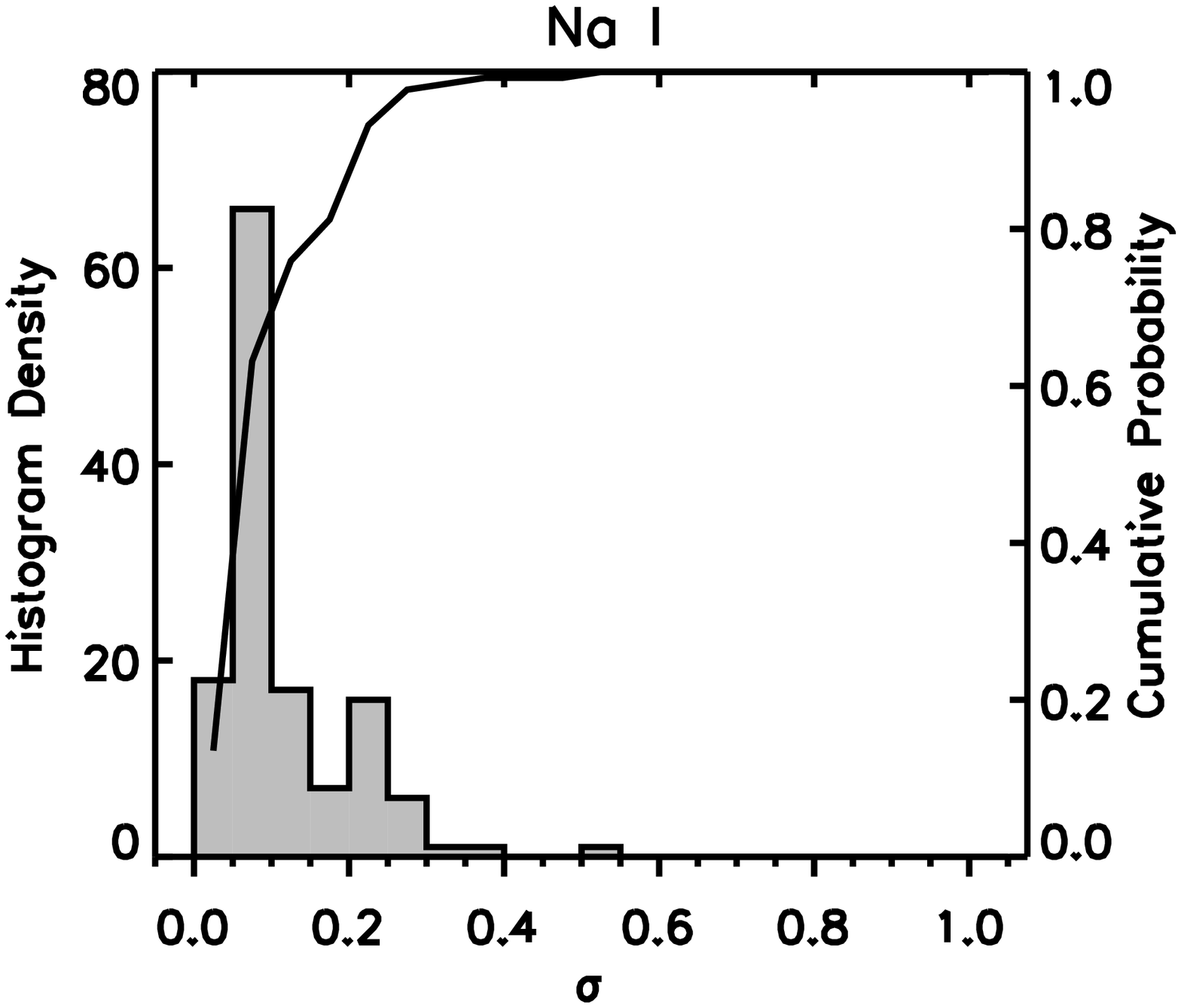}
\includegraphics[width=40mm]{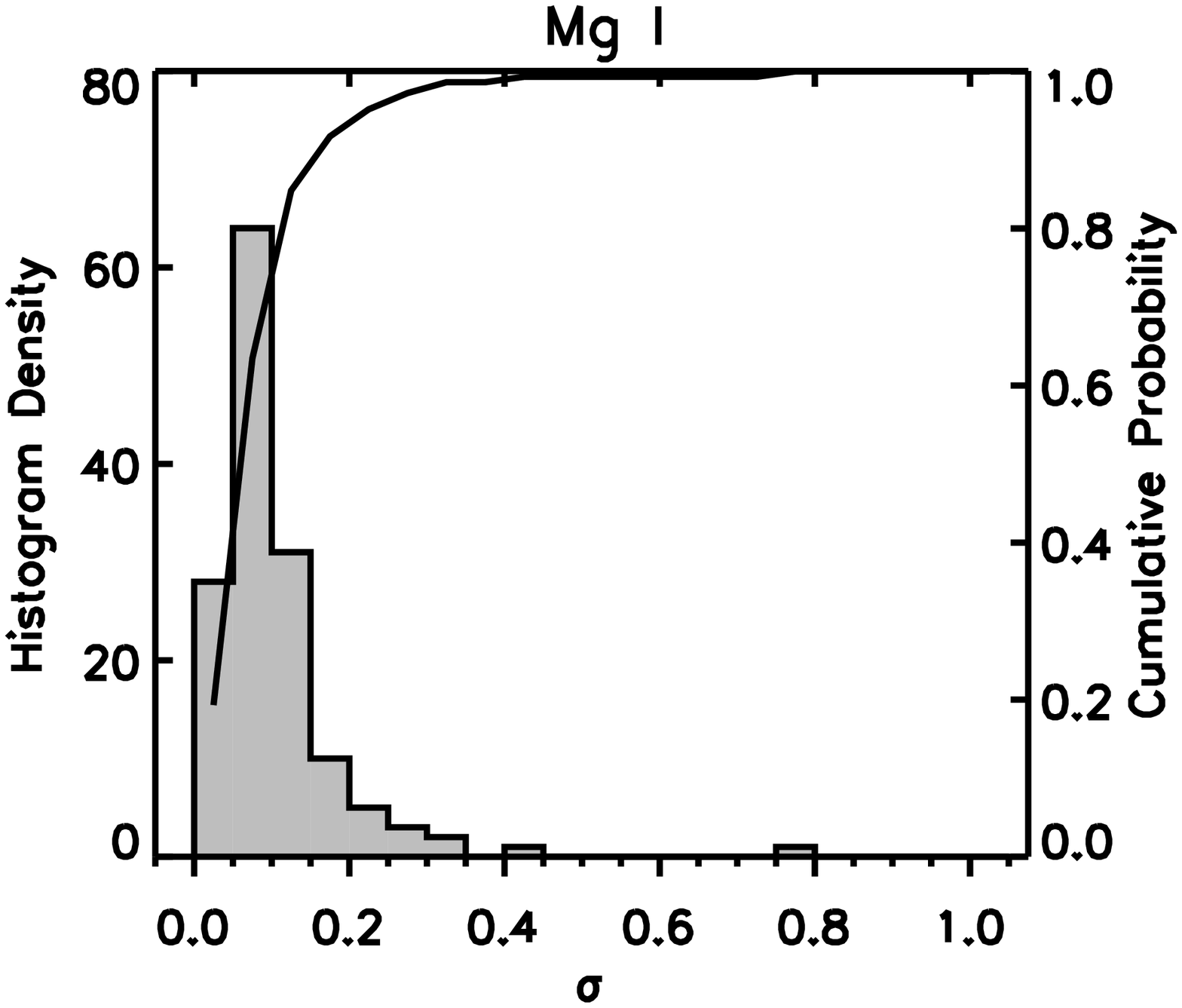}
\includegraphics[width=40mm]{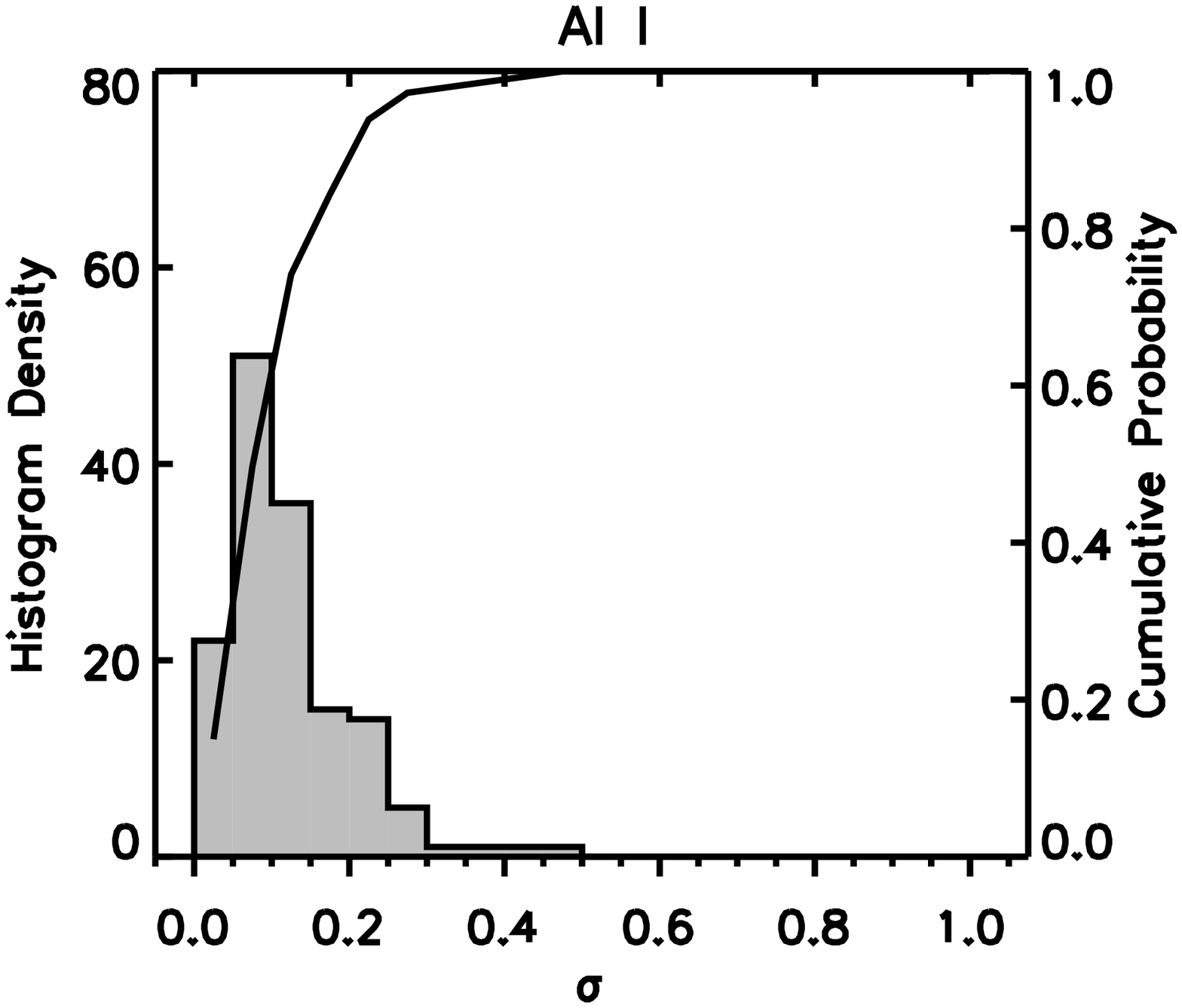}
\includegraphics[width=40mm]{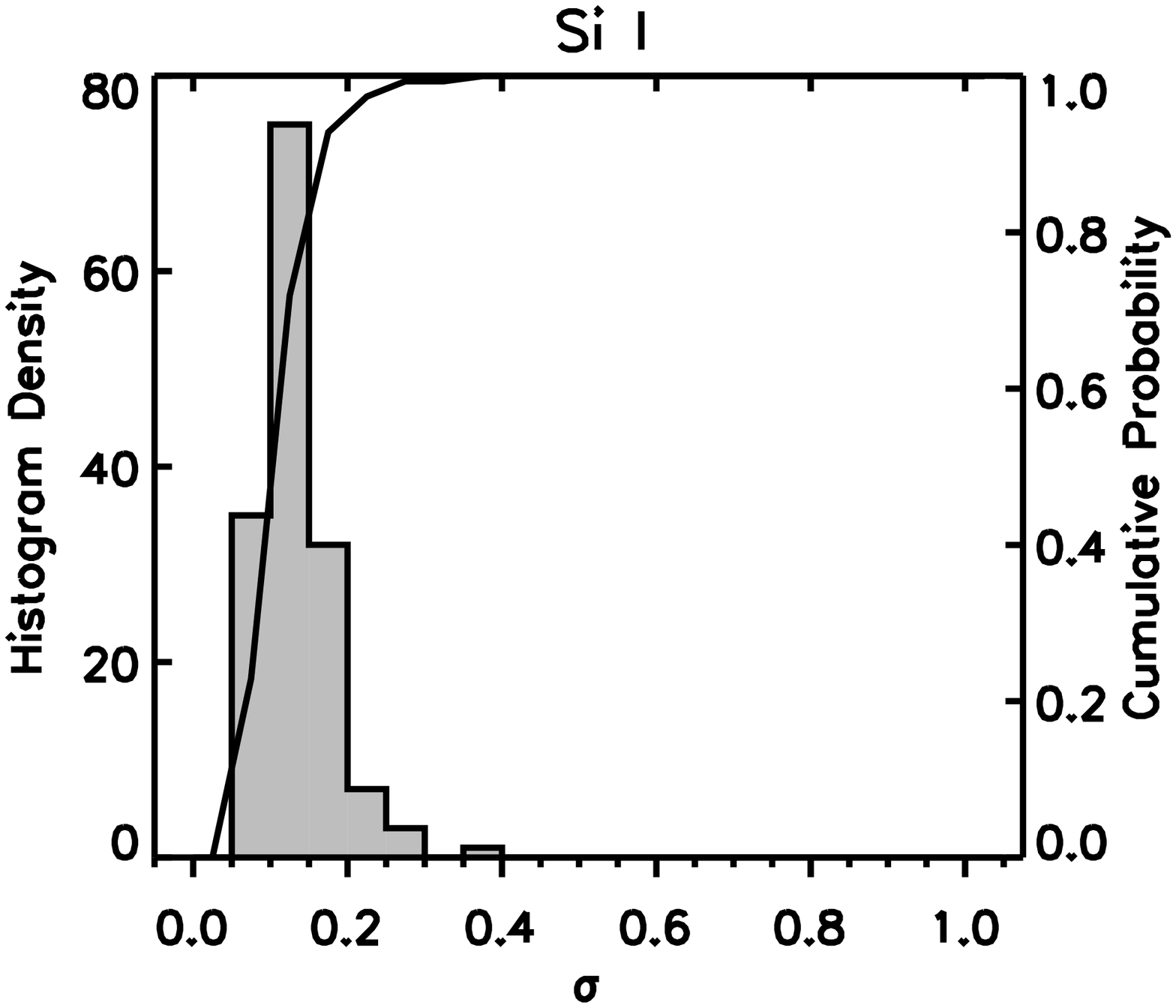}
\includegraphics[width=40mm]{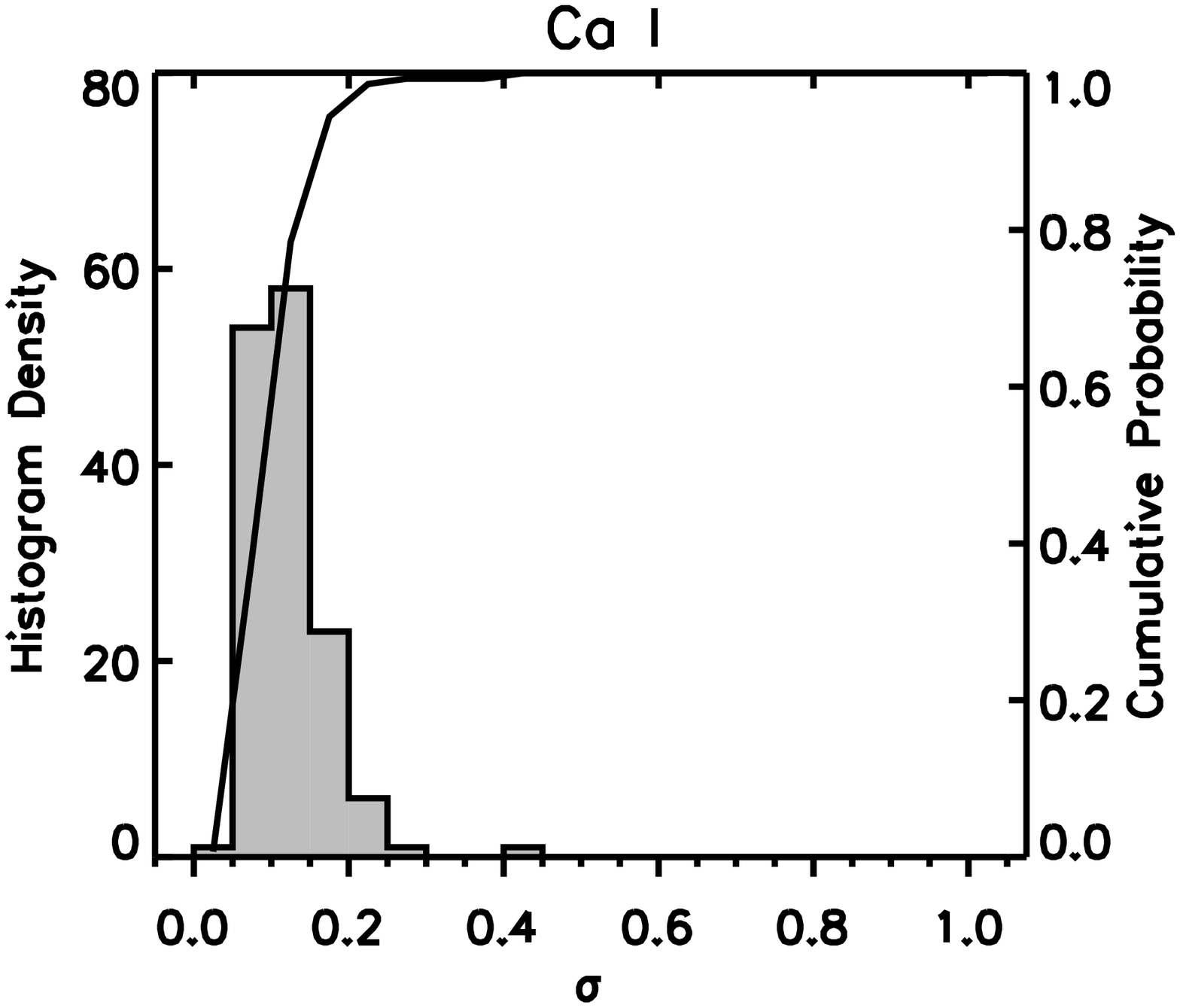}
\includegraphics[width=40mm]{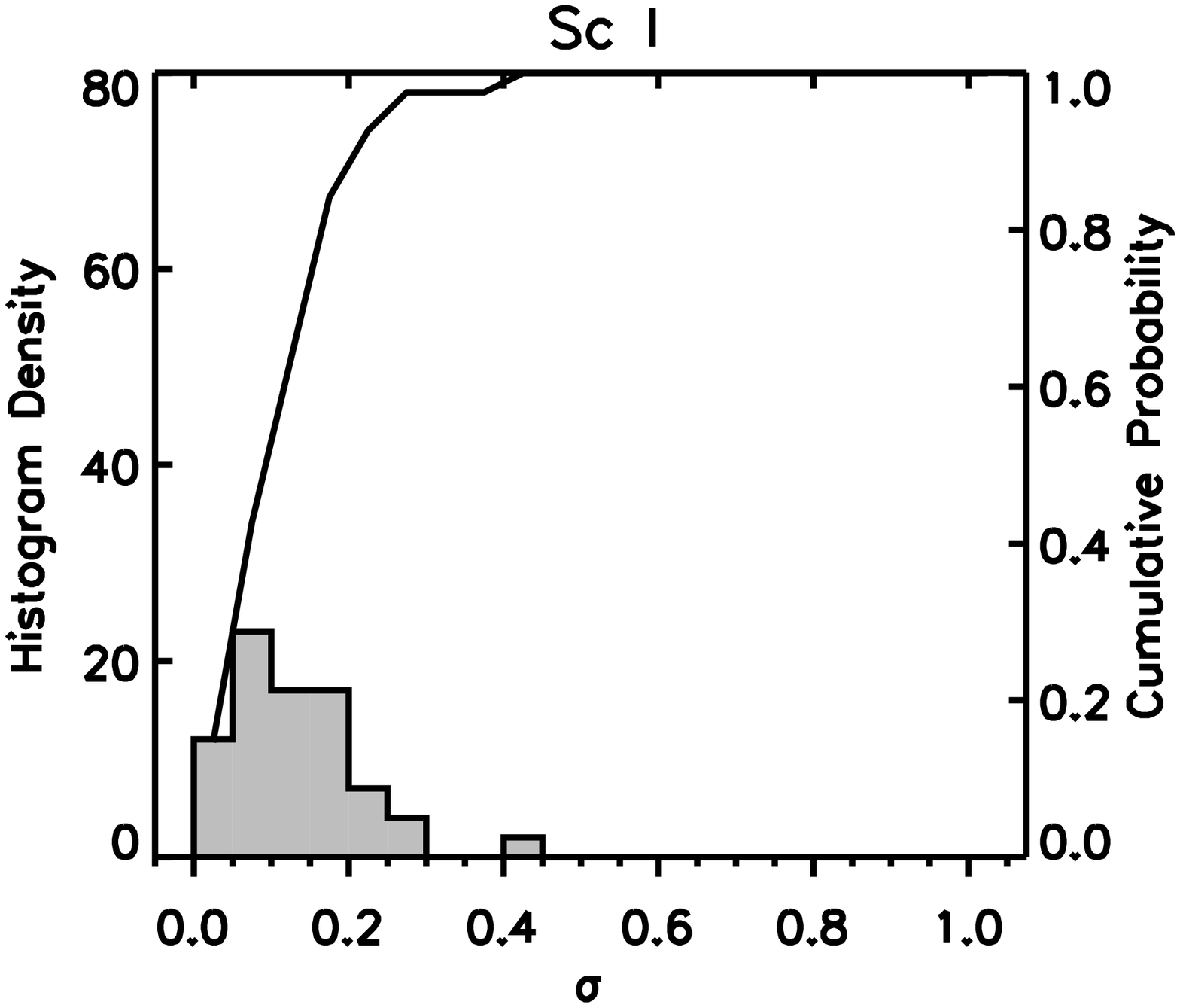}
\includegraphics[width=40mm]{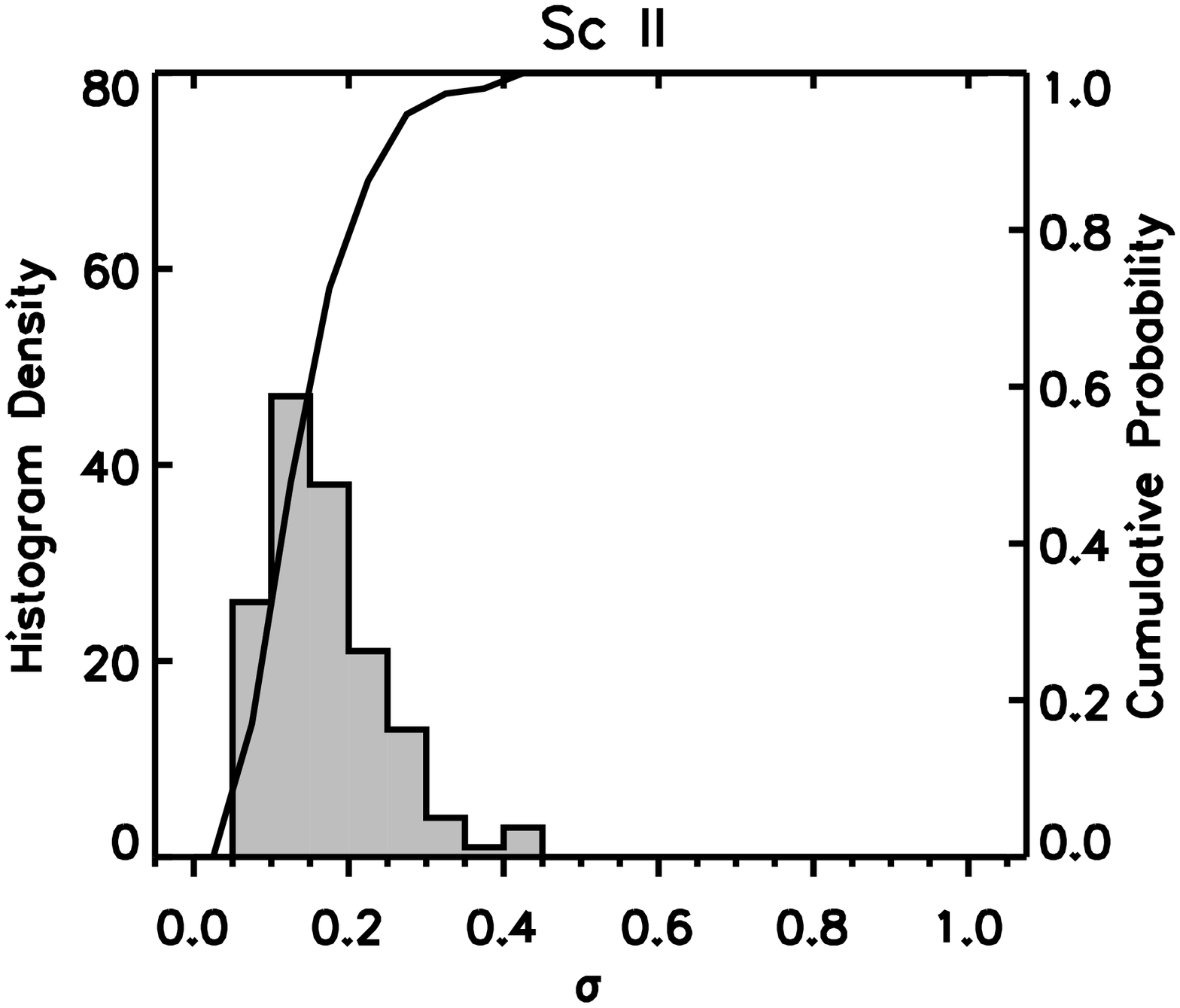}
\includegraphics[width=40mm]{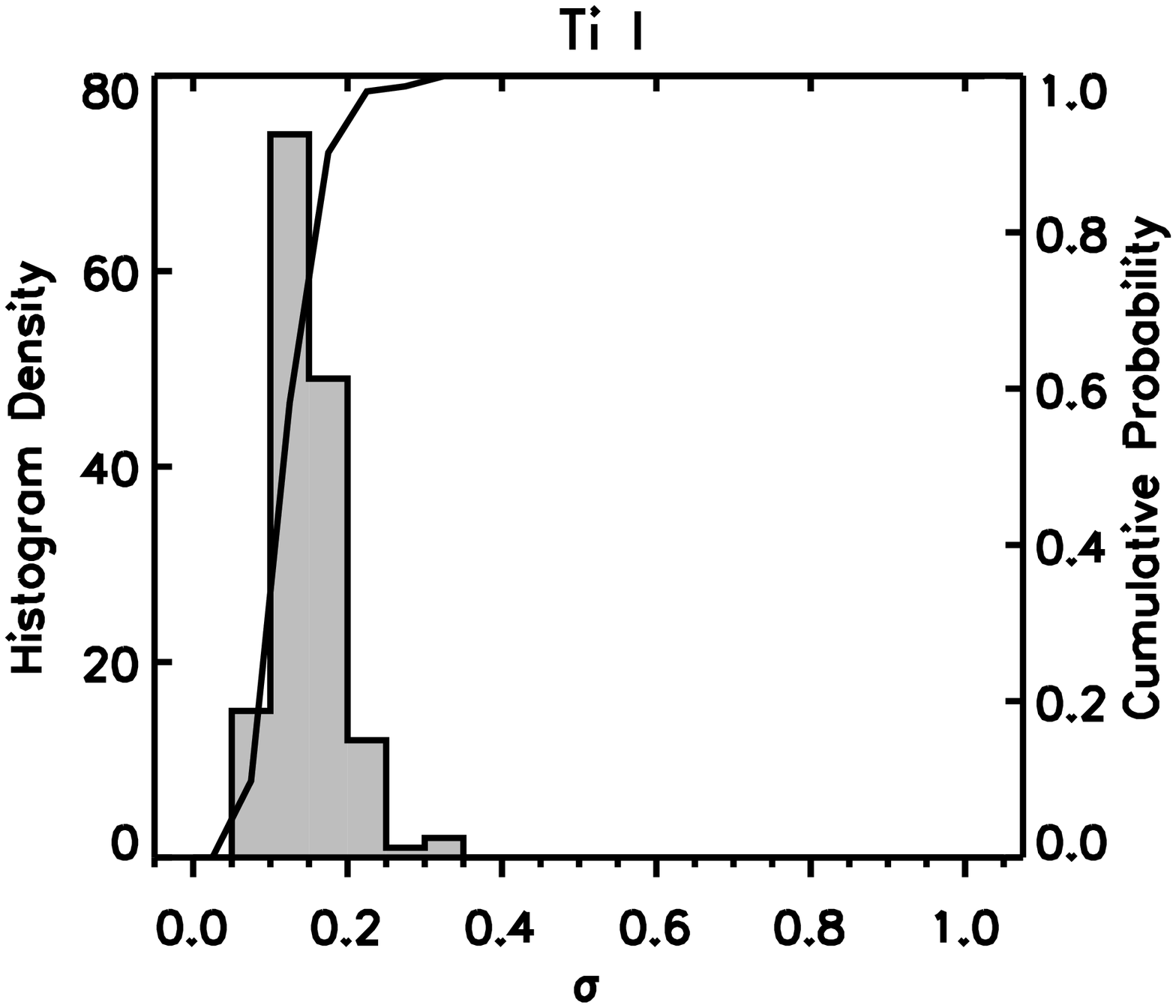}
\includegraphics[width=40mm]{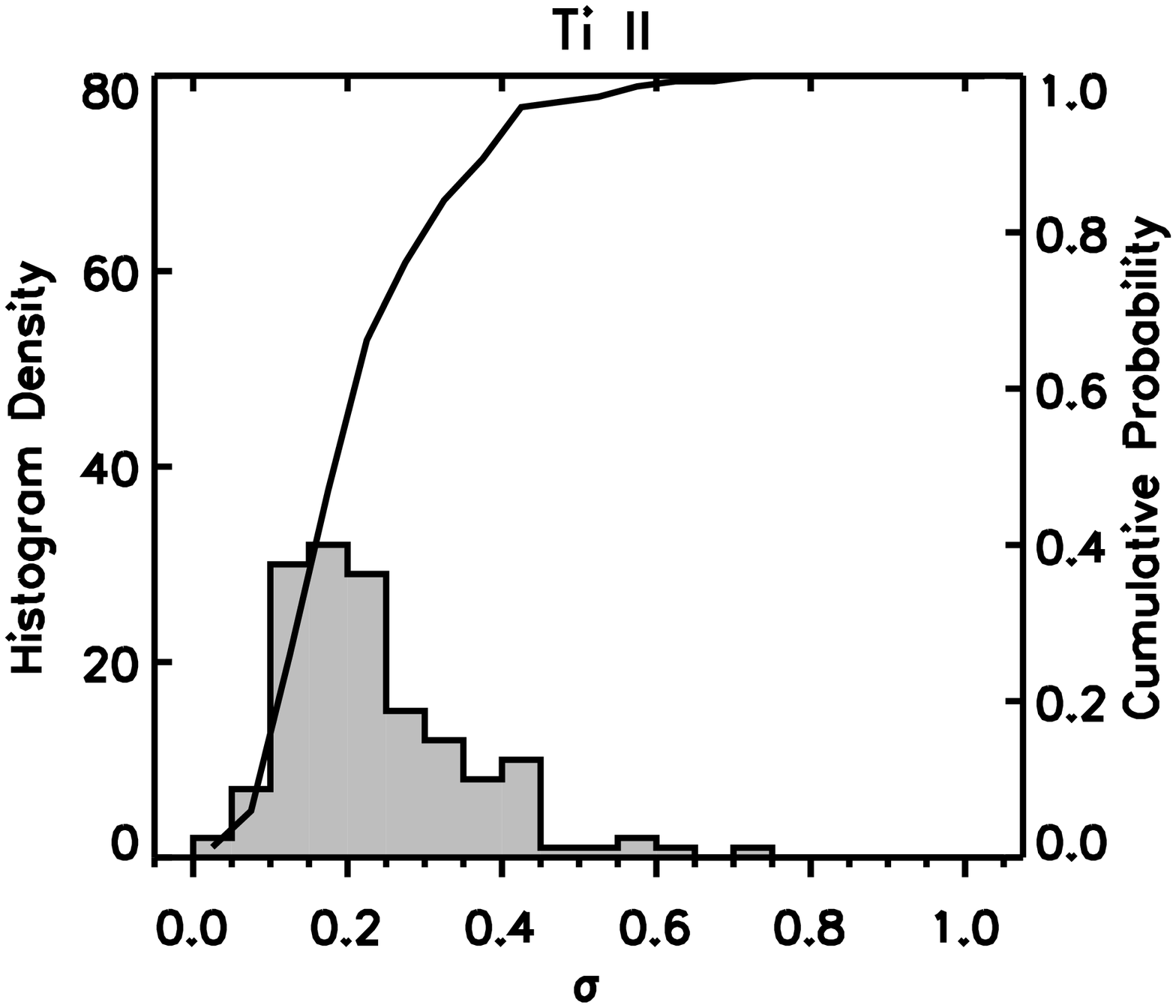}
\includegraphics[width=40mm]{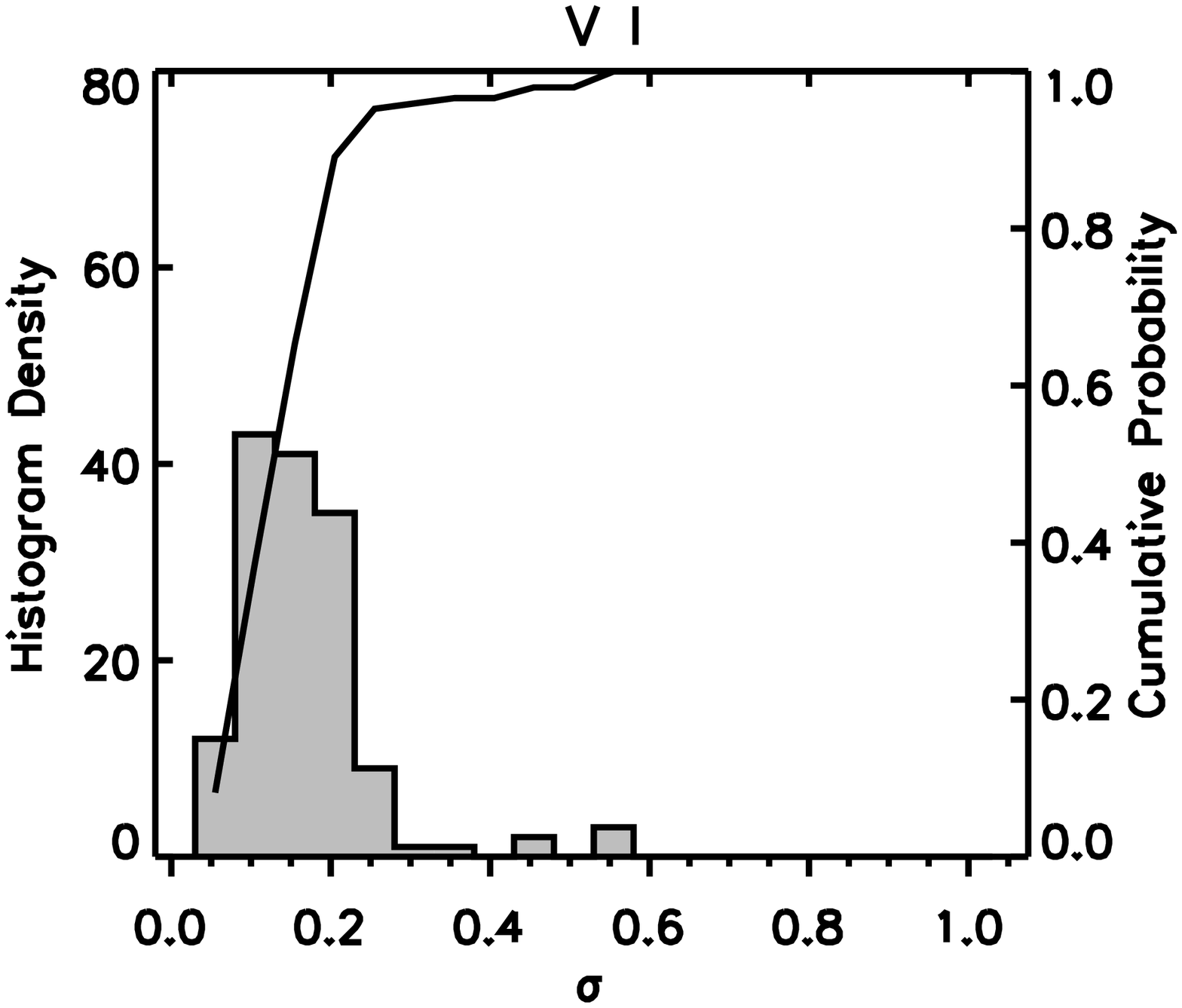}
\includegraphics[width=40mm]{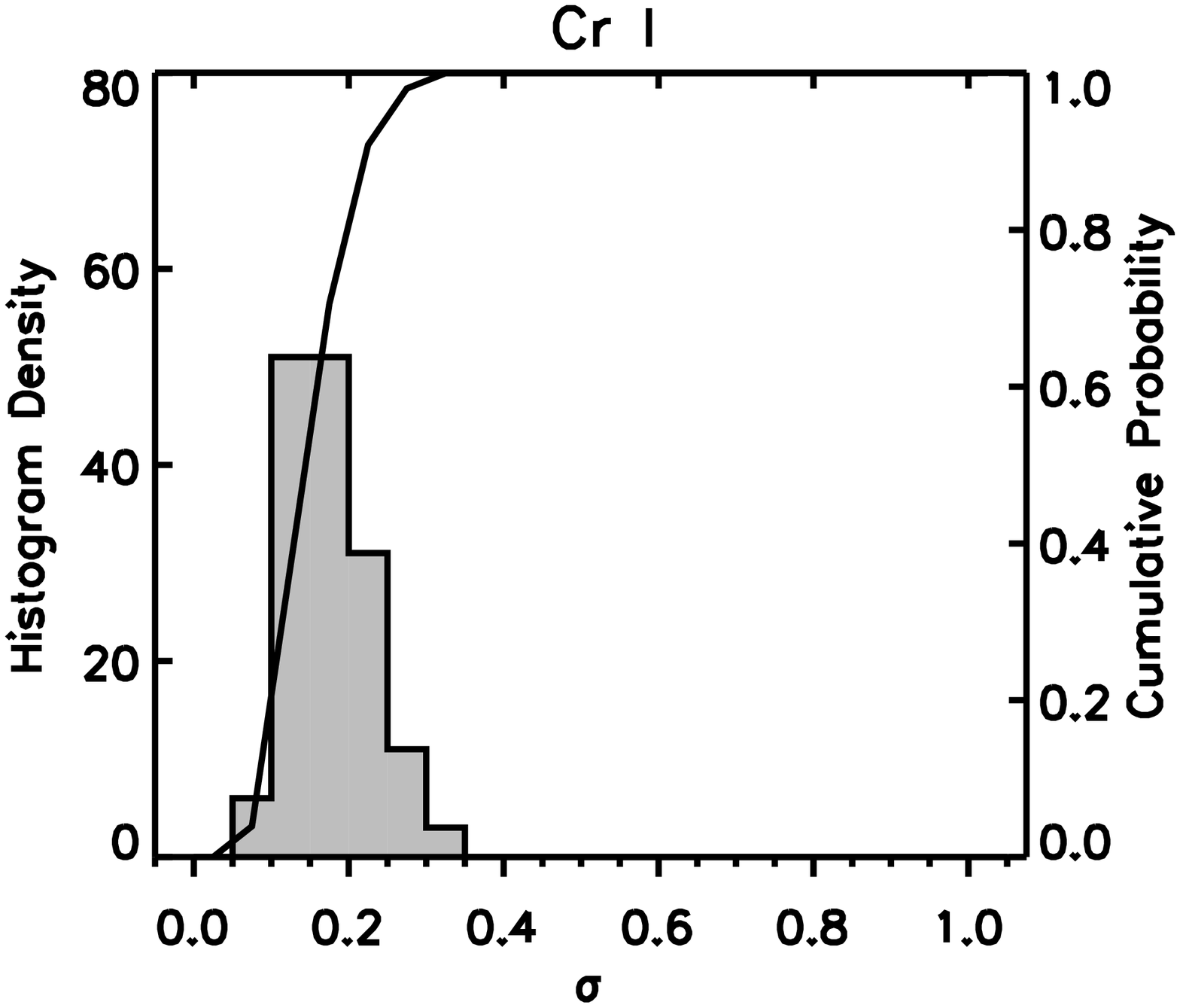}
\includegraphics[width=40mm]{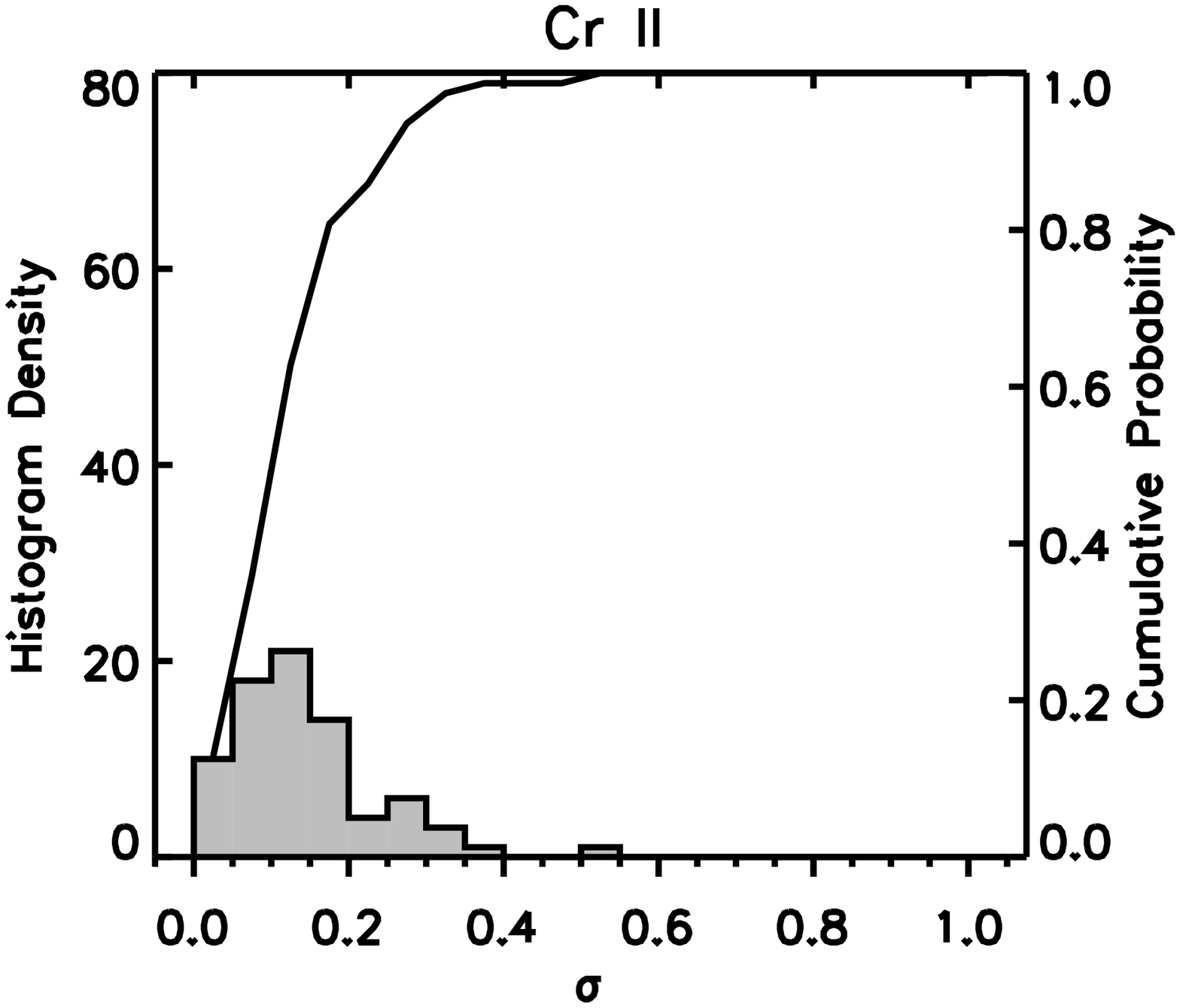}
\includegraphics[width=40mm]{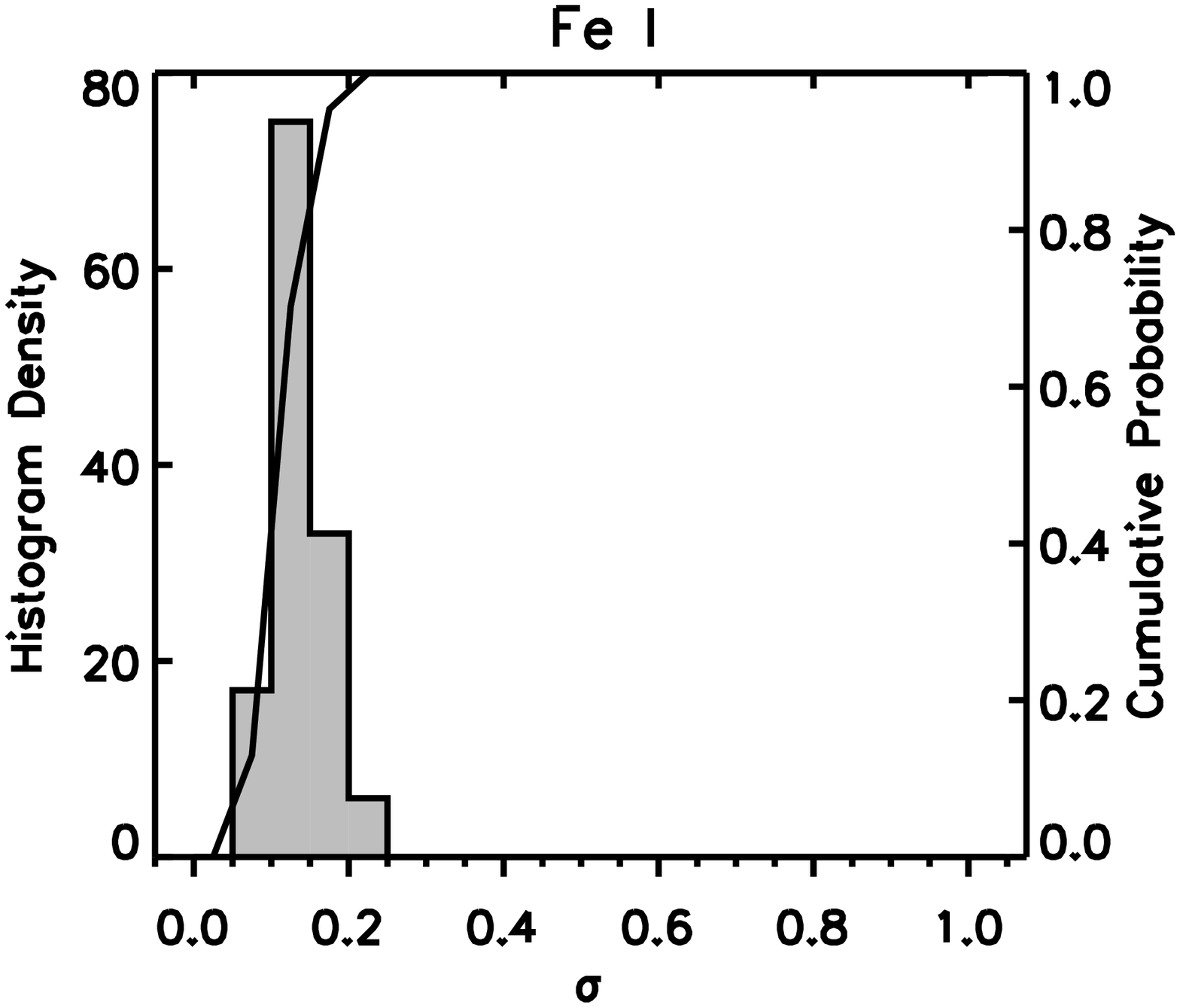}
\includegraphics[width=40mm]{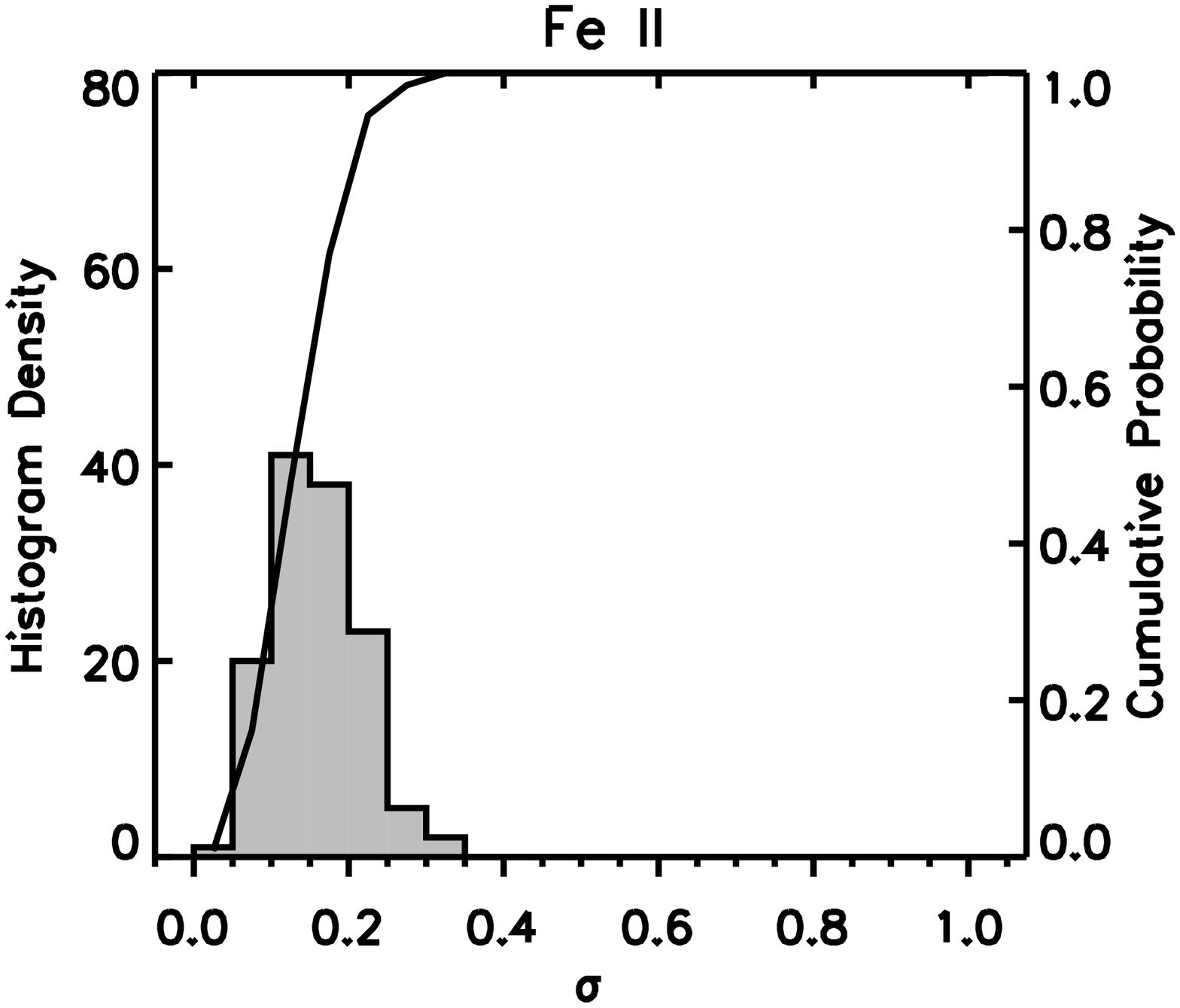}
\includegraphics[width=40mm]{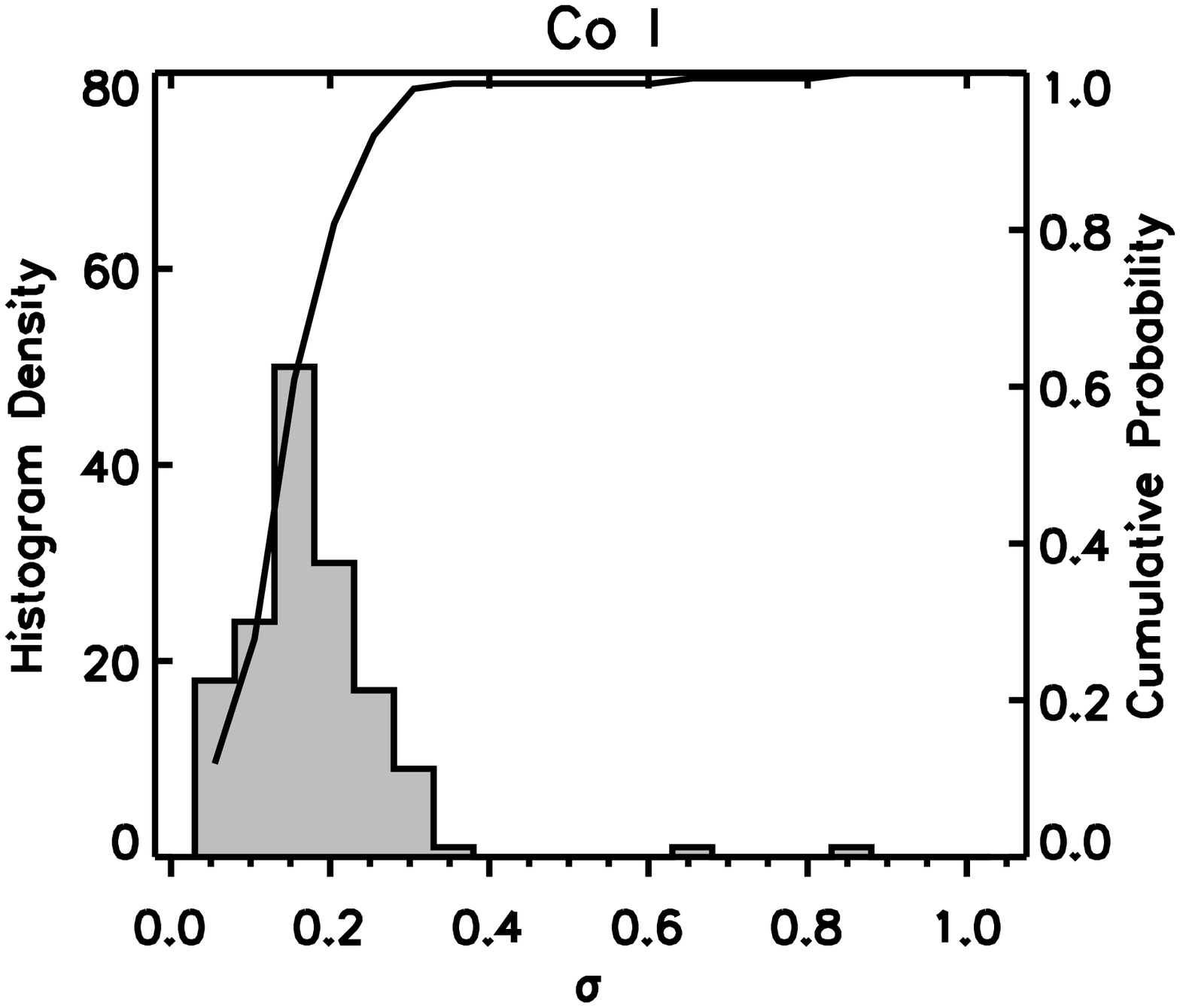}
\includegraphics[width=40mm]{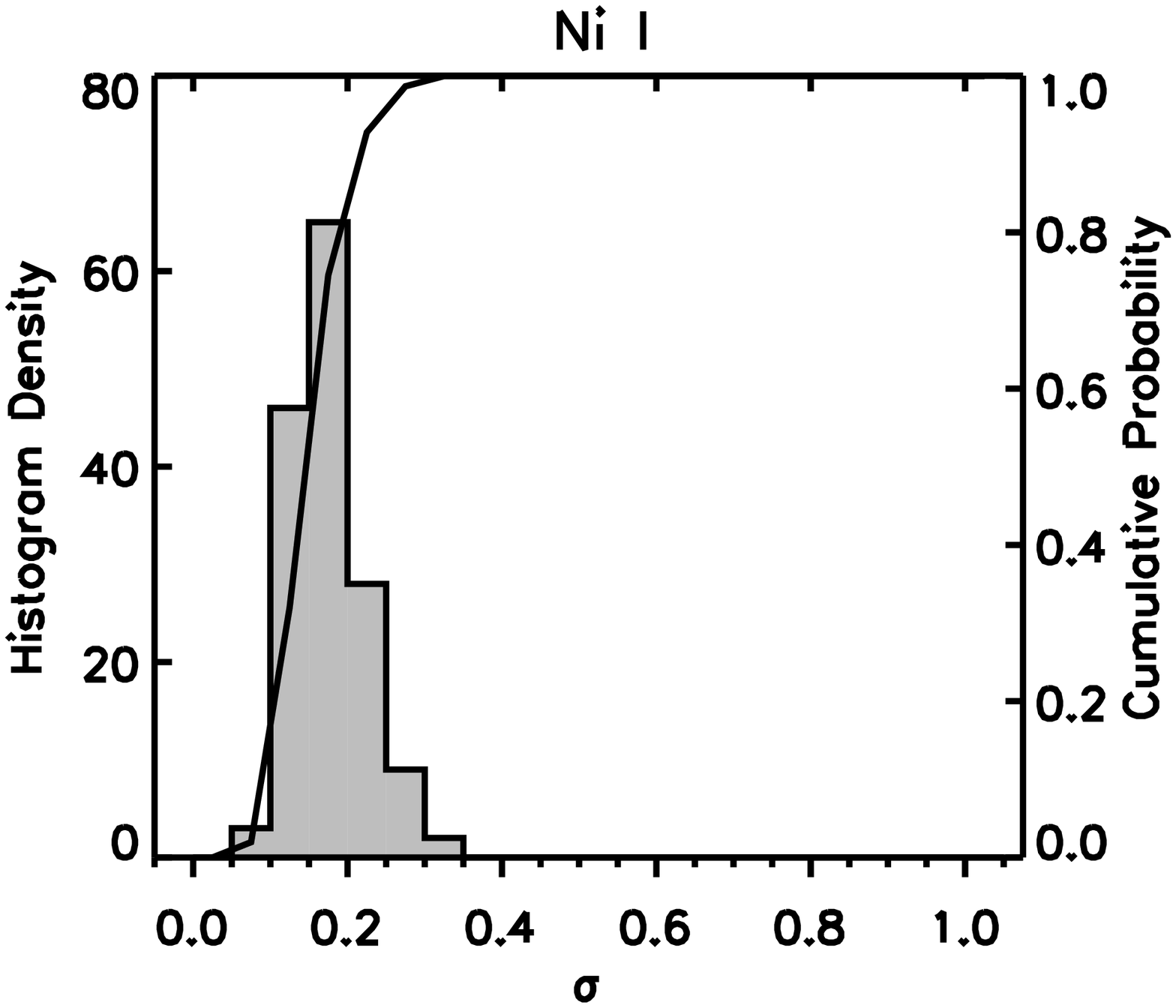}
\includegraphics[width=40mm]{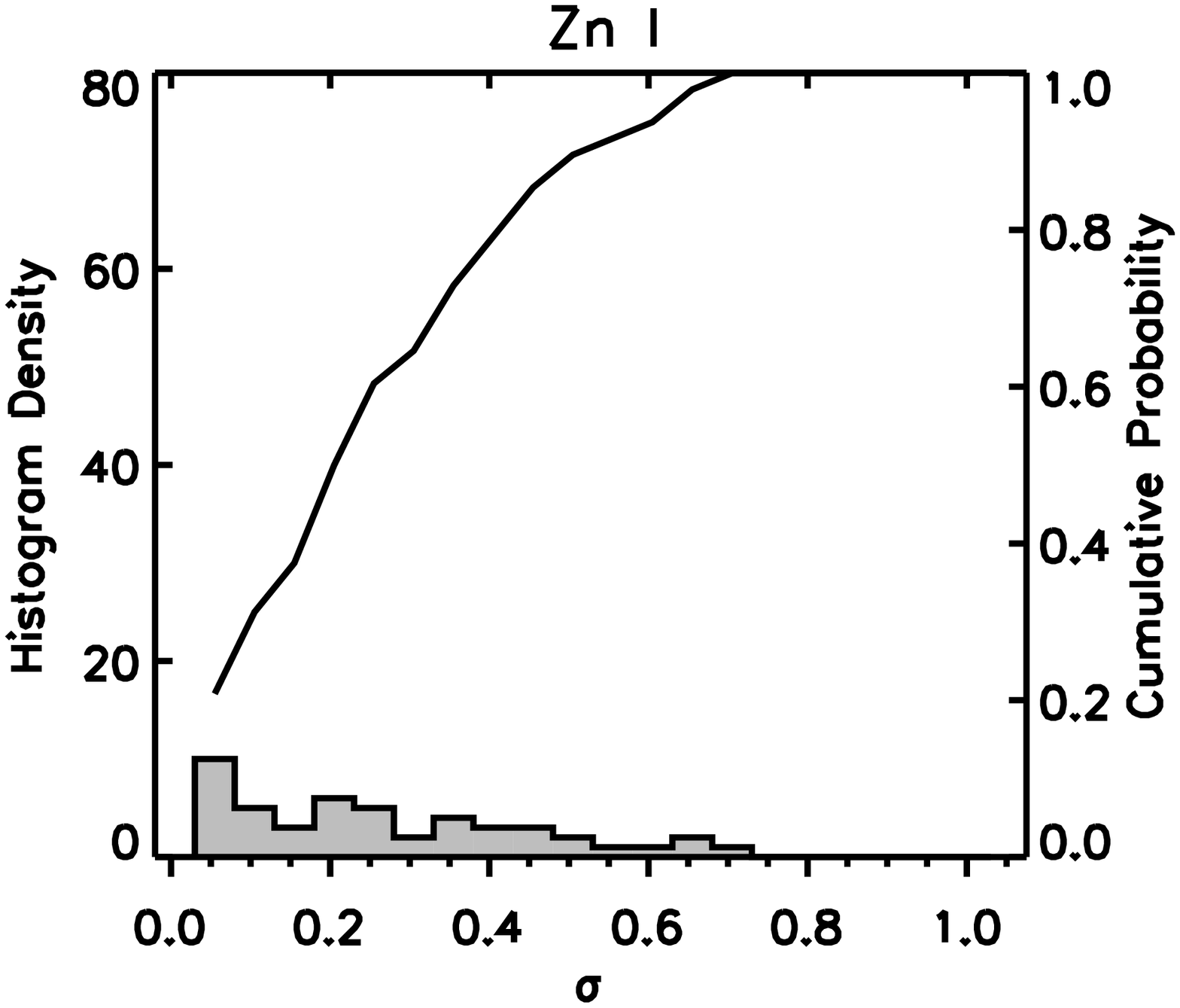}
\includegraphics[width=40mm]{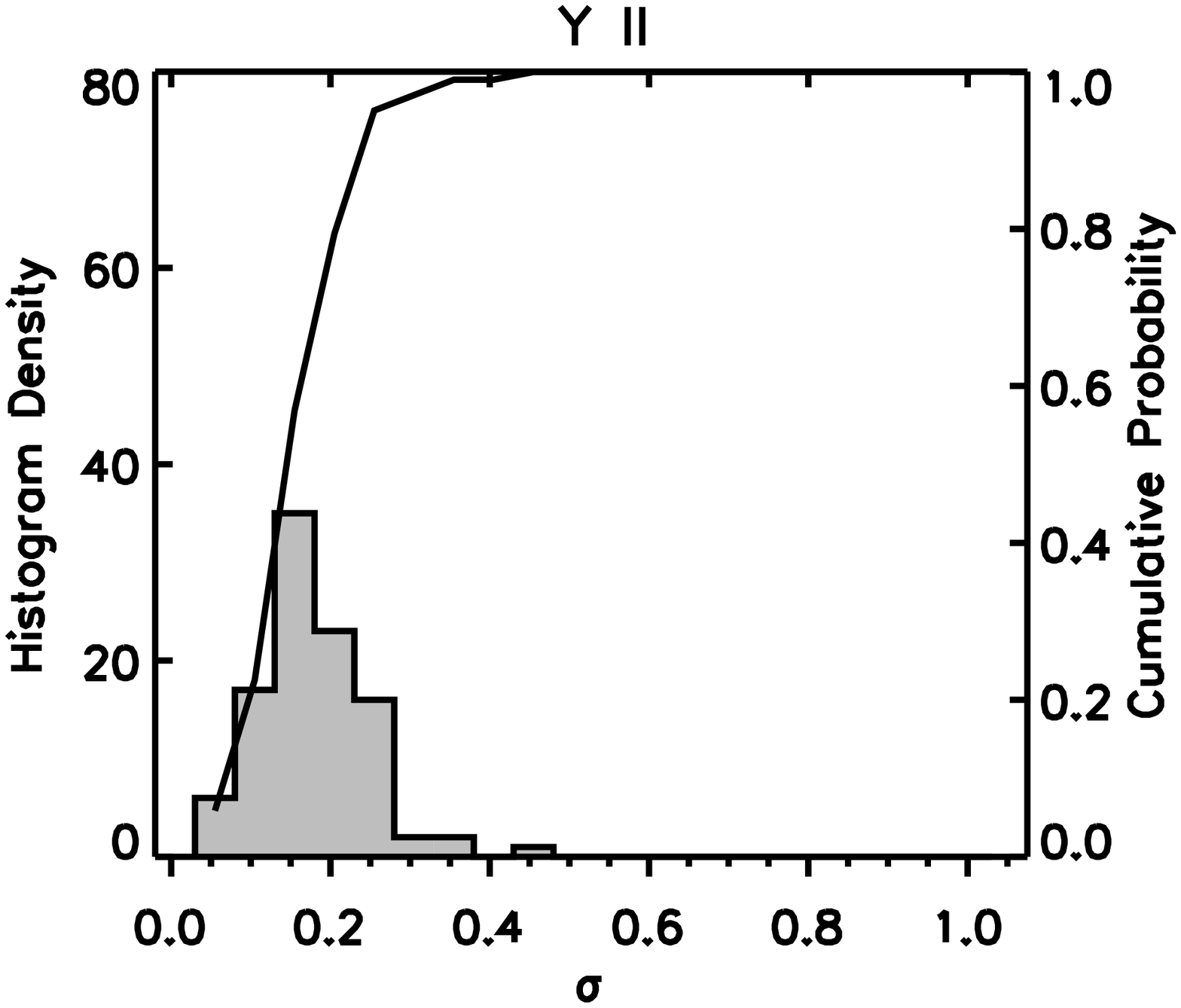}
\includegraphics[width=40mm]{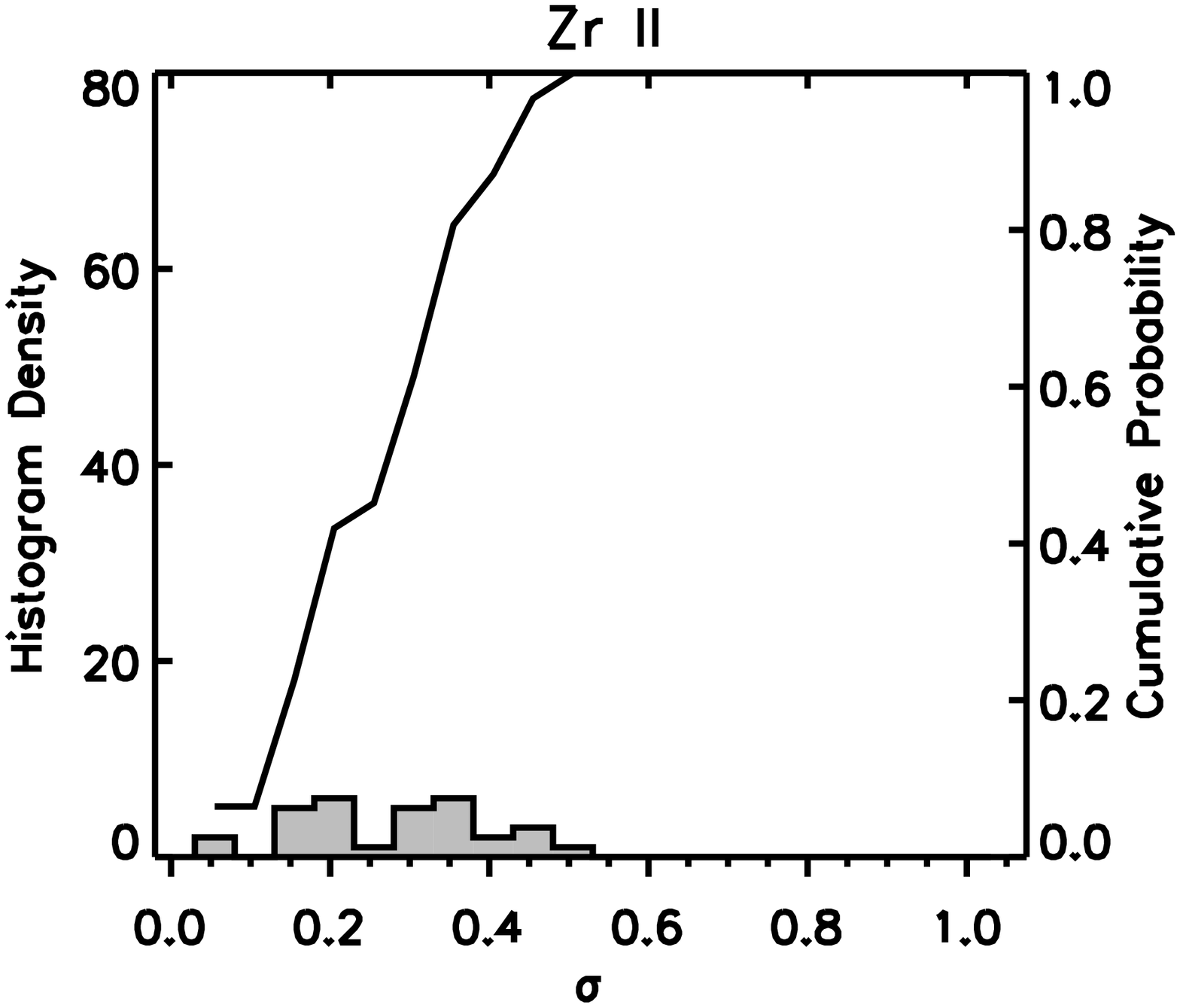}
\caption{iDR2 node-to-node dispersion of elemental abundances.}
\label{fig:UVES_abund_dispersion}
\end{figure*}

The node-to-node dispersions of elemental abundances in iDR2 is shown in Fig.\,\ref{fig:UVES_abund_dispersion}.
In general, $\approx$ 90\% of the results for each elements have dispersions below $\approx0.2$ dex. 
However the tail of the distributions extends to higher values in more difficult cases for which differences that arise from the different $W$ measurements \citep[see][]{Smiljanic_etal:2014} and line selection strategies play a role.
The dispersion tends to be higher also for abundances of ions like \ion{Ti}{II} and \ion{Cr}{II}.
Poor agreement is found for \ion{Zn}{I} and \ion{Zr}{II}.
Note that abundances for elements which require hyper-fine splitting were provided by the CAUP node only.

The internal precision is comparable with that of the UVES spectra of FGK-type analysis \citep[excluding stars in the field of young open cluster;][]{Smiljanic_etal:2014}.
Note that \cite{Smiljanic_etal:2014} make use of the {\it median of the absolute deviations from the median of the data} (MAD) to quantify the node-to-node dispersion, but this cannot be used here because of the small number of nodes providing abundances.
The dispersion from the mean used here should overestimate the node-to-node dispersion with respect to the MAD, although this is mitigated by the $\sigma$-clipping applied.
Overall, all this indicates that our internal precision for elemental abundances is roughly at the same level of the \cite{Smiljanic_etal:2014} one.

A survey inter-comparison with \cite{Smiljanic_etal:2014} results on the common calibration open cluster \object{NGC6705} was carried out for all elements except \element{Ce}, \element{La}, \element{Pr}, \element{Sm}, for which results did not pass the \cite{Smiljanic_etal:2014} quality control criteria.
The inter-comparison was satisfactory and confirmed the comparable precision with the \cite{Smiljanic_etal:2014} results.

\subsection{Mass accretion rate}

\begin{figure}[htp]
\centering
\includegraphics[width=90mm]{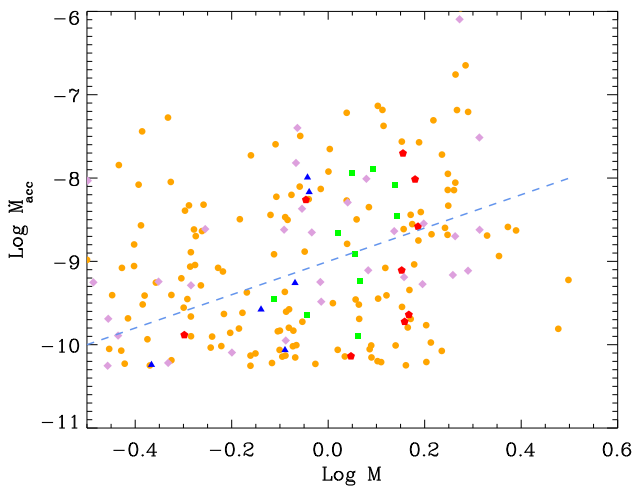}
\caption{Mass accretion rates vs. mass for all clusters in iDR2. Symbols and colours as in Fig.\,\ref{fig:giraffe_Ha10_EWHaAcc_pms}. The dashed line represents the $\dot{M} \propto M^2$ relationship. }
\label{fig:mdot_m}
\end{figure}

Mass accretion rates are estimated from the \ha10\ using the \citet[][Eq.\,(1)]{2004A&A...424..603N} formula.
The use of alternative methods, i.e. making use of the \wha, is discussed in 
\cite{Frasca_etal:2014} and will be implemented in the Gaia-ESO PMS analysis in future data releases.

The use of the \cite{2004A&A...424..603N} relationship has, undoubtedly, the advantage of allowing a simple estimate of \mdot\ from just the \ha10.
The accuracy and validity of this empirical relationship has, however, been questioned \citep[see, e.g.,][and references therein]{2012MNRAS.427.1344C}, especially in cases when only single epoch observations are available.

Recently, \cite{2014A&A...561A...2A} have computed the accretion rate by modelling the excess emission from the UV to the near-IR and provided empirical relationships between accretion luminosity and the luminosity of 39 emission lines from X-Shooter spectra.
In particular, they have shown that the comparison between \mdot\ derived through primary diagnostics (like the UV-excess) and that obtained with the \cite{2004A&A...424..603N} relationship has a large scatter, with this latter tending to underestimate \mdot\ for \ha10$<400$ \kms\ and to overestimate \mdot\ for \ha10$>400$ \kms.

A comparison of the iDR1 \mdot\ of $\gamma$ Vel and Cha\,I with mass accretion rates derived from line luminosity and the \cite{1998apsf.book.....H} relationship has been presented in \cite{Frasca_etal:2014}. 
They found discrepancies of $\sim 0.8$ dex for Cha I and $\sim 0.7$ dex for $\gamma$ Vel on average.  
\cite{Frasca_etal:2014} also compared the results obtained for Cha\,I with literature values, finding a fair agreement, with differences that can be ascribed to variability, different methodologies and the use of different evolutionary models.

Mass accretion rates derived for all clusters in the first 18 months of observations vs. stellar mass are shown in Fig.\,\ref{fig:mdot_m}.
Stellar mass is estimated from the recommended \teff\ and the age of the cluster using the \cite{Baraffe_etal:1998} models\footnote{The results shown here are for a mixing length parameter $\alpha=1.5$; in this analysis, however, the choice of $\alpha$ is uninfluential.}.
The expectations are that $\dot{M} \propto M^\alpha$ with $\alpha \sim 2$ \citep[e.g.,][]{2005ApJ...625..906M,2008ApJ...681..594H,2014A&A...561A...2A}.
As for the $\gamma$ Vel and Cha\,I cases discussed in \cite{Frasca_etal:2014}, however, the large scatter in $\dot{M}$ prevents us to make a meaningful comparison with such a relationship.
The Spearman's rank correlation analysis for Cha\,I gives, in the iDR2 case, $\rho = 0.43$ and $\sigma = 0.005$, i.e. a higher significance than found by \cite{Frasca_etal:2014} in the iDR1 case ($\rho = 0.26, \sigma = 0.16$ for $\dot{M}$ derived from \ha10), which indicates a better accuracy of our recommended iDR2 \ha10\ parameter. 
Amongst the younger clusters in our sample, we find $\rho = 0.47$ and $\sigma=0.14$ for $\rho$ Oph, while the correlation is rather poor for NGC2264 ($\rho = 0.19$, $\sigma=0.022$) possibly because of the larger uncertainties due to the residual nebular emission in the spectra of this cluster.
Note that the scatter in Fig.\,\ref{fig:mdot_m} is dominated by NGC2264. Ignoring this cluster, the scatter is consistent with what found by \cite{2014A&A...561A...2A} in their validation of the \cite{2004A&A...424..603N} relationship.
Interestingly, for the older clusters in our sample we find a not significant correlation in $\gamma$ Vel ($\rho = 0.29$, $\sigma=0.247$) but a well defined correlation in NGC2547 ($\rho = 0.89$, $\sigma=0.018$).
In both such cases, two kinematically distinct populations with different ages have been discovered \citep[][]{2014A&A...563A..94J,2015arXiv150101330S}, whose possible consequences in the $\dot{M}$ vs. $M$ relationship still need to be explored.

\subsection{Chromospheric \halpha\ and \hbeta\ flux}

After the \texttt{ROTFIT} determination of the fundamental parameters, a best matching template within the library of slowly-rotating inactive stars is identified.
The chromospheric excesses \hachr\ and \hbchr\ are derived using a spectral subtraction method \citep[see, e.g.,][and references therein]{1985ApJ...295..162B,1994A&A...284..883F,1995A&AS..114..287M} that has been extensively used in the past.
The photospheric flux is removed by subtraction of the spectrum of an {\it inactive} template star with very close fundamental parameters, rotationally broadened at the target \vsini, over the line wavelength range.
Such chromospheric $W$ excesses, \hachr\ and \hbchr, are then converted to flux, \fachr\ and \fbchr, by multiplying it by the theoretical continuum flux at the line's wavelength \citep[see, e.g.,][and references therein]{Frasca_etal:2014}.
It may be argued that even the templates may have some chromospheric {\it basal} flux \citep[see, e.g.,][and references therein]{1998ApJ...494..828J}, also variable in time following the stellar cycles \citep[see, e.g.,][]{2012A&A...540A.130S} which a detailed semi-empirical NLTE chromospheric modelling \citep[e.g.,][]{1990A&A...231..459H,1995A&A...302..839L} could take into account.
This latter is, however, unpractical for applications to large datasets like the Gaia-ESO one.
Furthermore, the chromospheric flux in young stars is much larger than the basal flux, so that this latter can be safely neglected.

\begin{figure}[htp]
\centering
\includegraphics[width=90mm]{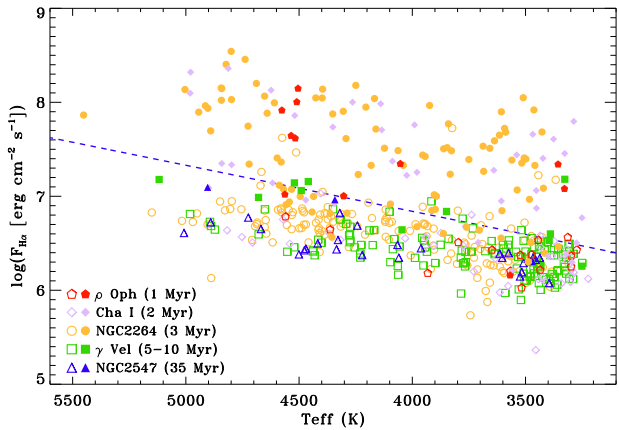}
\caption{Chromospheric \halpha\ flux vs. \teff\ for all young clusters observed in the first 18 months of observations. Symbols and colours as in Fig.\,\ref{fig:giraffe_Ha10_EWHaAcc_pms}, with filled (open) symbols used for CTTS (WTTS). The dashed line represents the chromospheric activity -- accretion dividing line of \cite{Frasca_etal:2014}.}
\label{fig:fluxhachr_teff_wg12_pms}
\end{figure}

Results for $\gamma$ Vel and Cha\,I (iDR1) are discussed in \cite{Frasca_etal:2014}, who were able to discriminate between chromospheric-dominated and accretion-dominated \halpha\ flux.
\hachr\ vs. \teff\ for all clusters observed in the first 18 months of observations (iDR2) is shown in Fig.\,\ref{fig:fluxhachr_teff_wg12_pms}.
We note that the chromospheric activity -- accretion dividing line proposed by \cite{Frasca_etal:2014} ($\log F_{H\alpha} = 6.35 + 0.00049 (T_{\rm eff}-3000)$) delimits quite neatly the two regimes in this larger sample as well, with some larger uncertainties in the case of NGC2264 likely due to residual nebular emission. 
This dividing line was also found by \cite{Frasca_etal:2014} to be in remarkable agreement with the saturation limit adopted by \cite{2003AJ....126.2997B} to separate CTTS and WTTS.

\begin{figure}[htp]
\centering
\includegraphics[width=90mm]{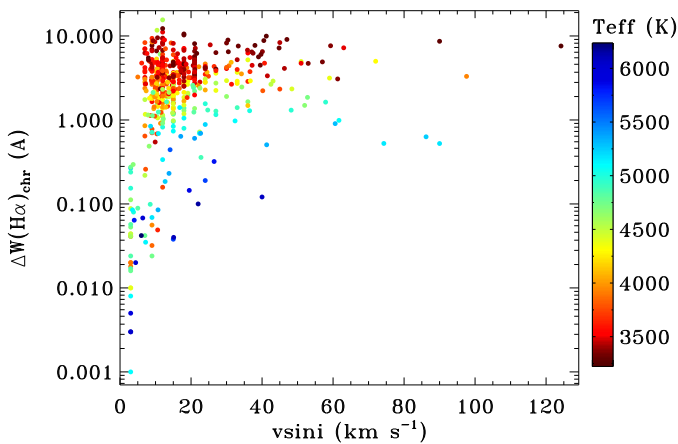}
\caption{Chromospheric \halpha\ equivalent width excess vs. \vsini\ for all young clusters observed in the first 18 months of observations. Colour coding is used for \teff.}
\label{fig:giraffe_EWHaChr_vsini_teff_pms_rotfit}
\end{figure}

Finally, in Fig.\,\ref{fig:giraffe_EWHaChr_vsini_teff_pms_rotfit} we show \hachr\ vs. \vsini\ for all young clusters observed in the first 18 months of observations.
While a full discussion on the activity--rotation relationship is deferred to future work, we note that our data display a \teff--dependent activity--rotation correlation regime at low \vsini, followed by a \teff--dependent saturation regime at high \vsini, as expected.
The behaviour at different \teff\ is quite neatly distinguishable, which further confirm the overall consistency of our results.

\section{Summary and conclusions}
\label{sec:Conclusions}

The Gaia-ESO PMS spectrum analysis provides an extensive list of stellar parameters from spectra acquired in the FLAMES/GIRAFFE/HR15N and FLAMES/UVES/580 setups in the field of young open clusters.
These include raw parameters that are directly measured on the input spectra (\wha, \ha10, and \wli), fundamental parameters (\teff, \logg, \feh, $\xi$, \vsini, and $r$), and derived parameters (\ali, \abund), \mdot, \hachr, \hbchr, \fachr, and \fbchr) which require prior knowledge of the former.
Our analysis strategy is devised to deal with peculiarities of PMS stars and young stars in general such as veiling, large broadening due to fast-rotation, emission lines due to accretion and/or chromospheric activity, and molecular bands.
The analysis is also made robust against residual sky-background or foreground features that cannot be completely removed as in the case of inhomogeneous nebular emission.

The availability of different methods for deriving stellar parameters increases the confidence on the output of our analysis.
It allows us to efficiently identify and discard outliers, like those deriving from failed fits or problems in the input spectra, as well as deriving realistic uncertainties from the internal dispersion of the data.
For \teff\ and \logg\ the external precision is estimated by comparison with results from interferometric angular diameter measurements.
These are estimated to be $\approx$ 120\,K r.m.s. in \teff\ and $\approx$0.3 dex r.m.s. in \logg\ for both the UVES and GIRAFFE setups.
The comparison with \teff\ derived from photometry for a selected group of stars in $\gamma$ Vel with the same foreground extinction and free from accretion signatures gives an agreement of $\approx260$\,K r.m.s.
Our recommended \feh\ results agree with assessed literature values for such a set of {\it benchmark} stars within $\approx$0.15 dex r.m.s.
A comparison with previous \feh\ determination for Cha\,I is discussed in \cite{2014A&A...568A...2S}.
Weakness or limitations of the methods used were identified by the node-to-node comparisons and by comparison with benchmark stars.

The observation strategy poses significant challenges to the analysis, since, for optimising the observation time, most of the relevant observations are carried out in just the FLAMES/GIRAFFE/HR15N setup. 
For our purposes, the wavelength range of this setup is the best available in the optical, as it contains very important diagnostics for young stars like the \halpha\ and \li\ line.
At the same time, surface gravity diagnostics in the HR15N setup are poorer than in other wavelength ranges and still not modelled with sufficient accuracy. 
\teff\ determination for spectral types earlier than early-G is also challenging since it is based mostly on the \halpha\ wings. 
For such a wavelength range, two methods based on the comparison with spectra or spectral indices of template stars have proved effective in providing fundamental parameters.
A satisfactory self-consistency of the results have been achieved, at the expense of discarding \logg\ values when a sufficient agreement between the two methods cannot be reached.
In such cases, however, it is still possible to provide an {\it evolutionary flag}, as it can be established with confidence whether the star is in a PMS, a MS or a post-MS stage.
An uncalibrated gravity-sensitive spectral index is also provided, useful for a rank order in age.

The reproducibility of the parameters obtained with the higher resolution and larger wavelength coverage from UVES using a much smaller wavelength range and a lower resolution as in the GIRAFFE/HR15N setup, together with the comparable accuracy and precision achieved in the two setups, is a remarkable achievement of this work.
This allows us to provide with confidence parameters for the much larger GIRAFFE sample.

The Gaia-ESO is an ongoing project and this paper describes the PMS spectrum analysis carried out on the first two data releases.
Work is ongoing to improve further our analysis for the next releases.
The tables with the public release results will be available through the ESO data archive\footnote{\url{http://archive.eso.org/wdb/wdb/adp/phase3_spectral/form?phase3_collection=GaiaESO}} and through the Gaia-ESO Survey science archive\footnote{\url{http://ges.roe.ac.uk/index.html}} hosted by the Wide Field Astronomy Unit (WFAU) of the Institute for Astronomy, Royal Observatory, Edinburgh, UK


\begin{acknowledgements}

Based on data products from observations made with ESO Telescopes at the La Silla Paranal Observatory under programme ID 188.B-3002.
This work was partly supported by the European Union FP7 programme through ERC grant number 320360 and by the Leverhulme Trust through grant RPG-2012-541.
We acknowledge the support from INAF and Ministero dell' Istruzione, dell' Universit\`a e della Ricerca (MIUR) in the form of the grant "Premiale VLT 2012" and the grant ``The Chemical and Dynamical Evolution of the Milky 
Way and Local Group Galaxies'' (prot. 2010LY5N2T).
The results presented here benefit from discussions held during the Gaia-ESO workshops and conferences supported by the ESF (European Science Foundation) through the GREAT Research Network Programme.

H.M.T. acknowledges the financial support from BES-2009-012182 and the ESF and GREAT for an exchange grant 4158. 
H.M.T. and D.M. acknowledges the financial support from the Spanish Ministerio de Econom\'ia y Competitividad (MINECO) under grant AYA2011-30147-C03-02.

J.I.G.H. acknowledges financial support from the MINECO under grants AYA2011-29060, and  2011 Severo Ochoa Program SEV-2011-0187.

S.G.S, EDM, and V.Zh.A. acknowledge support from the Funda\c{c}\~ao para a Ci\^encia e Tecnologia (Portugal) in the form of grants SFRH/BPD/47611/2008, SFRH/BPD/76606/2011, SFRH/BPD/70574/2010, respectively.

T.B. was funded by grant No. 621-2009-3911 from The Swedish Research Council.

\end{acknowledgements}

\bibliographystyle{aa}
\bibliography{GesWg12_2013}

\appendix

\section{\texttt{ROTFIT} templates}
\label{sec:ROTFIT_templates}

\begin{figure}[htp]
\centering
\includegraphics[width=8.0cm]{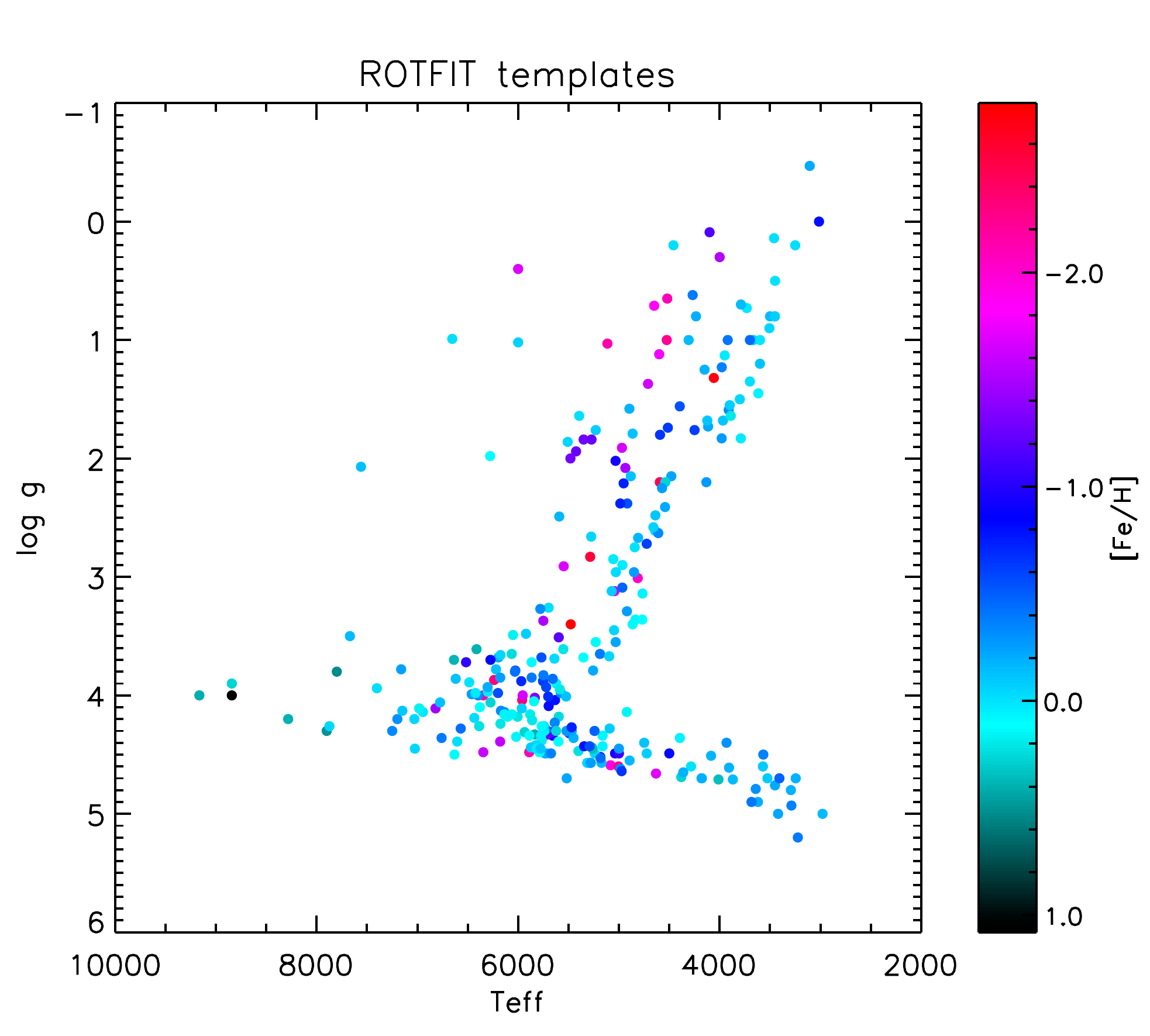}
\caption{Parameters of the whole set of \texttt{ROTFIT} templates adopted for the Gaia-ESO analysis.}
\label{fig:HRD_ROTFIT_templates}
\end{figure}
\begin{figure}[htp]
\centering
\includegraphics[width=8.0cm]{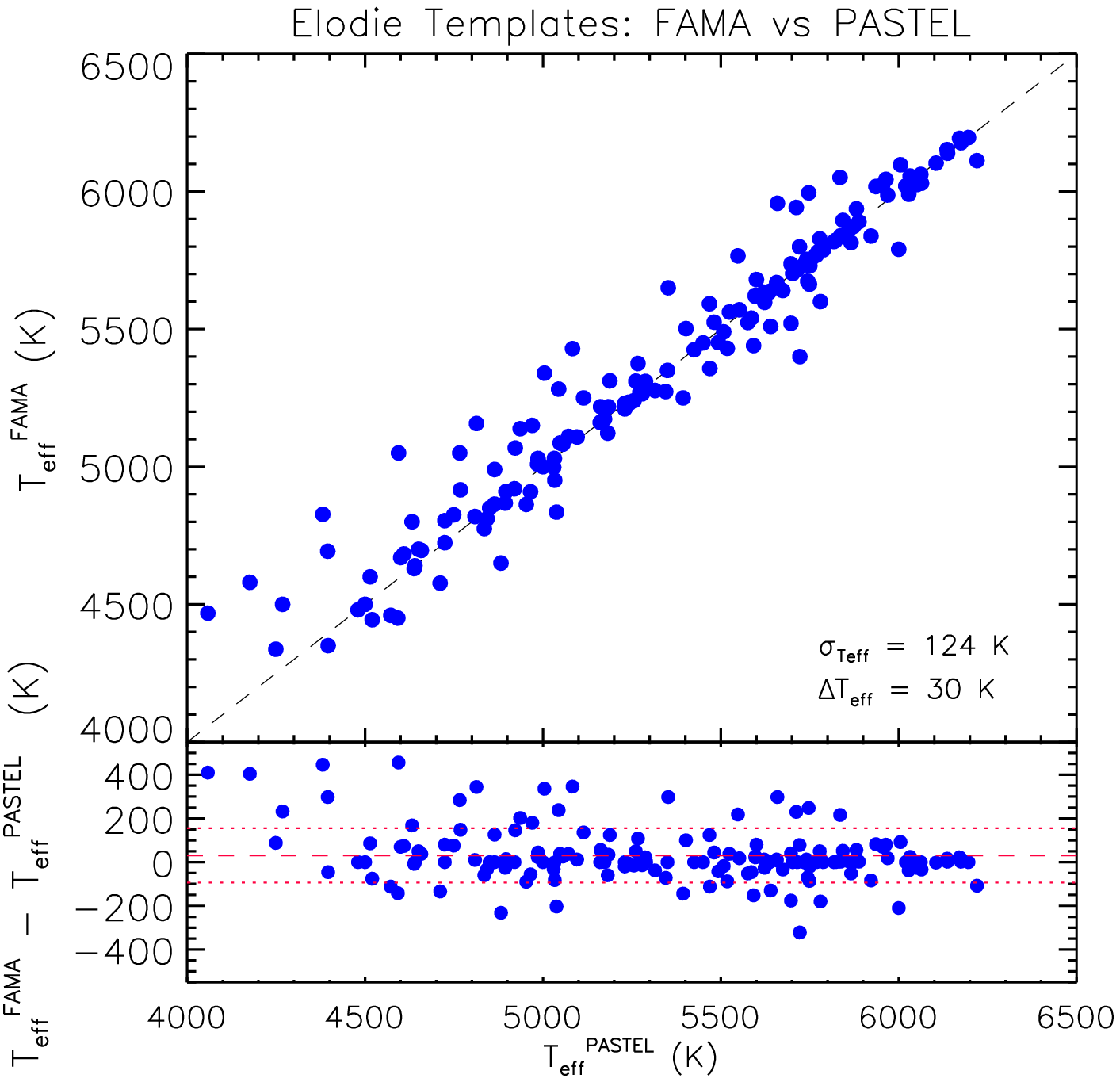}
\includegraphics[width=8.0cm]{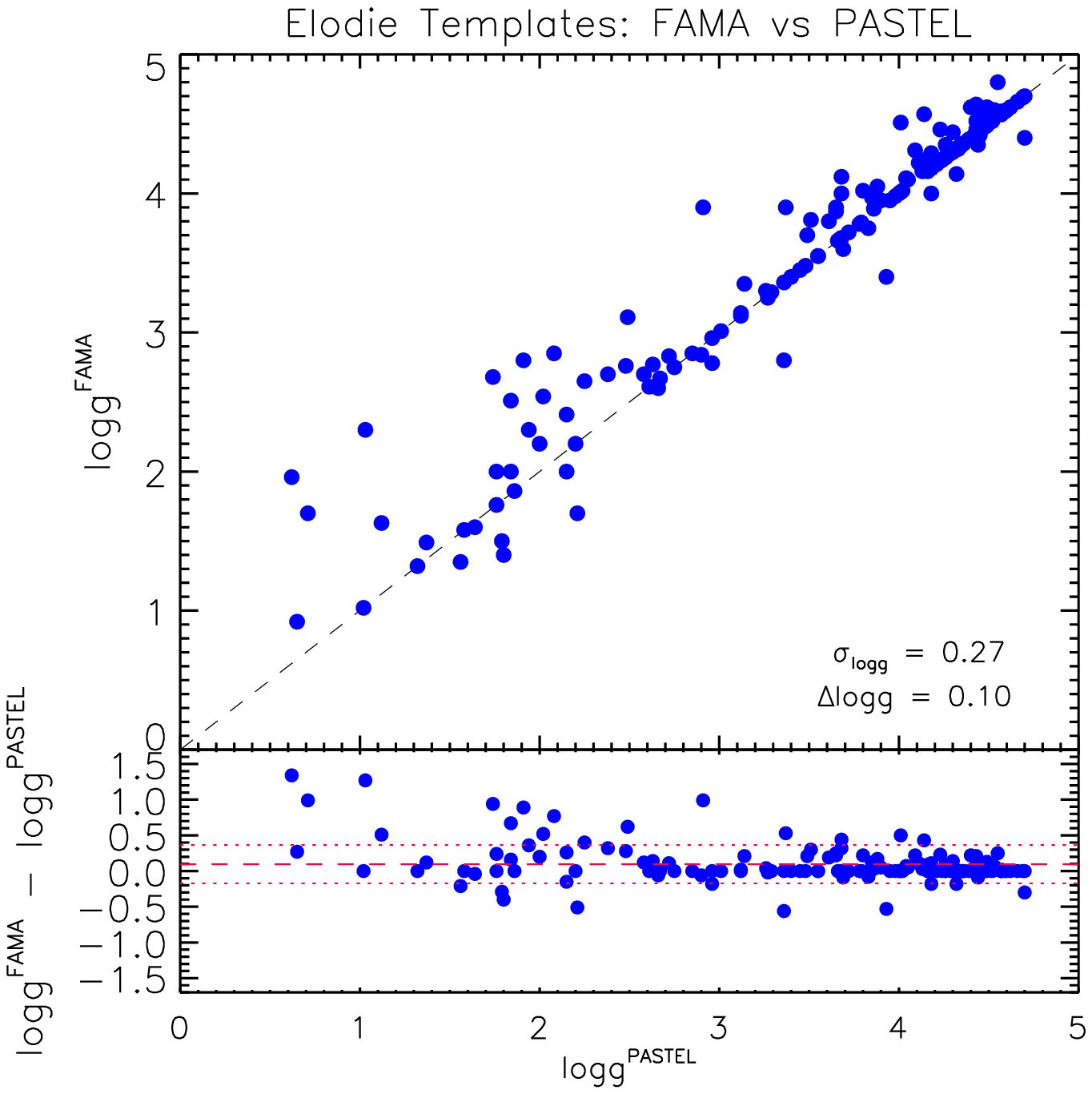}
\includegraphics[width=8.0cm]{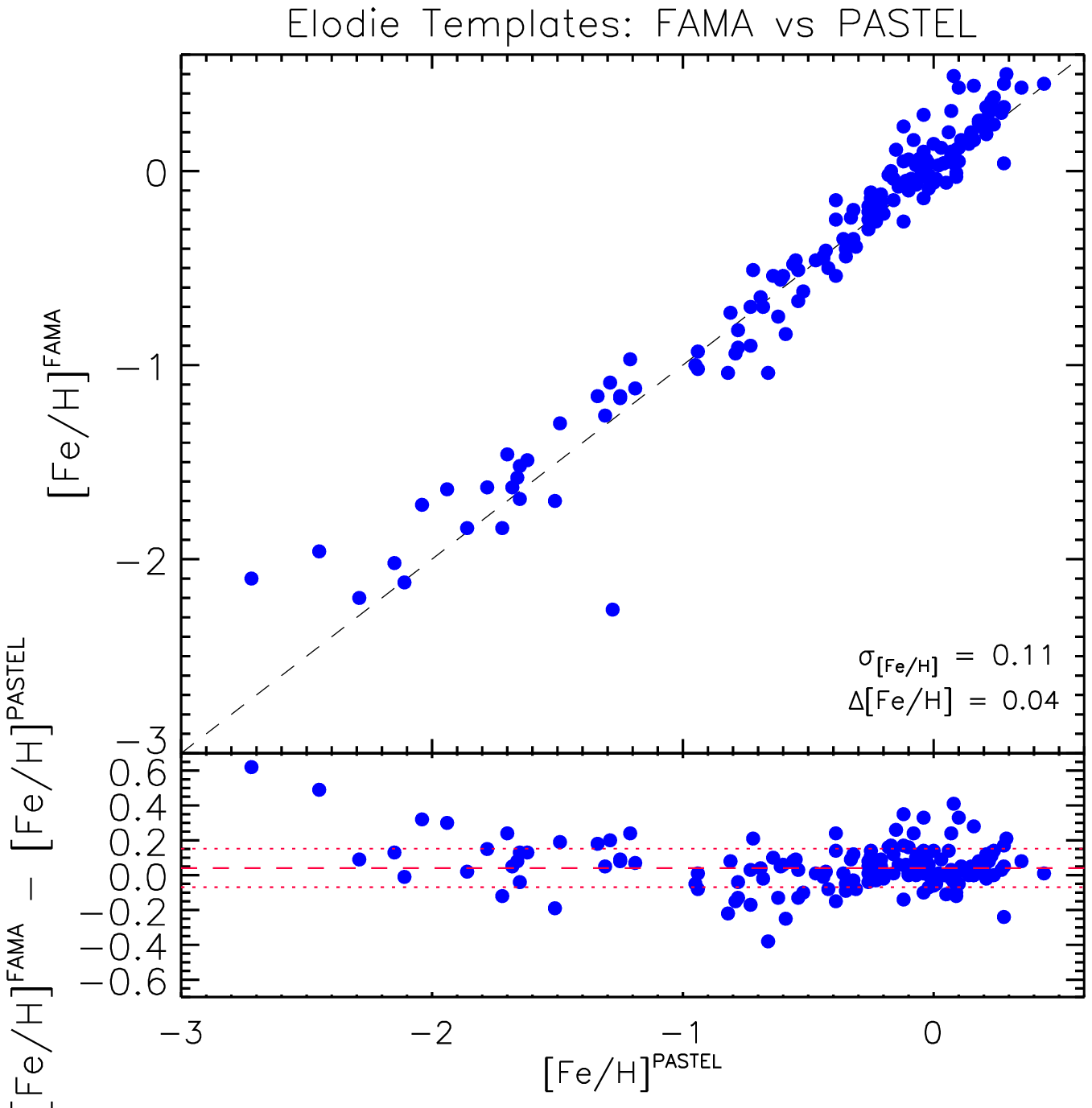}
\caption{Comparison between PASTEL and \texttt{FAMA} fundamental parameters of the ELODIE templates used by \texttt{ROTFIT}. See text for details.}
\label{fig:pastel_fama}
\end{figure}

The method implemented in \texttt{ROTFIT} relies on existing determination of fundamental parameters for the template stars and, in the past, the PASTEL catalogue \citep{2010A&A...515A.111S} has been used as input.
To ensure homogeneity amongst the Gaia-ESO spectrum analysis, however, the templates' parameters have been re-determined using Fast Automatic MOOG analysis \citep[\texttt{FAMA},][]{2013A&A...558A..38M} adopting the Gaia-ESO recommended model atmospheres and atomic parameters.
The templates' parameters were updated for most of the stars in the range from mid-F to late K, while for the M dwarf we adopted the parameters recently determined by \citet{2012ApJ...748...93R} and \citet{2012ApJ...757..112B}.
The parameters of the whole set of \texttt{ROTFIT} templates adopted for the Gaia-ESO analysis is illustrated in Fig.\,\ref{fig:HRD_ROTFIT_templates}. 
In Fig.\,\ref{fig:pastel_fama} we show the comparison between the PASTEL and \texttt{FAMA} parameters.
Average differences are 30\,K, 0.10\,dex, and 0.04\,dex for \teff, \logg, and \feh, respectively.
Standard deviations are 123\,K, 0.27\,dex, and 0.11\,dex for \teff, \logg, and \feh, respectively.
A table with the \texttt{ROTFIT} template parameters used for the Gaia-ESO analysis is reported in \cite{Frasca_etal:2014}.

\section{\texttt{ROTFIT} masks}
\label{sec:ROTFIT_masks}

Spectra of accreting stars or embedded in a dense cloud requires wavelength masks to exclude the residual nebular emission features that still remain after the data reduction process.
The \halpha\ profile in non-accreting young stars is nonetheless affected by a significant chromospheric contribution, which also varies in time.
The \halpha\ core must therefore be masked out to avoid considering the non-photospheric contribution to the line profile.
On the other hand, the \halpha\ wings are essential fundamental parameters' diagnostics, especially in GIRAFFE/HR15N spectra of F- and G-type stars for which the rest of the pass-band offers very poor constraints.
In some cases (e.g., old and inactive stars in the field) the whole of the \halpha\ profile can be used. 
Because of the Li depletion occurring in the stellar interior, \wli\ decreases rapidly with age in PMS stars of later spectral type and can be very different in stars with similar fundamental parameters; therefore this line must also be masked out.
The \halpha\ and \lif\ lines must be masked in the spectra of accreting stars too, but in such cases the \halpha\ mask must be wide enough to include the wings of the lines that can be very broad.

Therefore, the measurement of the {\em raw} parameters as a first step in the analysis process allows us to divide the spectra into three classes:
\begin{itemize}
\item NHL. Spectra with negligible \li\ absorption and no \halpha\ emission for which only a narrow \halpha-core mask ($\pm 2$\,\AA) is required and no $r$ evaluation is carried out;
\item HL. Spectra with significant \li\ absorption, \halpha\ core emission, or both, for which accretion, and therefore $r$ evaluation, can be excluded, but require a slightly larger \halpha-core mask ($\pm 5$\,\AA); 
\item HLV. Spectra with accretion signatures, which require a mask for the entire \halpha\ profile ($\pm 20$\,\AA), plus the evaluation of $r$.
\end{itemize}
In the case of HL and HLV classes, a mask of $\pm 3$\,\AA\ is applied around the \li\ line core.

\section{Comparison with \teff\ from photometry}
\label{sec:photometry}

A more extensive comparison can be made, at this stage, with \teff\ derived from photometry for cluster's members that have all the same foreground extinction and that are not significantly affected by colour excesses due to the presence of circumstellar material.
Fortunately, this turned out to be feasible for $\gamma$ Vel, as the extinction is fairly uniform and many likely members are free from large colour excess. 

BVI photometry of $\gamma$ Vel was presented in \citet{2000MNRAS.313L..23P} and in \citet{Jeffries_etal:2009}.
The photometry was taken in the Harris B, V and Kron-Cousins I filters and was converted to the standard Johnsons-Cousins photometric system by \cite{Jeffries_etal:2009}.
This data have been supplemented with 2MASS \citep{2003yCat.2246....0C} and Spitzer data \citep{2008ApJ...686.1195H} where available.

We used two different methods for deriving \teff\ from photometry, one based on the $(V-I)$ vs. $(B-V)$ colour-colour diagram 
($T_{\rm CC}$),
the other one based on the simultaneous fit of all available magnitudes, from the optical to the Spitzer bands
($T_{\rm SED}$).

\begin{figure}[t]
\centering
\includegraphics[width=9.0cm]{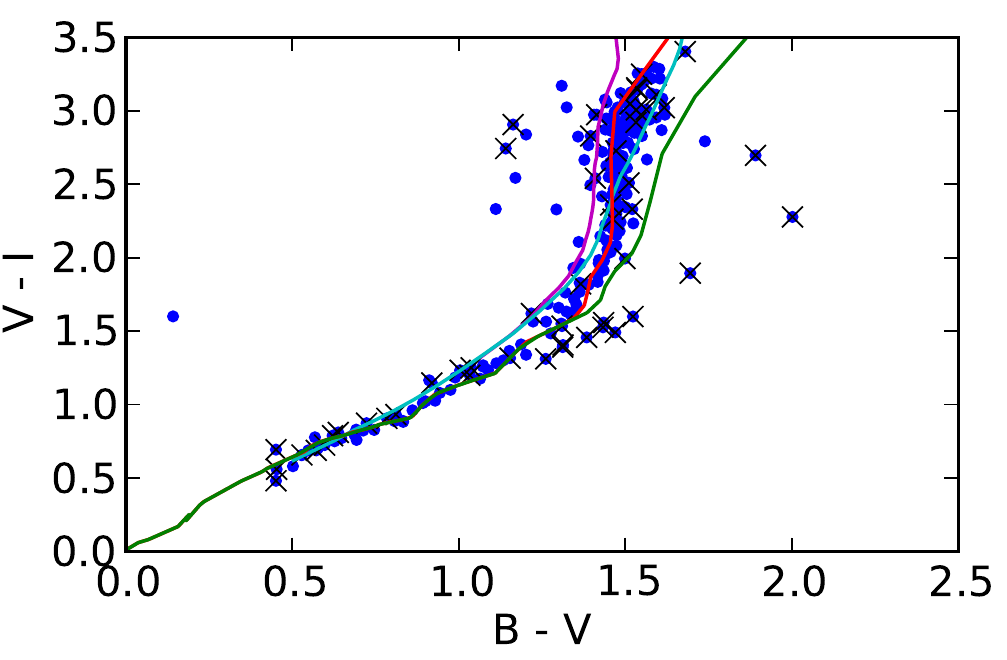}
\caption{
$V-I$ versus $B-V$ colour-colour diagram for likely members of $\gamma$ Vel.
Stars with radial velocities larger than 9 \kms\ from the mean velocity of $\gamma$ Vel are crossed out.
The empirical main sequence locus relationship from \cite{1995ApJS..101..117K} is over-plotted in green, the theoretical main sequence locus from BT-Settl \citep{Allard_etal:2011} in cyan, the theoretical pre-main sequence locus for an age of 7 Myr from BT-Settl in purple, and finally the locus obtained from Eq.\,(\ref{eq:col-col}) in red. All the colour loci plotted as lines have been shifted assuming an extinction of $E(B-V) = 0.038$ \citep{Jeffries_etal:2009} and $E(V-I)/E(B-V ) = 1.6$ \citep{1985ApJ...288..618R}.}
\label{fig:gammaVel_col_col}
\end{figure}

The $(V-I)$ vs. $(B-V)$ colour-colour diagram for likely members of $\gamma$ Vel, corrected for the foreground extinction as estimated by \citet{Jeffries_etal:2009}, is shown in Fig.\,\ref{fig:gammaVel_col_col}. 
The membership is based on Li and radial velocity and about 85\% of the stars in Fig.\,\ref{fig:gammaVel_col_col} are expected to be actual cluster members \citep[see][]{2014A&A...563A..94J}. 
The narrowness of this locus indicates that there is little differential extinction or veiling in $\gamma$ Vel. 
When excluding stars with radial velocities larger than 9 \kms\ from the mean velocity of $\gamma$ Vel, this locus narrows even further. 
As Fig.\,\ref{fig:gammaVel_col_col} shows, neither the empirical \teff-to-colour conversion for the ZAMS \citep[][hereafter KH95]{1995ApJS..101..117K} nor the theoretical BT-Settl ZAMS and 7 Myr PMS isochrones \citep{Allard_etal:2011} overlap with the observed locus in the colour-colour diagram.

The theoretically computed colours from BT-Settl do not reproduce accurately the stellar magnitudes possibly because of a still incomplete description of the opacity.
However, we can use them to define a new locus based on the assumption that the models correctly predict the colour difference due to gravity effects. 
If this assumption holds, we can add the colour shift between the theoretical ZAMS and PMS isochrones to the empirical ZAMS colours to get the actual PMS intrinsic colours, a procedure conceptually similar to that used by \cite{2013MNRAS.434..806B}. 
For every effective temperature, a new colour has been calculated using:

\begin{eqnarray}
\label{eq:col-col}
(B-V) &=& (B-V)_{\rm KH95, ZAMS} + \\ \nonumber 
      & & (B-V)_{\rm BT-S, PMS} - (B-V)_{\rm BT-S, ZAMS}
\end{eqnarray}
with a similar equation for $(V-I)$.
This new locus clearly describes the observed locus in $\gamma$ Vel much better than the alternatives (see Fig.\,\ref{fig:gammaVel_col_col}). 
The fit can probably be improved further by varying the age and extinction within the ranges given by \cite{Jeffries_etal:2009}, but we do not explore that here. 

Assuming a constant extinction of $E(B-V) = 0.038$ \citep{Jeffries_etal:2009} and zero veiling, which is likely to be approximately true for most stars lying on the observed narrow locus, we can fit $T_{\rm CC}$ to the observed colours through a least-square minimisation, which has been generalised for two dimensions with correlated uncertainties \citep{2010arXiv1008.4686H}:
\begin{equation}
\chi_{\rm CC}^2 = 
\left(
\begin{array}{cc}
\Delta_{BV}  & \Delta_{VI}
\end{array}
\right)
\left(
\begin{array}{cc}
\sigma_B^2 + \sigma_V^2  & - \sigma_V^2  \\
\sigma_V^2  & \sigma_V^2+\sigma_I^2   
\end{array}
\right)
\left(
\begin{array}{c}
\Delta_{BV} \\
\Delta_{VI}
\end{array}
\right)
\end{equation}
where $\Delta_{BV} \equiv \Delta(B-V)$ and $\Delta_{VI} \equiv \Delta(V-I)$ give the difference between the observed colours and those predicted
from our new locus as a function of the temperature of the star. 
These colour differences are multiplied with the inverse of the covariance matrix, which is given here as a function of the photometric uncertainties $\sigma_B$, $\sigma_V$, and $\sigma_I$. 
The temperatures computed in this way are based on the KH95 colour-temperature conversion of ZAMS stars 
with the colours adjusted for the lower surface gravity of PMS stars.

\begin{figure*}[t]
\centering
\includegraphics[width=7.6cm]{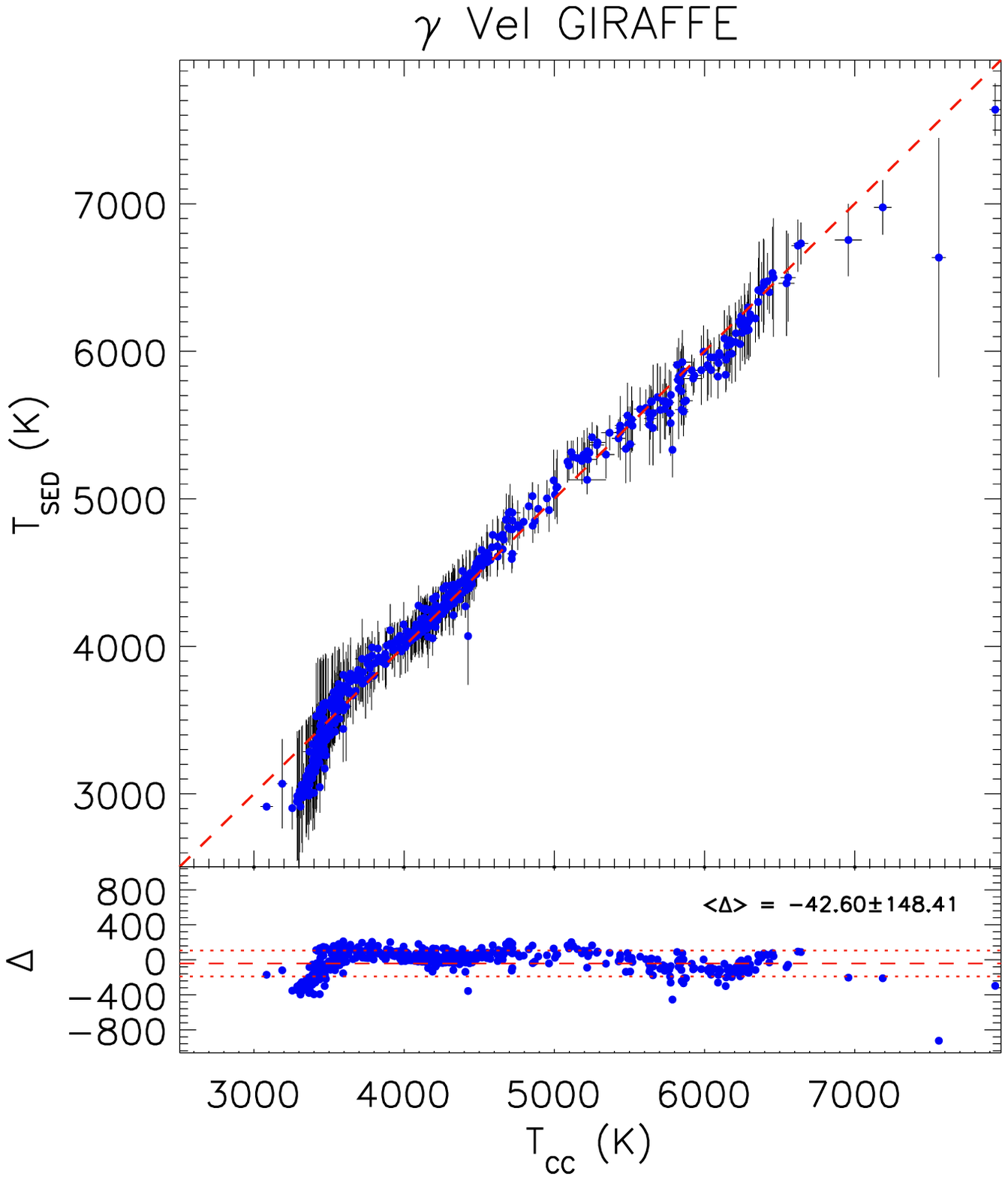}
\includegraphics[width=7.6cm]{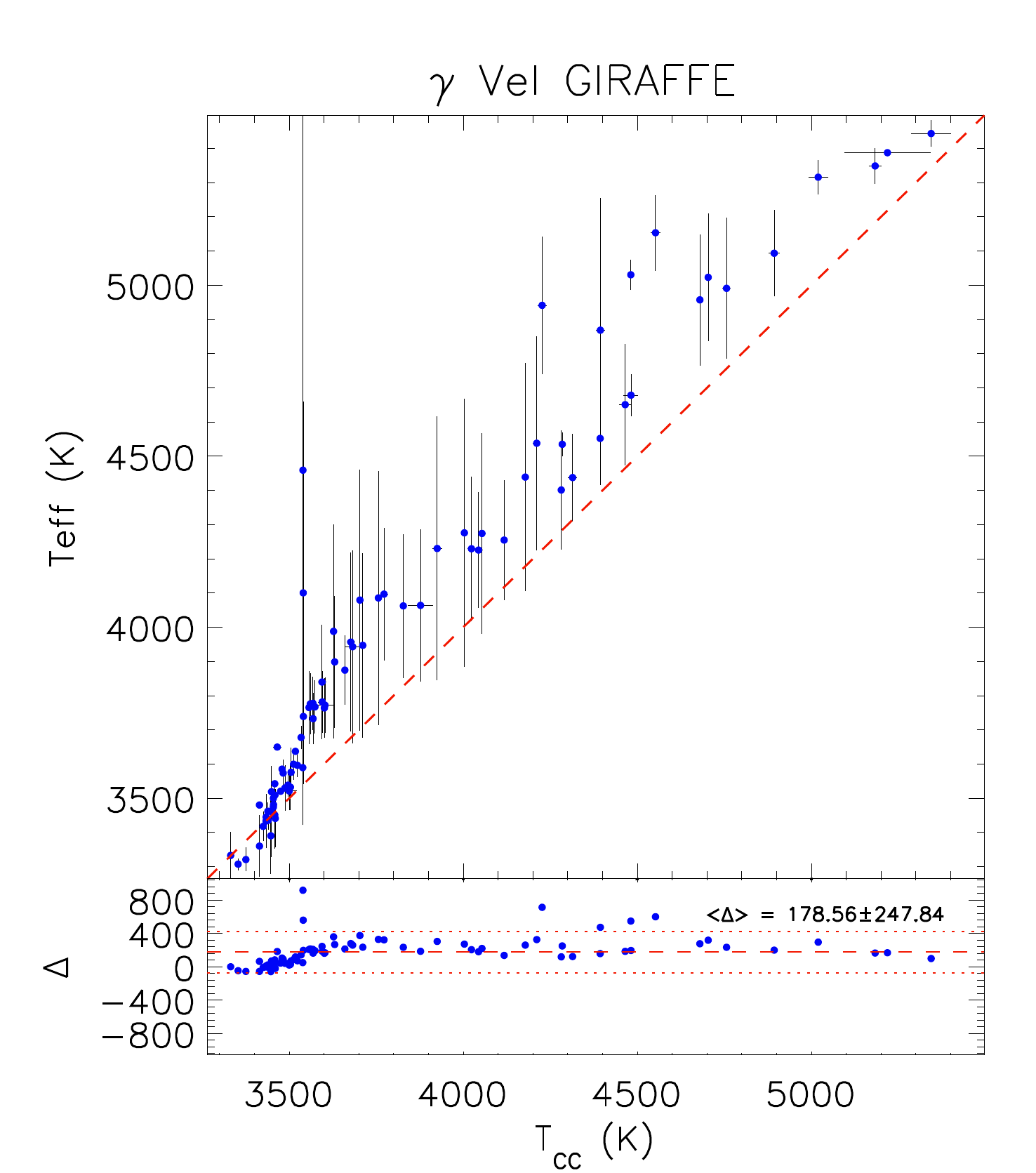}\\
\includegraphics[width=7.6cm]{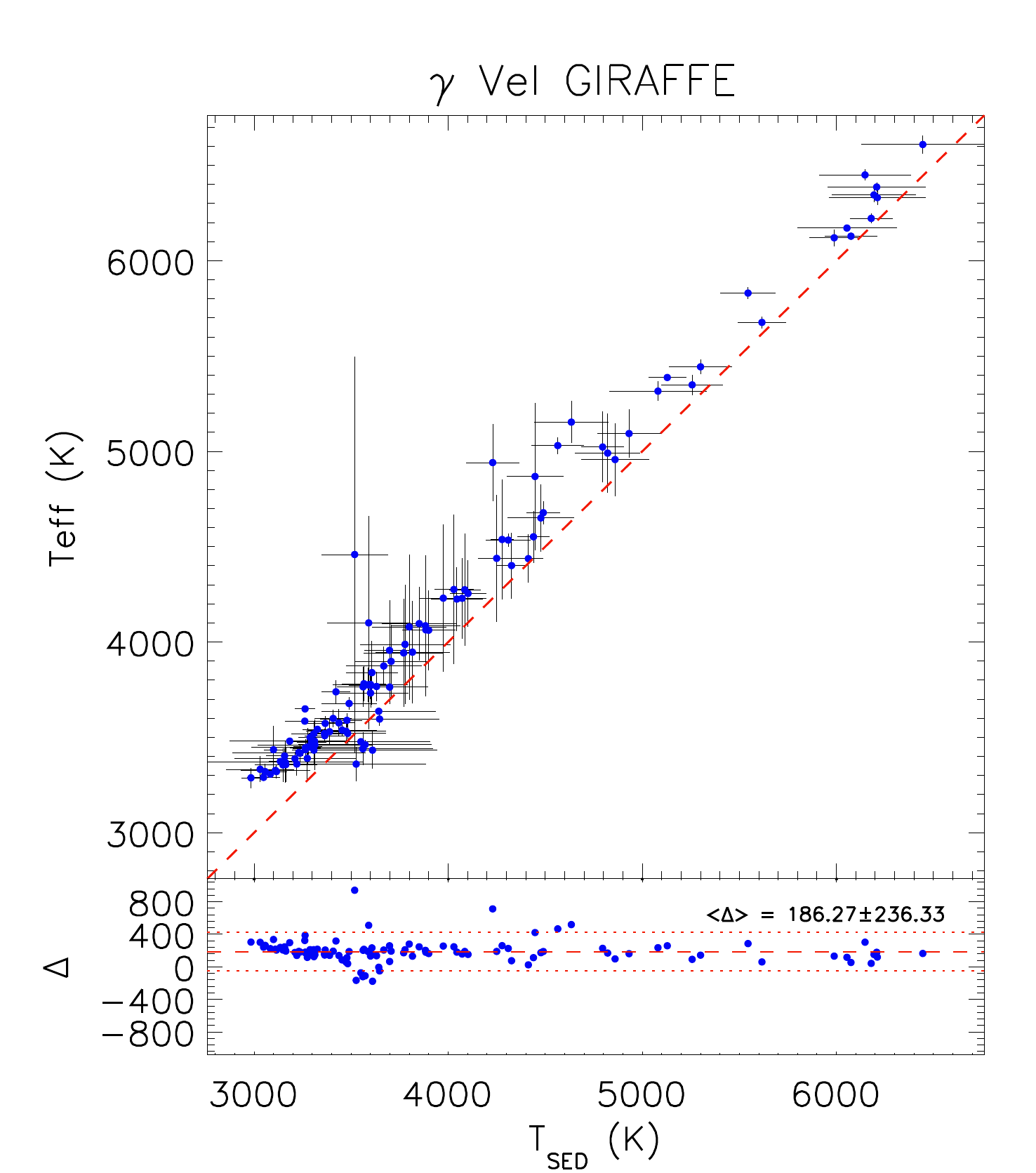}
\includegraphics[width=7.6cm]{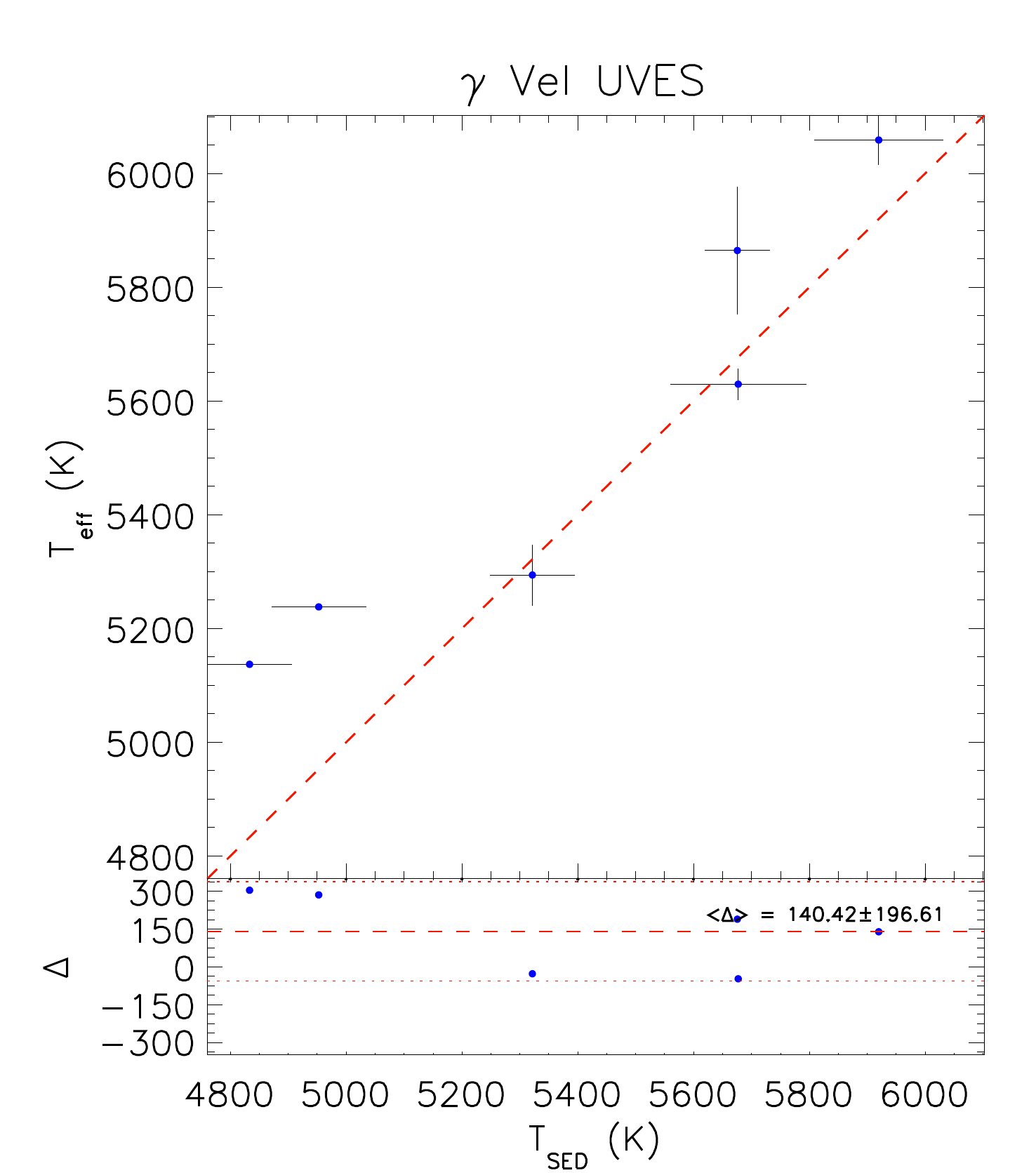}
\caption{Top left panel: comparison of $T_{\rm SED}$ and $T_{\rm CC}$ for all GIRAFFE's spectra in the $\gamma$ Vel field for which $\chi_{\rm CC}^2 < 7$, irrespective of Li- and RV-membership. 
 Top right panel: recommended \teff\ vs. $T_{\rm CC}$
 for GIRAFFE's spectra of likely members of $\gamma$ Vel. 
 Bottom left panel: recommended \teff\ vs. $T_{\rm SED}$
 for GIRAFFE's spectra of likely members of $\gamma$ Vel. 
 Bottom right panel: 
 recommended \teff\ vs. $T_{\rm SED}$ for all UVES's spectra in the $\gamma$ Vel field.}
\label{fig:Teff_phot_gamma2Vel}
\end{figure*}

In Fig.\,\ref{fig:Teff_phot_gamma2Vel} (top right panel) we compare the recommended \teff\ with 
$T_{\rm CC}$
for likely $\gamma$ Vel members. 
A further selection has been applied by considering 
$T_{\rm CC}$
with $\chi_{\rm CC}^2$ $< 7$ (i.e., consistent with being drawn by chance at the 99\% level) to avoid considering stars that may be significantly affected by colour excess due to the presence of circumstellar material.
Note that the formal uncertainties of 
$T_{\rm CC}$
turned out to be excessively small as a result of the definition of the locus in the ($V-I$,$B-V$) plane and the insensitivity of $T_{\rm CC}$ to $B-V$ at low temperatures.
The comparison shows that the agreement is mostly within the estimated uncertainties, although the spectroscopic \teff\ is systematically higher than $T_{\rm CC}$ above $\approx 3600$\,K, with an average difference of $\approx 180$\,K and standard deviation $\sigma \approx 250$\,K.

As an alternative approach, \teff\ from photometry has also been derived by taking all photometry available from optical, 2MASS, and Spitzer into account.
In this case we fit $T_{\rm SED}$ by a downhill simplex multidimensional minimisation \citep{Nelder_Mead:1965} of
\begin{equation}
\chi_{\rm SED}^2 = \sum_i \left( {x_i - w_i}\over{\sigma_i} \right)^2
\label{eq:SED-chisq}
\end{equation}
where $x_i \equiv (V-M_{\lambda})$ is the observed colour derived from the $V$-band magnitude and each of the photometric magnitudes available ($M_{\lambda}$), $\sigma_i$ is its uncertainty, and $w_i=w_i(T_{\rm eff}, \log g, {\rm[Fe/H]})$ is the theoretical colour from the BT-Settl models \citep{Allard_etal:2011} that are interpolated in \teff\ and \logg\ with \feh\ fixed at the solar value.
Standard deviation has been estimated using Monte-Carlo simulations with 1000 random synthetic realisations for each star.
This method is, effectively, a fit of the stellar spectral energy distribution (SED) and has the advantage of considering colours more sensitive to \teff\ than just $(V-I)$ and $(B-V)$ in the temperature range of interest.
The weakness of this method lies mostly in the theoretical model that, although being amongst the most advanced available to-date, yet do not accurately reproduce observed PMS colour \cite[see, e.g., ][]{Bell_etal:2012,Stauffer_etal:2007}.

Figure\,\ref{fig:Teff_phot_gamma2Vel} (top left panel) shows the comparison between 
$T_{\rm SED}$ and $T_{\rm CC}$
for all GIRAFFE's spectra in the $\gamma$ Vel field for which $\chi_{\rm CC}^2 < 7$ (no significant IR-excess), irrespective of Li- and RV-membership.
Although some systematic deviations in some temperature ranges are present 
(viz., $T_{\rm CC}<$3800\,K), mainly due to the way in which $T_{\rm CC}$ is derived, the two models agree within the error bars, the mean difference is $\simeq -40$\,K, and the standard deviation $\sigma\simeq 150$\,K.  
The comparison of the recommended \teff\ and $T_{\rm SED}$ for GIRAFFE is shown in the bottom left panel of Fig.\,\ref{fig:Teff_phot_gamma2Vel}.
The two sets generally agree within the error bars, the mean difference is $\simeq 180$\,K and $\sigma \simeq 240$\,K.
Finally, the comparison of the UVES recommended \teff\ with $T_{\rm SED}$ for $\gamma$ Vel radial velocity members \citep[as in][]{2014A&A...567A..55S} is shown in the bottom right panel of Fig.\,\ref{fig:Teff_phot_gamma2Vel}.

Note, finally, that the intrinsic variability of the targets together the non-simultaneity of the spectroscopic and photometric observations must also play a role in the comparisons presented here.
Considering also uncertainties and likely spread in age, which are not reliably estimated as yet, the comparison with photometry can be considered quite satisfactory. 

\end{document}